\pdfoutput=1
\newcommand*{\ATLASLATEXPATH}{latex/}
\documentclass[cernpreprint,UKenglish,texlive=2011,txfonts=true]{\ATLASLATEXPATH atlasdoc}

\usepackage[block=space]{\ATLASLATEXPATH atlaspackage}
\usepackage{\ATLASLATEXPATH atlasbiblatex}

\usepackage{\ATLASLATEXPATH atlascontribute}

\usepackage{\ATLASLATEXPATH atlasphysics}

\graphicspath{{logos/}{figures/}}

\usepackage{main-defs}
\usepackage{subfig} 
\usepackage{dcolumn}
\usepackage{longtable}
\usepackage{booktabs}
\usepackage{tabularx}

\AtlasTitle{Search for flavour-changing neutral current top quark decays $\boldsymbol{t\to Hq}$ in $\boldsymbol{pp}$ collisions at $\sqrt{s}=8\tev$ with the ATLAS detector}

\author{The ATLAS Collaboration}

\AtlasRefCode{TOPQ-2014-14}

\PreprintIdNumber{CERN-PH-EP-2015-229}

\AtlasJournal{JHEP}

\AtlasAbstract{
A search for flavour-changing neutral current decays of a top quark to an up-type quark ($q=u, c$) and the 
Standard Model Higgs boson, where the Higgs boson decays to $b\bar{b}$, is presented. 
The analysis searches for top quark pair events in which one top quark 
decays to $Wb$, with the $W$ boson decaying leptonically, and the other top quark decays to $Hq$.
The search is based on $pp$ collisions at $\sqrt{s}=8\tev$ recorded in 2012 with the ATLAS detector at the CERN Large Hadron Collider 
and uses an integrated luminosity of 20.3 fb$^{-1}$.
Data are analysed in the lepton-plus-jets final state, characterised by an isolated electron or muon 
and at least four jets. The search exploits the high multiplicity of $b$-quark jets characteristic of signal events, and employs a likelihood discriminant that uses the kinematic differences between the signal and the background, which is dominated by $t\bar{t} \to WbWb$ decays. 
No significant excess of events above the background expectation is found, and observed (expected) 95\% CL upper limits 
of 0.56\% (0.42\%) and 0.61\% (0.64\%) are derived for the $t\to Hc$ and $t\to Hu$ branching ratios respectively.
The combination of this search with other ATLAS searches in the $H \to \gamma\gamma$ and $H\to WW^*, \tau\tau$ decay modes significantly improves the sensitivity, yielding
observed (expected) 95\% CL upper limits on the $t\to Hc$ and $t\to Hu$ branching ratios of 0.46\% (0.25\%) and 0.45\% (0.29\%) respectively.
The corresponding combined observed (expected) upper limits on the $|\lambda_{tcH}|$ and $|\lambda_{tuH}|$ couplings are 0.13 (0.10) and 0.13 (0.10) respectively. 
These are the most restrictive direct bounds on $tqH$ interactions measured so far.
}

\AtlasCoverSupportingNote{Search for $\ttbar \to WbHq$, $H\to\bbbar$}{https://cds.cern.ch/record/1742309}
\AtlasCoverSupportingNote{Search for $\ttbar \to WbHq$, $H\to\gamma\gamma$}{https://cds.cern.ch/record/1613813}
\AtlasCoverSupportingNote{Search for $\ttbar \to WbHq$, $H\to W^+W^-, \tau^+\tau^-$}{https://cds.cern.ch/record/2026565}
\AtlasCoverCommentsDeadline{Comments deadline}

\AtlasCoverAnalysisTeam{
$H \to b\bar{b}$: Sebastian Grinstein, Aurelio Juste, Eve Le Menedeu, Javier Montejo, Francesco Rubbo, and Shota Tsiskaridze; 
$H \to WW^*, \tau\tau$: David Hohn, Marine Kuna, Rohin Narayan, Peter Onyisi, and Frank Seifert;
$H \to \gamma\gamma$: Jean-Baptiste de Vivie and Daniel Fournier;
Combination: Hongtao Yang}

\AtlasCoverEdBoardMember{Pascal Pralavorio~(chair), Oleg Brandt, Fabrice Hubaut, Filipe Veloso}

\AtlasCoverEgroupEditors{atlas-TOPQ-2014-14-editors@cern.ch}

\AtlasCoverEgroupEdBoard{atlas-TOPQ-2014-14-editorial-board@cern.ch}

\hypersetup{pdftitle={ATLAS draft},pdfauthor={The ATLAS Collaboration}}

\begin{document}

\maketitle

\tableofcontents


\section{Introduction}
\label{sec:intro}

Following the observation of a Higgs boson by the ATLAS and CMS collaborations~\cite{Aad:2012tfa,Chatrchyan:2012ufa},
a comprehensive programme of measurements of its properties is underway looking for deviations from the Standard Model (SM)
predictions.  An interesting possibility is the presence of flavour-changing neutral current (FCNC) interactions between the Higgs boson, 
the top quark, and a $u$- or $c$-quark, $tqH$ ($q=u,c$). Since the Higgs boson is lighter than the top quark, with a measured mass 
$m_H=125.09 \pm 0.24\gev$~\cite{Aad:2015zhl}, such interactions would manifest themselves as FCNC top quark decays, $t\to H q$.  
In the SM, such decays are extremely suppressed relative to the dominant $t\to Wb$ decay mode, since $tqH$ 
interactions are forbidden at the tree level and even suppressed at higher-orders in the perturbative expansion due to the 
Glashow--Iliopoulos--Maiani (GIM) mechanism~\cite{Glashow:1970gm}.  As a result, the SM predictions for the $t \to Hq$ branching 
ratios are exceedingly small: $\BR(t\to Hu) \sim 10^{-17} $ and $\BR(t\to Hc) \sim 10^{-15}$~\cite{Eilam:1990zc,Mele:1998ag,AguilarSaavedra:2004wm,Zhang:2013xya}.
On the other hand, large enhancements in these branching ratios are possible in some beyond-SM scenarios, where the GIM suppression can be relaxed
and/or new particles can contribute to the loops, yielding effective couplings orders of magnitude larger than those of the SM. Examples
include quark-singlet models~\cite{AguilarSaavedra:2002kr}, two-Higgs-doublet models (2HDM) of type I, with explicit flavour conservation,
and of type II, such as the minimal supersymmetric SM (MSSM)~\cite{Bejar:2000ub, Guasch:1999jp,Cao:2007dk}, or supersymmetric models
with R-parity violation~\cite{Eilam:2001dh}. In those scenarios, typical branching ratios can be as high as $\BR(t\to Hq) \sim 10^{-5}$. 
An even larger branching ratio of  $\BR(t\to Hc) \sim 10^{-3}$ can be reached in 2HDM without explicit flavour conservation (type III),
since a tree-level FCNC coupling is not forbidden by any symmetry~\cite{Cheng:1987rs,Baum:2008qm,Chen:2013qta}. While other FCNC top couplings, $tq\gamma$, $tqZ$, $tqg$, are also
enhanced relative to the SM prediction in those scenarios beyond the SM, the largest enhancements are typically for the $tqH$ couplings, and in
particular the $tcH$ coupling. See ref.~\cite{AguilarSaavedra:2004wm} for a review.

Searches for $t \to Hq$ decays have been performed  by the ATLAS and CMS collaborations, taking advantage of the large samples
of $t\bar{t}$ events collected during Run 1 of the LHC. In these searches, one of the top quarks is required to decay into $Wb$,
while the other top quark decays into $Hq$, yielding $\ttbar \to WbHq$.\footnote{ 
In the following $WbHq$ is used to denote both $W^+b H\bar{q}$ and its charge conjugate, $HqW^- \bar{b}$. Similarly, 
$WbWb$ is used to denote $W^+b W^- \bar{b}$.}  Assuming SM decays for the Higgs boson and $m_H=125\gev$, the most sensitive single-channel 
searches have been performed in the $H\to\gamma\gamma$ decay mode which, despite the tiny branching ratio of
$\BR(H\to \gamma\gamma)\simeq 0.2\%$, is characterised by very small background and excellent diphoton mass re\-so\-lu\-tion.
The resulting observed (expected) 95\% confidence level (CL) upper limits on $\BR(t\to Hq)$ are 0.79\% (0.51\%) and 0.69\% (0.81\%), respectively from  
the ATLAS~\cite{Aad:2014dya} and CMS~\cite{Khachatryan:2014jya} collaborations. These searches are insensitive to the difference between
$t \to Hu$ and $t \to Hc$, and thus the above limits can be interpreted as applying to the sum $\BR(t\to Hu)$+$\BR(t\to Hc)$.
The CMS Collaboration has also reinterpreted searches in multilepton (three or four leptons)
final states~\cite{Khachatryan:2014jya} in the context of $t\bar{t}\to WbHq$ with $H\to WW^*, \tau\tau$, resulting in an observed (expected) 
upper limit of $\BR(t\to Hc)<1.28\%\;(1.17\%)$ at the 95\% CL. 
Multilepton searches are able to exploit a significantly larger branching ratio for the Higgs boson decay compared to the $H\to\gamma\gamma$ decay mode, and
are also characterised by relatively small backgrounds.
 However, in general they do not have good mass resolution,\footnote{An exception is the $H\to ZZ^*\to \ell^+\ell^- \ell^{\prime +}\ell^{\prime -}$ 
($\ell, \ell^\prime = e, \mu$) decay mode, which has a very small branching ratio and thus is not promising for this search.} 
so any excess would be hard to interpret as originating from $t \to Hq$ decays.
The combination of CMS searches in diphoton and multilepton (three or four leptons) final states yields an observed (expected) 
upper limit of $\BR(t\to Hc)<0.56\%\;(0.65\%)$ at the 95\% CL~\cite{Khachatryan:2014jya}.

Upper limits on the branching ratios $\BR(t\to Hq)$ ($q=u,c$) can be translated to upper limits on the non-flavour-diagonal Yukawa couplings $\lamHq$ 
appearing in the following Lagrangian:
\begin{equation}
{\cal L}_{\rm FCNC} = \lambda_{tcH} \bar{t}Hc + \lambda_{tuH} \bar{t}Hu + h.c.
\end{equation}
The branching ratio $\BR(t\to Hq)$ is estimated as the ratio of its partial width~\cite{Zhang:2013xya} to the SM $t \to Wb$ partial width~\cite{Denner:1990ns}, 
which is assumed to be dominant. Both predicted partial widths include next-to-leading-order (NLO) QCD corrections.
Using the expression derived in ref.~\cite{Aad:2014dya}, the coupling $|\lamHq|$ can be extracted as $| \lamHq | = (1.92 \pm 0.02) \sqrt{\BR(t\to Hq)}$.

The results presented in this paper fill a gap in the current programme of searches for $t \to Hq$ decays at the LHC by considering the dominant decay mode $H\to b\bar{b}$,  
which has $\BR(H\to b\bar{b})\simeq 58\%$. This search is focused on the $t\bar{t}\to Wb Hq$ ($q=u,c$) process, with $W\to\ell\nu$ ($\ell = e, \mu, \tau$) and $H\to b\bar{b}$, 
resulting in a lepton-plus-jets final state with high $b$-jet multiplicity, which can be effectively exploited to suppress the overwhelming 
$t\bar{t}$ background. Early studies of the prospects for this search at the LHC were performed in ref.~\cite{AguilarSaavedra:2000aj}.
Only events with an electron or muon, including those produced via leptonically decaying taus, are considered.
The lepton-plus-jets final state also allows the kinematic reconstruction of the final state and in particular the dijet invariant mass spectrum from the 
$H\to b\bar{b}$ decay, providing additional handles that would help in detecting $t\bar{t}\to Wb Hq$ events.
Most of this paper is devoted to the discussion of this particular search, for which 
background estimation techniques, systematic uncertainties and statistical treatment closely follow those used
in recent ATLAS searches using the same final-state signature~\cite{Aad:2015gra,Aad:2015kqa}.
This paper also includes a reinterpretation of the ATLAS search for $t\bar{t}H$ associated production, with $H \to WW^*, ZZ^*, \tau\tau$, 
resulting in multilepton final states~\cite{Aad:2015iha}. This reinterpretation only considers the final states with a significant expected
contribution from $t\bar{t}\to WbHq$, $H \to WW^*, \tau\tau$ signal, namely two same-charge leptons with and without an identified 
hadronic tau lepton and three leptons. A combination of the 
three ATLAS searches for $t\bar{t}\to WbHq$, probing the  $H\to b\bar{b}$, $H \to WW^*, \tau\tau$, and $H \to \gamma\gamma$ decay modes, 
is performed and bounds are set on $\BR(t\to Hc)$ and $\BR(t\to Hu)$, as well as on the corresponding non-flavour-diagonal 
Yukawa couplings. 

This paper is organised as follows. A brief description of the ATLAS detector is provided in section~\ref{sec:detector}.
Subsequent sections are devoted to a detailed discussion of the $\Hqbb$ search, covering the object reconstruction 
(section~\ref{sec:object_reco}), the data sample and event preselection (section~\ref{sec:data_presel}), the modelling of the
backgrounds and the signal (section~\ref{sec:simulated_samples}), the analysis strategy (section~\ref{sec:analysis_strategy}),
and the systematic uncertainties (section~\ref{sec:systematics}). Section~\ref{sec:stat_analysis} provides a discussion of the
statistical methods used. Section~\ref{sec:result}  presents the results obtained by the three individual ATLAS searches as well as their combination.
Finally, the conclusions are given in section~\ref{sec:conclusion}.

\section{ATLAS detector}
\label{sec:detector}

The ATLAS detector~\cite{PERF-2007-01} consists of the following main subsystems: an inner tracking system, 
electromagnetic and hadronic calorimeters, and a muon spectrometer.
The inner detector provides tracking information from silicon pixel and microstrip detectors in the pseudorapidity\footnote{ATLAS 
uses a right-handed coordinate system with its origin at the nominal interaction point (IP) in the 
centre of the detector and the $z$-axis coinciding with the axis of the beam pipe.  The $x$-axis points from
the IP to the centre of the LHC ring, and the $y$-axis points upward. Cylindrical coordinates ($r$,$\phi$) are used 
in the transverse plane, $\phi$ being the azimuthal angle around the beam pipe. The pseudorapidity is defined in 
terms of the polar angle $\theta$ as $\eta = - \ln \tan(\theta/2)$.} 
range $|\eta|<2.5$ and from a straw-tube transition radiation tracker covering $|\eta|<2.0$, all immersed in a 2 T axial magnetic field provided 
by a superconducting solenoid.  The electromagnetic (EM) sampling calorimeter uses lead as the absorber material 
and liquid-argon (LAr) as the active medium, and is divided into barrel ($|\eta|<1.475$) and end-cap ($1.375<|\eta|<3.2$) regions.  
Hadron calorimetry is also based on the sampling technique, with either scintillator tiles or LAr as the active medium, and with 
steel, copper, or tungsten as the absorber material. The calorimeters cover $|\eta|<4.9$. The muon spectrometer measures the deflection 
of muons with $|\eta|<2.7$ using multiple layers of high-precision tracking chambers located in a toroidal field of 
approximately 0.5~T and 1~T in the central and end-cap regions of ATLAS, respectively. The muon spectrometer is also 
instrumented with separate trigger chambers covering $|\eta|<2.4$.
A three-level trigger system~\cite{PERF-2011-02} is used to select interesting events.
The first-level trigger is implemented in custom electronics and uses a subset of detector information to reduce the event rate to at most 75~kHz.
This is followed by two software-based trigger levels exploiting the full detector information and yielding a 
typical recorded event rate of 400~Hz during 2012.

\section{Object reconstruction}
\label{sec:object_reco}

Electron candidates~\cite{Aad:2014fxa} are reconstructed from energy clusters in the EM
calorimeter that are matched to reconstructed tracks in the inner detector.  
Electron clusters are required to have a transverse energy  $\et$ greater than $25\gev$ 
and $|\eta_{\rm cluster}| < 2.47$, excluding the transition region $1.37 < |\eta_{\rm cluster}| < 1.52$ 
between sections of the EM calorimeter. 
The longitudinal impact parameter of the electron track with respect to the event's primary
vertex (see section~\ref{sec:data_presel}), $z_{0}$, is required to be less than 2 mm. 
Electrons are required to satisfy ``tight'' quality requirements~\cite{Aad:2014fxa} based 
on calorimeter,  tracking and combined variables that provide good separation between prompt electrons and jets.
To reduce the background from non-prompt electrons resulting from semileptonic decays of $b$- or $c$-hadrons, and 
from jets with a high fraction of their energy deposited in the EM calorimeter, 
electron candidates must also satisfy calorimeter- and track-based isolation requirements.
The calorimeter isolation variable is based on the energy sum of cells  within a cone of size   
$\Delta R = \sqrt{(\Delta\phi)^2 + (\Delta\eta)^2} = 0.2$ around the direction of each electron candidate, 
and an $\eta$-dependent requirement is made, giving an average efficiency of 90\% across $\eta$ for prompt electrons from $Z$ boson decays.
This energy sum excludes cells associated with the electron cluster and is corrected for
leakage from the electron cluster itself as well as for energy deposits from additional $pp$ 
interactions within the same bunch crossing (``pileup'').
A further 90\%-efficient isolation requirement is made on the track transverse momentum ($\pt$) sum around the
electron (excluding the electron track itself) in a cone of size $\Delta R = 0.3$.

Muon candidates~\cite{Aad:2014zya,Aad:2014rra} are reconstructed from track segments in the various layers of the muon spectrometer 
that are matched with tracks found in the inner detector.  The final candidates are refitted using the complete
track information from both detector systems and are required to have $\pt > 25\gev$ and $|\eta|<2.5$. 
The longitudinal impact parameter of the muon track with respect to the primary vertex, $z_{0}$, is required to be less than 2 mm.
Muons are required to satisfy a $\pt$-dependent track-based isolation
requirement: the scalar sum of the $\pt$ of the tracks within a cone of 
variable size $\Delta R=10\gev/\pt^\mu$ around the muon (excluding the muon track itself) must be less than 5\% of the muon $\pt$ ($\pt^\mu$). 
This requirement has good signal efficiency and background rejection even under high-pileup conditions, 
as well as in boosted configurations where the muon is close to a jet.  For muons from $W$ boson decays in simulated $\ttbar$ events, 
the average efficiency of the isolation requirement is about 95\%.

Jets are reconstructed with the anti-$k_t$
algorithm~\cite{Cacciari:2008gp,Cacciari:2005hq,Cacciari:2011ma} with a
radius parameter $R=0.4$, using ca\-li\-bra\-ted topological
clusters~\cite{Cojocaru:2004jk,topoclusters} built from energy deposits in the
calorimeters.  Prior to jet finding, a local cluster calibration scheme~\cite{Aad:2011he}
is applied to correct the topological cluster energies for the non-compensating response of the calorimeter,
as well as for the energy lost in dead material and via out-of-cluster leakage. The corrections are obtained
from simulations of charged and neutral particles. 
After energy calibration~\cite{Aad:2014bia}, jets are required to have 
$\pt > 25\gev$ and $|\eta| < 2.5$.  
To reduce the contamination due to jets originating from pileup interactions,
a requirement on the absolute value of the jet vertex fraction (JVF) variable above 0.5 
is applied to jets with $\pt<50\gev$ and $|\eta|<2.4$.
This requirement ensures that at least 50\% of the scalar sum of the $\pt$ of the tracks 
with $\pt>1\gev$ associated with a jet comes from tracks originating from the primary vertex.
During jet reconstruction, no distinction is made between identified
electrons and jet energy deposits.  Therefore, if any of the jets lie 
within $\Delta R=0.2$ of a selected electron, the closest jet
is discarded in order to avoid double-counting of electrons as jets.
Finally, any electron or muon within $\Delta R=0.4$ of a selected jet is discarded.

Jets containing $b$-hadrons are identified ($b$-tagged) 
via an algorithm~\cite{CLTaggingEfficiency} that uses 
multivariate techniques to combine information from the impact
parameters of displaced tracks as well as topological properties of
secondary and tertiary decay vertices reconstructed within the jet.
For each jet, a value for the multivariate $b$-tagging discriminant is calculated.
The jet is considered $b$-tagged if this value is above a given threshold.
The threshold used in this search corresponds to 70\% efficiency to tag
a $b$-quark jet, with a light-jet\footnote{Light-jet denotes a jet originating from the hadronisation of a light quark ($u$, $d$, $s$) or gluon.} 
rejection factor of $\sim$130 and a charm-jet rejection factor of 5, as determined for jets with
$\pt >20\gev$ and $|\eta|<2.5$ in simulated $\ttbar$ events.

The missing transverse momentum ($\met$) is constructed~\cite{Aad:2012re} from the vector sum of all calorimeter energy deposits 
contained in topological clusters. All topological cluster energies are corrected
using the local cluster calibration scheme discussed previously in the context of the jet energy calibration.
Those topological clusters associated with a high-$\pt$ object (e.g. jet or electron)
are further calibrated using their respective energy corrections. 
In addition, contributions from the $\pt$ of selected muons are included in the calculation of $\met$.

\section{Data sample and event preselection}
\label{sec:data_presel}
This search is based on $pp$ collision data at $\sqrt{s}=8\tev$ collected by the ATLAS experiment between April and December 2012. 
Only events recorded with a single-electron or single-muon trigger under stable beam conditions and for which 
all detector subsystems were operational are considered.  The corresponding integrated luminosity is $20.3\pm 0.6$~$\ifb$~\cite{Aad:2013ucp}.
Single-lepton triggers with different $\pt$ thresholds are combined in a logical OR in order to increase the overall efficiency.
The $\pt$ thresholds are 24 or 60~\gev\ for the electron triggers and 24 or 36~\gev\ for the muon triggers.  The triggers with the lower $\pt$
threshold include isolation requirements on the candidate lepton, resulting in inefficiencies at high $\pt$ 
that are recovered by the triggers with higher $\pt$ threshold. 

Events satisfying the trigger selection are required to have at least one reconstructed vertex with at  least five associated tracks 
with $\pt>400\mev$, consistent with originating from the beam collision region in the $x$--$y$ plane. 
The average number of $pp$ interactions per bunch crossing is approximately 20, 
resulting in several vertices reconstructed per event. If more than one vertex is 
found, the hard-scatter primary vertex is taken to be the one which has the largest 
sum of the squared transverse momenta of its associated tracks. For the event 
topologies considered in this paper, this requirement leads to a probability to 
reconstruct and select the correct hard-scatter primary vertex larger than 99\%.

Preselected events are required to have exactly one electron or muon, as defined
in section~\ref{sec:object_reco}, that matches, within $\Delta R=0.15$, the lepton candidate reconstructed by the trigger.
In addition, at least four jets are required, of which at least two must be $b$-tagged.

\section{Background and signal modelling}
\label{sec:simulated_samples}

After the event preselection, the main background is $t\bar{t} \to WbWb$ production, possibly 
in association with jets, denoted by $\ttbar$+jets in the following.
Single top quark production and production of a $W$ boson in association with jets ($W$+jets) 
contribute to a lesser extent. Small contributions arise from multijet, $Z$+jets and diboson
($WW,WZ,ZZ$) production, as well as from the associated production of a vector boson $V$ ($V=W,Z$)
or a Higgs boson and a $\ttbar$ pair ($\ttbar V$ and $\ttbar H$). Signal and all backgrounds are estimated from 
simulation and normalised to their theoretical cross sections, with the exception of the multijet background,
which is estimated with data-driven methods~\cite{Aad:2010ey}.  

Simulated samples of $t\bar{t}$ events are generated with the NLO generator 
{\sc Powheg-Box} 2.0~\cite{Frixione:2007nw,Nason:2004rx,Frixione:2007vw,Alioli:2010xd} using the CT10~\cite{Lai:2010vv} set 
of parton distribution functions (PDF).
The nominal sample is interfaced to {\sc Pythia} 6.425~\cite{Sjostrand:2006za} for parton showering and hadronisation with 
the CTEQ6L1 PDF set and the Perugia2011C~\cite{Skands:2010ak} set of optimised parameters for the underlying event (UE) description, 
referred to as the ``UE tune''. An alternative sample, used to study the uncertainty related to the hadronisation model, is 
interfaced to {\sc Herwig} v6.520~\cite{Corcella:2000bw} with the CTEQ6L1 PDF set and {\sc Jimmy} v4.31~\cite{Butterworth:1996zw} 
to simulate the UE. All samples are generated assuming a top quark mass of $172.5\gev$ and top quark decays exclusively 
through $t \to Wb$. The $\ttbar$ process is normalised to a cross section of $253^{+15}_{-16}$~pb,
computed using {\sc Top++} v2.0~\cite{Czakon:2011xx} at next-to-next-to-leading 
order (NNLO) in QCD, including resummation of next-to-next-to-leading logarithmic (NNLL) soft gluon 
terms~\cite{Cacciari:2011hy,Baernreuther:2012ws,Czakon:2012zr,Czakon:2012pz,Czakon:2013goa}, 
and using the MSTW 2008 NNLO~\cite{Martin:2009iq,Martin:2009bu} PDF set. 
Theoretical uncertainties result from variations of the factorisation and renormalisation scales, as well as from uncertainties on the
PDF and $\alpha_{\rm S}$. The latter two represent the largest contribution to the overall theoretical uncertainty on the cross section 
and were calculated using the PDF4LHC prescription~\cite{Botje:2011sn} 
with the MSTW 2008 68\% CL NNLO, CT10 NNLO~\cite{Lai:2010vv,Gao:2013xoa} and NNPDF2.3 5f FFN~\cite{Ball:2012cx} PDF sets.
In the case where a non-zero $\BR(t \to Hq)$ is assumed, an additional factor of $[1-\BR(t \to Hq)]^2$ is applied to the sample normalisation.
It is not possible to generate the $\ttbar \to WbHq$ signal with {\sc Powheg-Box}, and a different event generator is used instead,
as discussed below.

The $\ttbar$ samples are generated inclusively, but events are categorised depending
on the flavour content of additional particle jets not originating from the decay of the $\ttbar$ system.\footnote{Particle jets are reconstructed 
by clustering stable particles excluding muons and neutrinos using the anti-$k_t$ algorithm with a radius parameter $R=0.4$.}
Details about this categorisation scheme can be found in ref.~\cite{Aad:2015gra}.
In this way, a distinction is made between $\ttbb$, $\ttcc$ and $\ttbar$+light-jets events.
The first two categories are generically referred to as $\ttbar$+HF events (with HF standing for ``heavy flavour''),
while the latter category also includes events with no additional jets.
The modelling of $\ttbar$+HF in {\sc Powheg-Box}+{\sc Pythia} is via the parton-shower evolution. To study 
uncertainties related to this simplified description, an alternative $\ttbar$+jets sample is generated
with {\sc Madgraph5} 1.5.11~\cite{Alwall:2011uj} using the CT10 PDF set. It includes tree-level diagrams with up to three 
additional partons (including $b$- and $c$-quarks) and is interfaced to {\sc Pythia} 6.425.

Since the best possible modelling of the $\ttbar$+jets background is a key aspect of this search, 
a correction is applied to simulated $\ttbar$ events in {\sc Powheg-Box}+{\sc Pythia}
based on the ratio of the differential cross sections measured in data and simulation at $\sqrt{s}=7\tev$ 
as a function of top quark $\pt$ and $\ttbar$ system $\pt$~\cite{Aad:2014zka}.
This correction significantly improves agreement between simulation and data at $\sqrt{s}=8\tev$ in 
distributions such as the jet multiplicity and the $\pt$ of decay products of the $\ttbar$ system~\cite{Aad:2015gra}, and 
is applied only to $\ttbar$+light-jets and $\ttcc$ events.
The modelling of the $\ttbb$ background is improved by reweighting the {\sc Powheg-Box}+{\sc Pythia} prediction to an NLO prediction 
of $\ttbb$ with massive $b$ quarks and including parton showering~\cite{Cascioli:2013era}, based on 
{\sc Sherpa+OpenLoops}~\cite{Gleisberg:2008ta, Cascioli:2011va} using the CT10 PDF set.  
Such treatment is not possible for the $\ttcc$ background since a corresponding NLO prediction is not currently available.
More details about the modelling of the $\ttbar$+jets background can be found in ref.~\cite{Aad:2015gra}.

Samples of single-top-quark backgrounds corresponding to the $t$-channel, $s$-channel, and $Wt$ production mechanisms 
are generated with {\sc Powheg-Box} 2.0~\cite{Alioli:2009je,Re:2010bp} 
using the CT10 PDF set and interfaced to {\sc Pythia} 6.425 with the CTEQ6L1 PDF set in combination with the Perugia2011C UE tune.  
Overlaps between the \ttbar\ and $Wt$ final states are avoided using the ``diagram removal'' scheme~\cite{Frixione:2005vw}.
The single-top-quark samples are normalised to the approximate NNLO theoretical cross sections~\cite{Kidonakis:2011wy,Kidonakis:2010ux,Kidonakis:2010tc}, calculated using the MSTW 2008 NNLO PDF set. 

Samples of $W/Z$+jets events are generated with up to five additional partons using the {\sc Alpgen} v2.14~\cite{Mangano:2002ea} 
LO generator with the CTEQ6L1 PDF set and interfaced to {\sc Pythia} v6.426.
To avoid double-counting of partonic configurations generated by both the matrix-element  calculation and the parton shower, 
a parton--jet matching scheme (``MLM matching'')~\cite{Mangano:2001xp} is employed. 
The $W$+jets samples are generated separately for $W$+light-jets, $Wb\bar{b}$+jets, $Wc\bar{c}$+jets, and $Wc$+jets. 
The $Z$+jets samples are generated separately for $Z$+light-jets, $Zb\bar{b}$+jets, and $Zc\bar{c}$+jets. Overlap between 
$VQ\bar{Q}$+jets ($V=W,Z$ and $Q=b,c$)  events generated from the matrix-element calculation and those generated from parton-shower 
evolution in the $W/Z$+light-jets samples is avoided via an algorithm based on the angular separation between the extra heavy quarks:
if $\Delta R(Q,\bar{Q})>0.4$, the matrix-element prediction is used, otherwise the parton-shower prediction is used. 
Both the $W$+jets and $Z$+jets background contributions are normalised to their inclusive NNLO theoretical cross sections~\cite{Melnikov:2006kv}.
Further corrections are applied to $W/Z$+jets events in order to better describe data in the preselected sample.
Normalisation factors for each of the $W$+jets categories ($Wb\bar{b}$+jets, $Wc\bar{c}$+jets, $Wc$+jets and $W$+light-jets) are derived
for events with one lepton and at least four jets by simultaneously analysing six different event categories, defined by the
$b$-tag multiplicity (0, 1 and $\geq$2) and the sign of the lepton charge~\cite{Aad:2015noh}. The $b$-tag multiplicity provides information 
about the heavy-flavour composition of the $W$+jets background, while the lepton charge is used to determine the normalisation 
of each component, exploiting the expected charge asymmetry for $W$+jets production in $pp$ collisions as predicted by {\sc Alpgen}.
In the case of $Z$+jets events, a correction to the heavy-flavour fraction is derived to reproduce the relative rates of $Z$+2-jets
events with zero and one $b$-tagged jet observed in data. In addition, the $Z$ boson $\pt$ spectrum is compared between
data and the simulation in $Z$+2-jets events, and a reweighting function is derived in order to improve the modelling.
This reweighting function is also applied to the $W$+jets simulated sample and it was verified that this correction further improves 
the agreement between data and simulation for $W$+jets events.
In any case, $W/Z$+jets events constitute a very small background in this analysis after final event selection. 

The $WW/WZ/ZZ$+jets samples are generated with up to three additional partons using {\sc Alpgen} v2.13 and the 
CTEQ6L1 PDF set, interfaced to {\sc Herwig} v6.520 and {\sc Jimmy} v4.31 for parton showering, hadronisation and UE modelling. 
The MLM parton--jet matching scheme is used. The $WW$+jets samples require at least
one of the $W$ bosons to decay leptonically, while the $WZ/ZZ$+jets samples require one $Z$ boson to decay leptonically and 
other boson decays inclusively. Additionally, $WZ$+jets samples requiring the $W$ boson to decay leptonically and 
the $Z$ boson to decay hadronically, are generated with up to three additional partons 
(including massive $b$- and $c$-quarks) using {\sc Sherpa} v1.4.1 and the CT10 PDF set.
All diboson samples are normalised to their NLO theoretical cross sections~\cite{Campbell:1999ah}.

Samples of $\ttbar V$ events, including $\ttbar WW$, are generated with up to two additional partons using {\sc Madgraph5} 1.3.28 
with the CTEQ6L1 PDF set, and interfaced to {\sc Pythia} 6.425 with the AUET2B UE tune~\cite{ATLASUETune1}.
A sample of $\ttbar H$ events is generated with the {\sc PowHel} framework~\cite{Garzelli:2011vp}, which combines the {\sc Powheg-Box} generator and
NLO matrix elements obtained from the HELAC-Oneloop package~\cite{Bevilacqua:2011xh}. The sample is generated
using the CT10nlo PDF set~\cite{Lai:2010vv}. Showering is performed with
{\sc Pythia} 8.1~\cite{Sjostrand:2007gs}  using the CTEQ6L1 PDF set and the AU2 UE tune~\cite{ATLASUETune1,ATLASUETune2}. 
Inclusive decays of the Higgs boson are assumed in the generation of the $\ttbar H$ sample.
The $\ttbar V$ samples are normalised to the NLO cross-section predictions~\cite{Garzelli:2012bn}.
The $\ttbar H$ sample is normalised using the NLO cross section~\cite{Dawson:2003zu,Beenakker:2002nc,Beenakker:2001rj}  
and the Higgs decay branching ratios~\cite{Djouadi:1997yw,Bredenstein:2006rh,Actis:2008ts,Denner:2011mq} collected in ref.~\cite{Dittmaier:2011ti}.

The multijet background contributes to the selected data sample via several
production and misreconstruction mechanisms.  In the electron
channel, it consists of non-prompt electrons (from semileptonic
$b$- or $c$-hadron decays) as well as misidentified photons 
(e.g.~from a conversion of a photon into an $e^+e^-$ pair) or jets with a high fraction of 
their energy deposited in the EM calorimeter.  
In the muon channel, the multijet background is predominantly from non-prompt muons.
Its normalisation and shape are estimated directly from data by using the ``matrix method''
technique~\cite{Aad:2010ey}, which exploits differences in lepton-identification-related 
properties between prompt and isolated leptons and leptons that are either non-isolated or result from the
misidentification of photons or jets. Further details can be found in ref.~\cite{Aad:2015kqa}.

The $\ttbar \to WbHq$ signal process is modelled using the {\sc Protos}~v2.2~\cite{protos,AguilarSaavedra:2009mx} 
LO generator with the {\sc CTEQ6L1} PDF set,
and interfaced to {\sc Pythia} 6.426 and the Perugia2011C UE tune. Two separate samples are generated corresponding to 
$\ttbar \to WbHc$ and $\ttbar \to WbHu$, with the $W$ boson forced to decay leptonically, $W \to \ell \nu$ ($\ell=e, \mu,\tau$),
The top quark and Higgs boson masses are set to $172.5\gev$ and $125\gev$, respectively. The Higgs boson is allowed 
to decay to all SM particles with branching ratios as given in ref.~\cite{Dittmaier:2011ti}. The signal sample is normalised to the
same NNLO cross section as used for the $t\bar{t}\to WbWb$ sample, and the corresponding branching ratios:
$\sigma(t\bar{t}\to W(\to \ell\nu)bHq)=2\BR(t \to Hq)[1-\BR(t \to Hq)]\BR(W \to \ell\nu)\sigma_{\ttbar}$, with
$\BR(W \to \ell\nu)=0.324$ and $\BR(t \to Hq)$ depending on the branching ratio being tested.
Typically a reference branching ratio of $\BR(t \to Hq)=1\%$ is used.
The case of both top quarks decaying into $Hq$ is neglected in the analysis given existing 
upper limits on $\BR(t \to Hq)$ (see section~\ref{sec:intro}).
In order to improve the modelling of the signal kinematics, a two-step reweighting procedure is applied:
the first step is designed to correct the spectrum of top quark $\pt$ and $\ttbar$ system $\pt$ to match that 
of the uncorrected $\ttbar \to WbWb$ {\sc Powheg-Box}+{\sc Pythia} sample; the second step involves the
same correction to the top quark $\pt$ and $\ttbar$ system $\pt$ applied to the $\ttbar$+jets background 
(see discussion above). 

Finally, all generated samples are processed through a simulation~\cite{Aad:2010ah} of the detector geometry and res\-pon\-se 
using {\sc Geant4}~\cite{Agostinelli:2002hh}. Additional minimum-bias $pp$ interactions are simulated with 
the {\sc Pythia} 8.1 generator with the MSTW 2008 LO PDF set and the A2 UE tune~\cite{ATLASUETune3}.
They are overlaid on the simulated signal and background events according to the luminosity profile of the recorded data. 
The contributions from these pileup interactions are modelled both within the same bunch crossing as the 
hard-scattering process and in neighbouring bunch crossings.  
All simulated samples are processed through the same reconstruction software as the data. 
Simulated events are corrected so that the object identification efficiencies, energy
scales, and energy resolutions match those determined from data control samples.

\section{Analysis strategy}
\label{sec:analysis_strategy}

This section presents an overview of the analysis strategy followed by the $\Hqbb$ search.

\subsection{Event categorisation}
\label{sec:event_categorisation}

Given the focus on the $W\to\ell\nu$ and $H\to b\bar{b}$ decay modes, 
the $\ttbar \to WbHq$ signal is expected to have typically four jets, of which three or four are $b$-tagged.
The latter case corresponds to the $\ttbar \to WbHc$ signal where the charm quark, as well as the three $b$-quark jets, are $b$-tagged.
Additional jets can also be present because of initial- or final-state radiation.
In order to optimise the sensitivity of the search, the selected events are categorised into different
channels depending on the number of jets (4, 5 and $\geq$6) and on the number of $b$-tagged jets (2, 3 and $\geq$4).
Therefore, the total number of analysis channels considered in this search is nine:
(4 j, 2 b), (4 j, 3 b), (4 j, 4 b), (5 j, 2 b), (5 j, 3 b), (5 j, $\geq$4 b), ($\geq$6 j, 2 b), ($\geq$6 j, 3 b), and ($\geq$6 j, $\geq$4 b), 
where ($n$ j, $m$ b) indicates $n$ selected jets and $m$ $b$-tagged jets.

The overall rate and composition of the $\ttbar$+jets background strongly depends on the jet and $b$-tag multiplicities,
as illustrated in figure~\ref{fig:Yields_BR001}.
The $\ttbar$+light-jets background is dominant in events with exactly two or three $b$-tagged jets, with the two $b$-quarks
from the top quark decays being tagged in both cases, and a charm quark from the hadronic $W$ boson decay also being 
tagged in the latter case. Contributions from $\ttcc$ and $\ttbb$ become significant as the jet and $b$-tag multiplicities
increase, with the $\ttbb$ background  being dominant for events with $\geq$6 jets and $\geq$4 $b$-tags. 

In the channels with four or five jets and three or at least four $b$-tags, which dominate the sensitivity of this search, 
selected signal events have a $H \to b\bar{b}$ decay in more than 95\% of the events.
The channels most sensitive to the $\ttbar \to WbHu$ and $\ttbar \to WbHc$ signals are (4 j, 3 b) and (4 j, 4 b) respectively.
Because of the better signal-to-background ratio in the (4 j, 4 b) channel, this analysis is expected to 
have better sensitivity for $\ttbar \to WbHc$ than for $\ttbar \to WbHu$ signal.
The rest of the channels have significantly lower signal-to-background ratios, but they are useful for calibrating 
the $t\bar{t}$+jets background prediction and constraining the related systematic uncertainties (see section~\ref{sec:systematics})
through a likelihood fit to data (see section~\ref{sec:stat_analysis}). This strategy was first used in the ATLAS search 
for $t\bar{t}H$ associated production, with $H \to b\bar{b}$~\cite{Aad:2015gra}, and is adopted in this analysis.
A table summarising the observed and expected yields before the fit to data in each of the analysis channels 
can be found in appendix~\ref{sec:prepostfit_yields_appendix}.

\begin{figure*}[t]
\begin{center}
\includegraphics[width=0.55\textwidth]{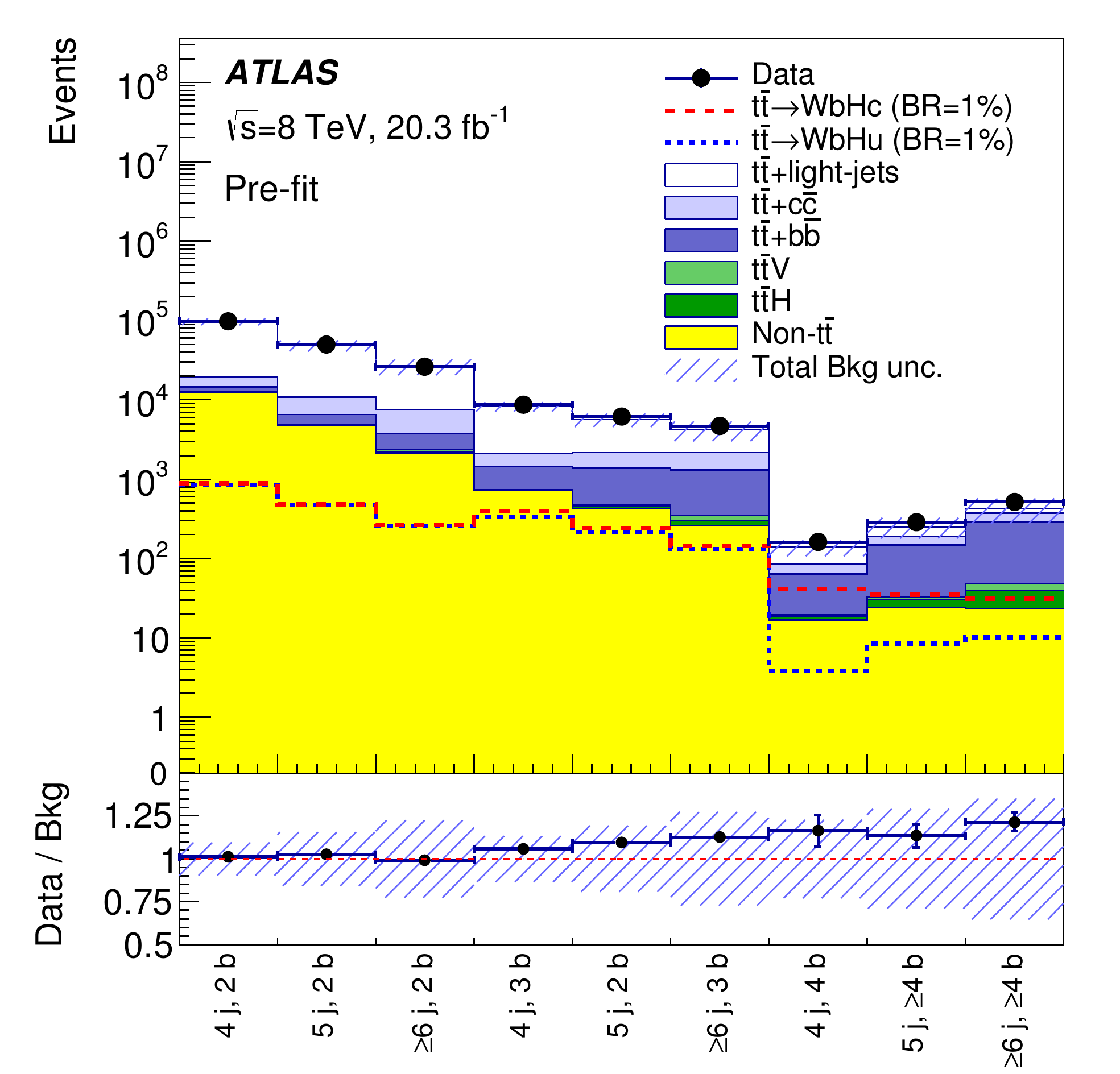}
\caption{Comparison between the data and background prediction for the yields in each of the analysis channels considered 
before the fit to data (pre-fit). Backgrounds are normalised to their nominal cross sections discussed in section~\ref{sec:simulated_samples}.
The expected $\Hc$ and $\Hu$ signals (dashed histograms) are shown separately normalised to $\BR(t\to Hq)=1\%$.
The $\ttbar \to WbWb$ background is normalised to the SM prediction.
The small contributions from $W/Z$+jets,  single top, diboson and multijet backgrounds are combined into a single background source 
referred to as ``Non-$\ttbar$''. 
The bottom panel displays the ratio of data to the SM background (``Bkg'') prediction. 
The hashed area represents the total uncertainty on the background.} 
\label{fig:Yields_BR001}
\end{center}
\end{figure*}

\subsection{Discrimination of signal from background}
After event categorisation, the signal-to-background ratio is very low even in the most sensitive analysis channels, and a suitable
discriminating variable between signal and background needs to be constructed in order to improve the sensitivity of the search.
A powerful discriminant between signal and background can be defined as:
\begin{equation}
D({\bf x}) = \frac{P^{\rm sig}({\bf x}) }{P^{\rm sig}({\bf x}) +P^{\rm bkg}({\bf x}) },
\label{eq:D}
\end{equation}
where $P^{\rm sig}({\bf x}) $ and $P^{\rm bkg}({\bf x}) $ represent the probability density functions (pdf) of a given event under
the signal hypothesis ($\ttbar \to WbHq$) and under the background hypothesis ($\ttbar \to WbWb$) respectively.
Both pdfs  are functions of ${\bf x}$, representing the four-momentum vectors of all final-state particles at the reconstruction level:
the lepton ($\ell$), the neutrino ($\nu$; reconstructed as discussed below), and the $N_{\rm jets}$ selected jets in a given analysis channel.

Since both signal and background result from the $\ttbar$ decay, there are few experimental handles available to discriminate
between them. The most prominent features are the different resonances present in the
decay (i.e. the Higgs boson in the case of $\Hq$ and a hadronically decaying $W$ boson in the case of $\ttbar \to WbWb$),
and the different flavour content of the jets forming those resonances.
This is the main information exploited in the construction of $P^{\rm sig}({\bf x})$ and $P^{\rm bkg}({\bf x})$ in this analysis, so that 
${\bf x}$ is extended to include not only the four-momenta of jets ${\bf p}_{\rm jet}$, but also the value of their multivariate $b$-tagging 
dis\-cri\-mi\-nant ${\bf w}_{\rm jet}$, i.e., ${\bf x}\equiv \{p_\ell, p_\nu, (p_{{\rm jet}_i}, w_{{\rm jet}_i})\}$ ($i=1, \ldots, N_{\rm jets}$).
There is also some angular information from the different spins of the daughter resonances (Higgs and $W$ boson) 
that could be exploited, but it is expected to be subleading in importance and is neglected in this analysis. 

The calculation of $P^{\rm sig}({\bf x})$ and $P^{\rm bkg}({\bf x})$ is discussed in detail in sections~\ref{sec:psig} and~\ref{sec:pbkg} respectively.
In the following, $b_\ell$ denotes the $b$-quark jet from the semileptonic top quark decay,  $q_h$ and $b_h$
denote the light-quark jet ($q_h = u$ or $c$)  and $b$-quark jet from the hadronic top quark decay in background and signal events respectively,
$q_1$ and $q_2$ denote the up-type-quark jet ($u$ or $c$) and down-type-quark jet ($d$ or $s$) from the $W$ boson decay respectively,
and $b_1$ and $b_2$ denote the two $b$-quark jets from the Higgs boson decay.
The level of separation achieved between signal and background with the resulting discriminant  $D$  is illustrated in section~\ref{sec:D}.

\subsubsection{Signal probability}
\label{sec:psig}
The construction of $P^{\rm sig}({\bf x})$ will now be described step by step to illustrate the method. 
If the partonic origin of each jet were known [see figure~\ref{fig:cartoon}(a)], $P^{\rm sig}({\bf x})$ would be defined in this analysis 
as the product of the normalised pdfs for each of the reconstructed invariant masses
in the event: the semileptonic top quark mass ($M_{\ell\nu b_{\ell}}$), the hadronic top quark mass ($M_{b_{1}b_{2}q_{h}}$) and the
Higgs boson mass ($M_{b_{1}b_{2}}$). Since $M_{b_{1}b_{2}q_{h}}$ and $M_{b_{1}b_{2}}$ are correlated, their
difference in quadrature, $X_{b_{1}b_{2}q_{h}} \equiv M_{b_{1}b_{2}q_{h}} \ominus M_{b_{1}b_{2}}$, is used instead of $M_{b_{1}b_{2}q_{h}}$.
Therefore the expression for $P^{\rm sig}$ just making use of the above kinematic information, denoted by $P^{\rm sig}_{\rm kin}$, is:
\begin{equation}
P^{\rm sig}_{\rm kin}({\bf x}) = P^{\rm sig}(M_{\ell\nu b_{\ell}})P^{\rm sig}(X_{b_{1}b_{2}q_{h}}) P^{\rm sig}(M_{b_{1}b_{2}}).
\label{eq:psig_kin}
\end{equation}

The distributions of these invariant masses are obtained from simulated signal events using the reconstructed lepton and/or jets 
corresponding to the correct parton--jet assignment, determined by matching a given quark (before final-state radiation) to the closest jet 
with $\Delta R<0.3$.  The corresponding pdfs are constructed as unit-normalised one-dimensional histograms. To compute 
$M_{\ell\nu b_{\ell}}$, the neutrino four-momentum is needed, which is reconstructed as follows.
Initially, the $x$ and $y$ components of the neutrino momentum, $p_{x,\nu}$ and $p_{y,\nu}$, are identified
with those of the reconstructed $\met$ vector. The $z$ component of the neutrino momentum, $p_{z,\nu}$, is inferred
by solving $M_{W}^{2} = (p_{\ell} + p_{\nu})^2$, with $M_W=80.4\gev$ being the $W$ boson mass.
If two real solutions (``2sol'') exist, they are sorted according to their absolute value of $|p_{z,\nu}|$ i.e., $|p_{z,\nu 1} |< |p_{z, \nu 2}|$. 
It is found that in 62\% of the cases $p_{z,\nu 1}$ is closer than $p_{z, \nu 2}$ to the generator-level neutrino $p_{z, \nu}$.
In this case, two different pdfs are constructed, one for each solution, and $P^{\rm sig}_{\rm 2sol} (M_{\ell\nu b_{\ell}})$ is defined as
the average of the two pdfs weighted by their fractions (0.62 for $p_{z,\nu 1}$ and 0.38 for $p_{z,\nu 2}$).
If no real solution  (``nosol'')  exists, which happens in about 30\% of the cases, the $p_{x,\nu}$ and $p_{y,\nu}$ components are scaled by a 
common factor until the discriminant of the quadratic equation is exactly zero, yielding only one solution for $p_{z,\nu}$. This solution 
for $p_{z,\nu}$ is used to compute $M_{\ell\nu b_{\ell}}$, from which the corresponding $P^{\rm sig}_{\rm nosol}(M_{\ell\nu b_{\ell}})$ is constructed.
In the calculation of  $P^{\rm sig}_{\rm kin}({\bf x})$ from equation~(\ref{eq:psig_kin}), $P^{\rm sig}(M_{\ell\nu b_{\ell}})$ is identified with $P^{\rm sig}_{\rm 2sol} (M_{\ell\nu b_{\ell}})$ or with 
$P^{\rm sig}_{\rm nosol}(M_{\ell\nu b_{\ell}})$, depending on how many neutrino solutions can be found for the event.

In practice, the partonic origin of the jets is not known, so it is necessary to evaluate $P^{\rm sig}({\bf x})$ by averaging over the
$N_p$ possible parton--jet assignments, which dilutes the kinematic information. At this point $b$-tagging information can
be used to suppress the impact from parton--jet assignments that are inconsistent with the correct parton flavours as follows:
\begin{equation}
P^{\rm sig}({\bf x}) = \frac{\sum\limits_{k=1}^{N_p} P^{\rm sig}_{\rm btag}({\bf x}^k)  P^{\rm sig}_{\rm kin}({\bf x}^k)}{\sum\limits_{k=1}^{N_p} P^{\rm sig}_{\rm btag}({\bf x}^k)},
\label{eq:psig}
\end{equation}
where $P^{\rm sig}_{\rm kin}({\bf x})$ is given by equation~(\ref{eq:psig_kin}) and $P^{\rm sig}_{\rm btag}({\bf x})$ is defined as:
\begin{equation}
P^{\rm sig}_{\rm btag}({\bf x}) = P_b({\rm jet}_1) P_b({\rm jet}_2) P_b({\rm jet}_3) P_{q_h}({\rm jet}_4), 
\label{eq:psig_btag}
\end{equation}
with ${\rm jet}_i$ ($i=1,\ldots, 4$) representing the parton--jet assignment being evaluated, and 
$P_f({\rm jet}_i)$ denoting the probability that jet $i$, characterised by its four-momentum $p_{{\rm jet}_i}$ and $b$-tagging weight value $w_{{\rm jet}_i}$, originates from a
parton with flavour $f$ ($b$, $c$, or $l$; $l$ for light parton). The calibration of the $b$-tagging algorithm is performed for fixed thresholds  on the 
multivariate $b$-tagging discriminant variable, corresponding to different average $b$-tagging efficiencies in $\ttbar$ events of 60\%, 70\%, and 80\%,
also referred to as ``operating points'' (OP). The corresponding thresholds are denoted by $w_{\rm cut}^{\rm OP}$, with ${\rm OP}= 60\%, 70\%$, or $80\%$. 
Pa\-ra\-me\-te\-ri\-sa\-tions of the $b$-tagging efficiencies for different jet flavours as functions of jet $\pt$ and $\eta$ are available for each of these 
operating points, $\epsilon_f^{\rm OP}(\pt, \eta)$, which can be used to compute $P_f$ as follows:
if the jet $b$-tagging weight falls between the thresholds for operating points ${\rm OP_1}$ and ${\rm OP_2}$, 
$w_{\rm cut}^{\rm OP_1} < w_{\rm jet} \leq w_{\rm cut}^{\rm OP_2}$, then 
$P_f =  \epsilon_f^{\rm OP_1} - \epsilon_f^{\rm OP_2}$;
alternatively, if the jet $b$-tagging weight is below (above) the threshold corresponding to the 80\% (60\%) operating 
point, then $P_f =  1 - \epsilon_f^{\rm 80\%}$ ($P_f =  \epsilon_f^{\rm 60\%}$).

\begin{figure*}[htbp]
\begin{center}
\subfloat[]{\includegraphics[width=0.40\textwidth]{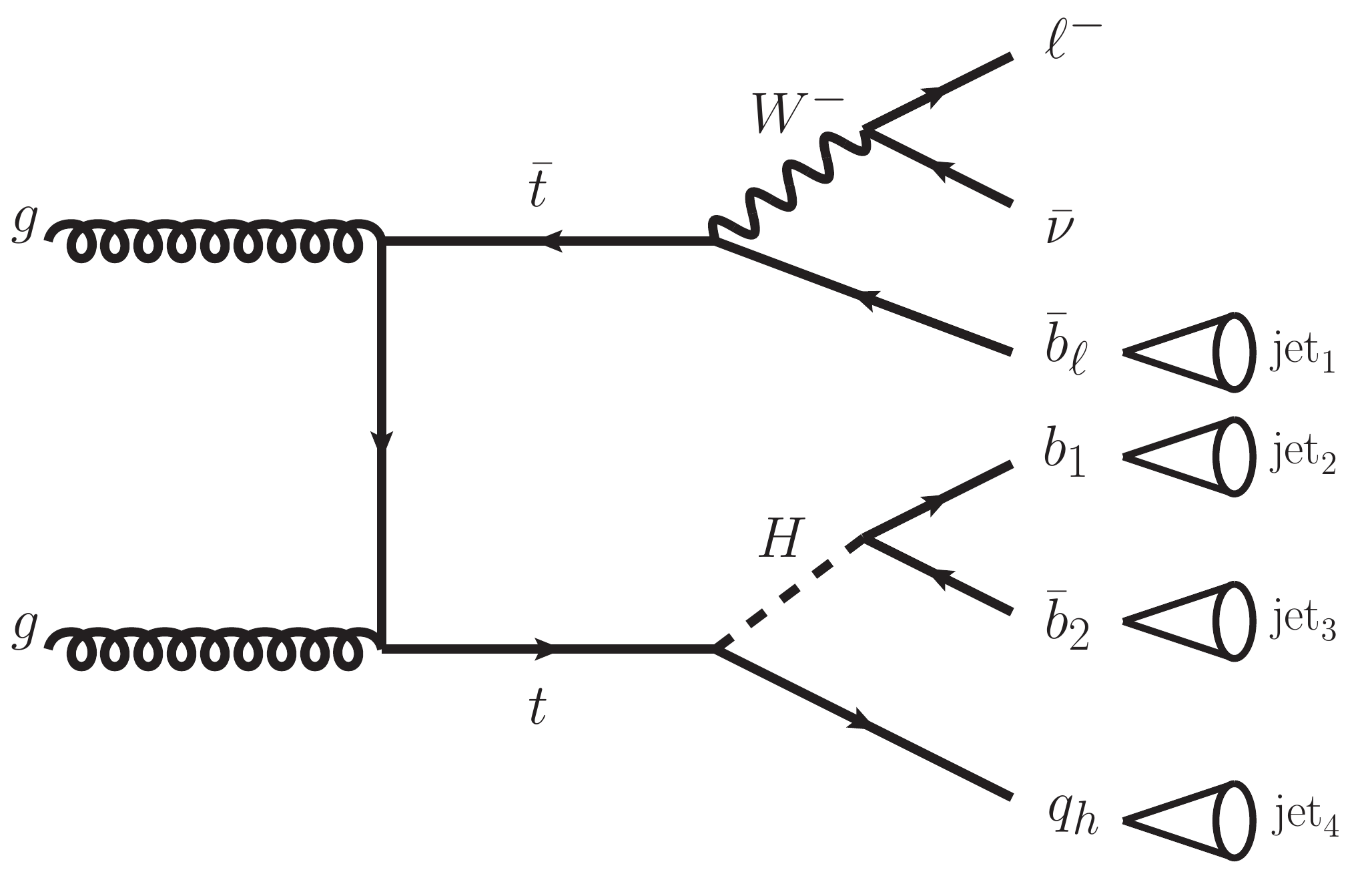}} \quad
\subfloat[]{\includegraphics[width=0.40\textwidth]{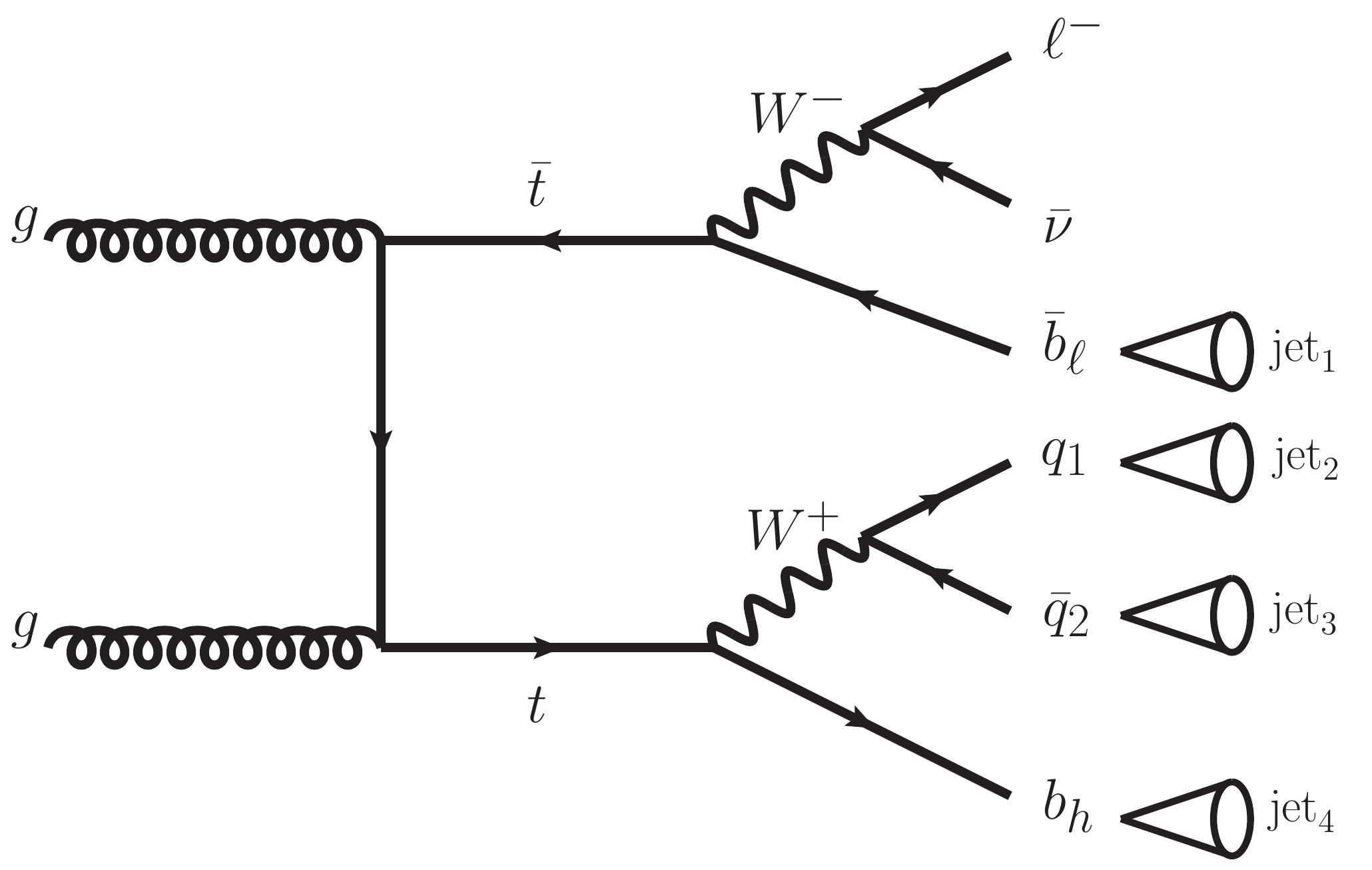}} \\
\subfloat[]{\includegraphics[width=0.40\textwidth]{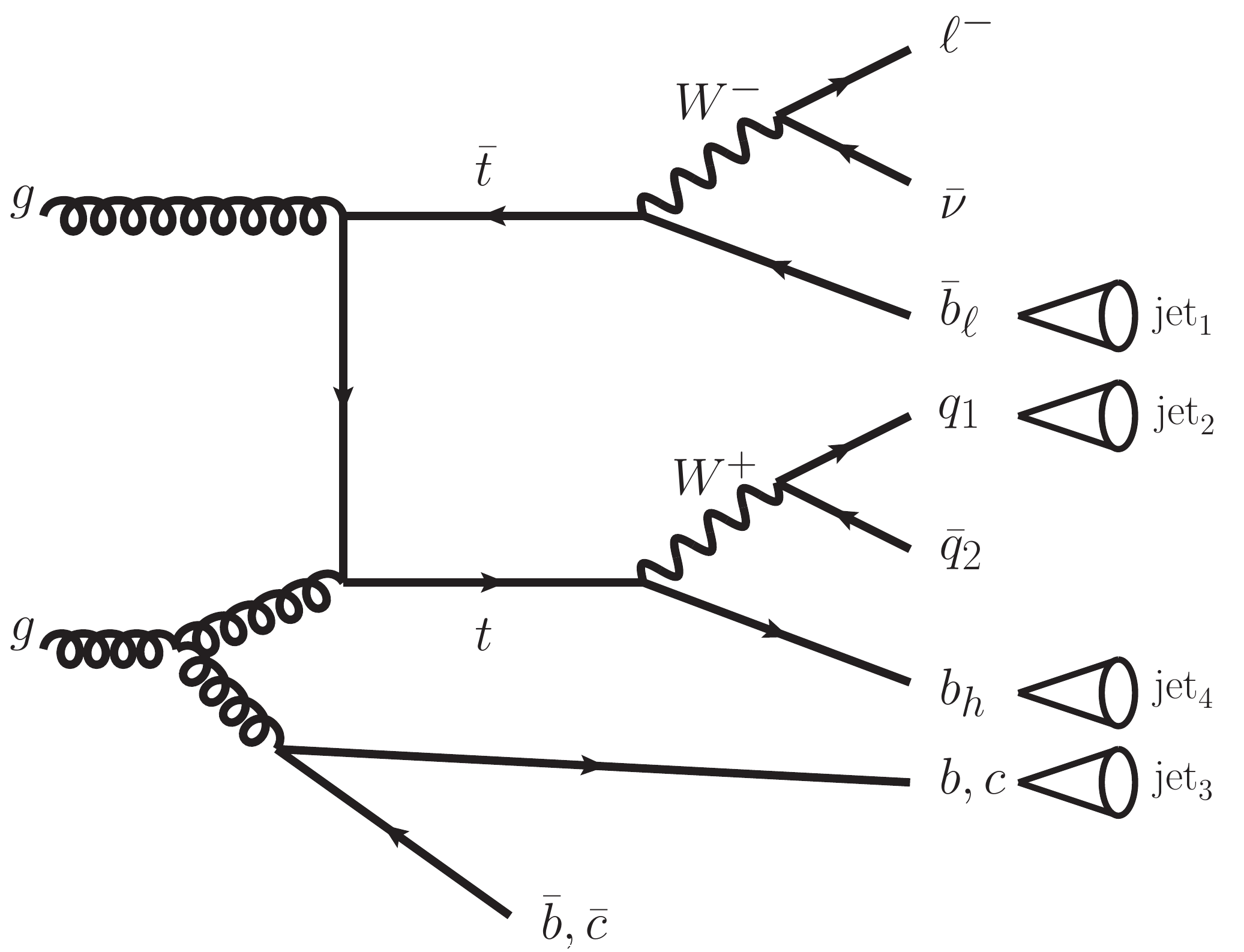}}
\caption{Representative Feynman diagrams illustrating the partonic configurations and parton--jet assignments considered
in the construction of (a) the signal probability and (b) and (c) the background probability used in the
definition of the final discriminant (see text for details).} 
\label{fig:cartoon}
\end{center}
\end{figure*}

\subsubsection{Background probability}
\label{sec:pbkg}

The calculation of $P^{\rm bkg}$ follows a similar approach to that discussed in section~\ref{sec:psig}, although it is 
slightly more complicated to account for the varying fraction and different kinematic features of the
$\ttbar$+light-jets, $\ttcc$ and $\ttbb$ backgrounds as a function of the analysis channel. This is particularly relevant
in the (4 j, 3 b) and (4 j, 4 b) channels, which dominate the sensitivity of the search. 
While $\ttbar$+light-jets events often have both jets from the hadronic $W$ boson decay among the four selected jets 
[see figure~\ref{fig:cartoon}(b)], 
this is seldom the case for $\ttbb$ and $\ttcc$ events, especially in the (4 j, 4 b)  channel. In this case 
the four $b$-tagged jets typically originate from the two $b$-quarks from the top quark decays, the charm quark from the
$W$ boson decay, and an extra heavy-flavour quark ($b$ or $c$) produced in association with the $\ttbar$ system,
while the jet associated with the down-type quark from the $W$ boson decay is not reconstructed [see figure~\ref{fig:cartoon}(c)].

To account for this, the following kinematic variables are considered: $M_{\ell\nu b_{\ell}}$, $X_{q_{1} j b_{h}}$ and $M_{q_{1} j}$, with 
$X_{q_{1} j b_{h}} \equiv M_{q_{1} j b_{h}} \ominus M_{q_{1} j}$, were $j$ denotes an extra quark-jet which
can either originate from the $W$ boson decay ($q_2$) or from an extra heavy-quark ($b$ or $c$) produced in association
with the $\ttbar$ system.  For each of these possibilities, occurring in a fraction $f_j$ of the cases, corresponding pdfs are constructed. 
As a generalisation of equation~(\ref{eq:psig}), the expression for $P^{\rm bkg}({\bf x})$ becomes:
\begin{equation}
P^{\rm bkg}({\bf x}) = \frac{\sum\limits_{k=1}^{N_p} \; \sum\limits_{j \in \{ b, c, q_2\}} f_j P^{{\rm bkg}, j}_{\rm btag}({\bf x}^k) P^{{\rm bkg}, j}_{\rm kin}({\bf x}^k)}{\sum\limits_{k=1}^{N_p}   \;  \sum\limits_{j \in \{ b, c, q_2\}} f_j P^{{\rm bkg}, j}_{\rm btag}({\bf x}^k)},
\label{eq:pbkg}
\end{equation}
with
\begin{equation}
P^{{\rm bkg}, j}_{\rm kin}({\bf x}) =  P^{\rm bkg}(M_{\ell\nu b_{\ell}}) P^{\rm bkg}(X_{q_{1} j b_{h}})  P^{\rm bkg}(M_{q_{1}j}),
\label{eq:pbkg_kin}
\end{equation}
and
\begin{equation}
P^{{\rm bkg}, j}_{\rm btag}({\bf x}) = P_b({\rm jet}_1) P_{q_1}({\rm jet}_2)  P_{j}({\rm jet}_3) P_{b}({\rm jet}_4).
\label{eq:pbkg_btag}
\end{equation}
where $P_f({\rm jet}_i)$ are computed as discussed in section~\ref{sec:psig}.
In the above expression, $P_j = P_l$ for $j=q_2$, the down-type quark in the $W$ boson decay,
and $P_{q_1} = f_c P_c + (1-f_c) P_l$, where $f_c$ is the fraction of events where the up-type quark 
from the $W$ boson decay assigned to the jet is a charm quark. This fraction is different in each analysis 
channel, primarily depending on the $b$-tag multiplicity requirements. It varies from $\sim 50\%$ for events 
in the (4  j, 2 b) channel to $\sim 90\%$ for events in the (4 j, 4 b) channel.

\subsubsection{Final discriminant}
\label{sec:D}
The final discriminant $D$ is computed for each event as given in equation~(\ref{eq:D}), 
using the definitions for $P^{\rm sig}$ and $P^{\rm bkg}$ given in equations~(\ref{eq:psig}) and~(\ref{eq:pbkg}), respectively.
Since this analysis has higher expected sensitivity to a $\ttbar \to WbHc$ signal than to a $\ttbar \to WbHu$ signal and, in order to allow probing 
of the $\BR(t\to Hu)$ versus $\BR(t\to Hc)$ plane, the discriminant optimised for $\ttbar \to WbHc$ is used for both the $Hc$ and $Hu$
decay modes. It was verified that using the $\Hc$ discriminant for the $\Hu$ search does not result in a significant sensitivity loss.
Figure~\ref{fig:DV_Shape_comparison} compares the shape of the $D$ distribution between the $\Hc$ and $\Hu$ signals and the 
$t\bar{t}\to WbWb$ background in each of the channels considered in this analysis.

\begin{figure*}[htbp]
\begin{center}
\subfloat[]{\includegraphics[width=0.33\textwidth]{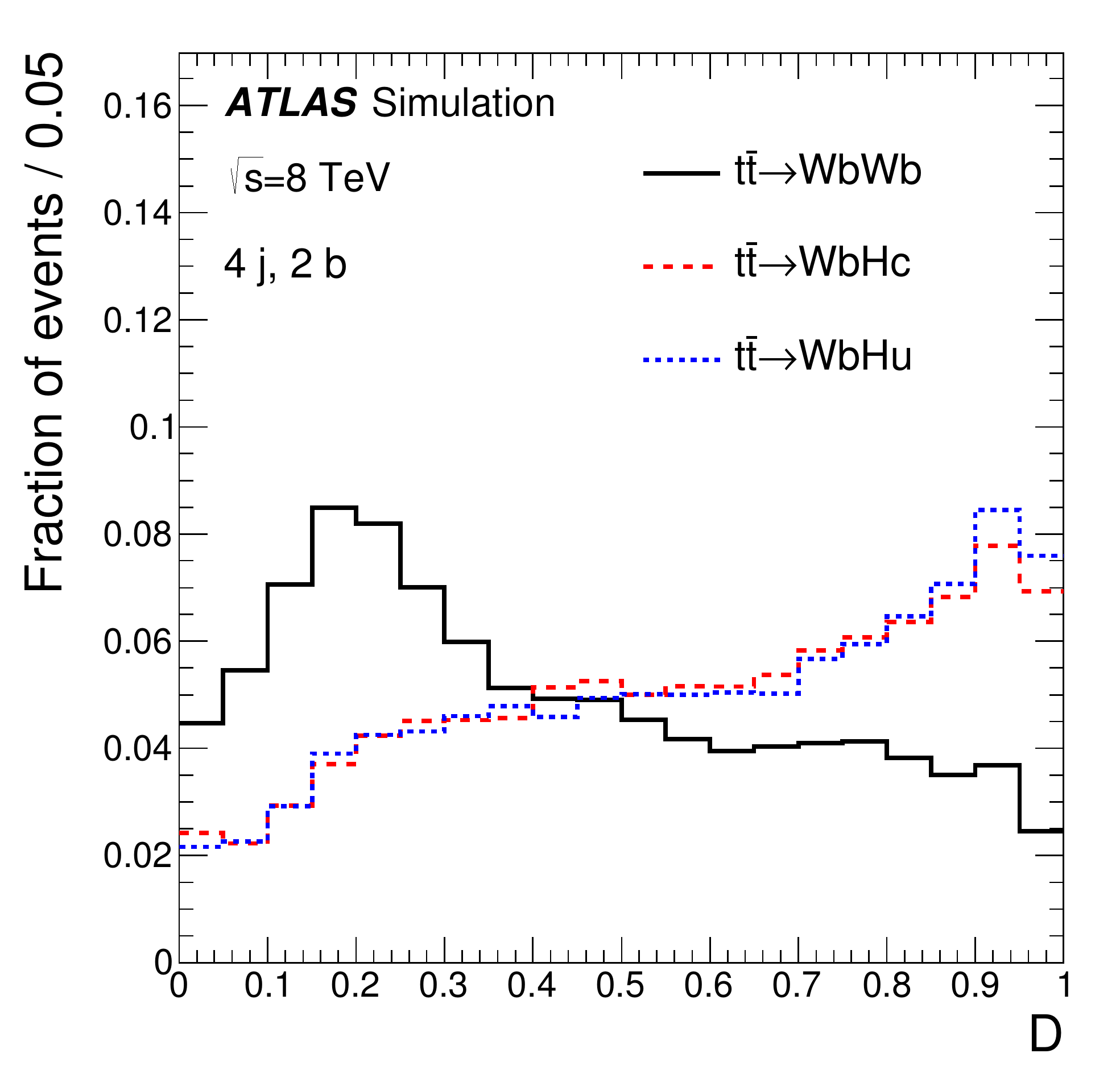}}
\subfloat[]{\includegraphics[width=0.33\textwidth]{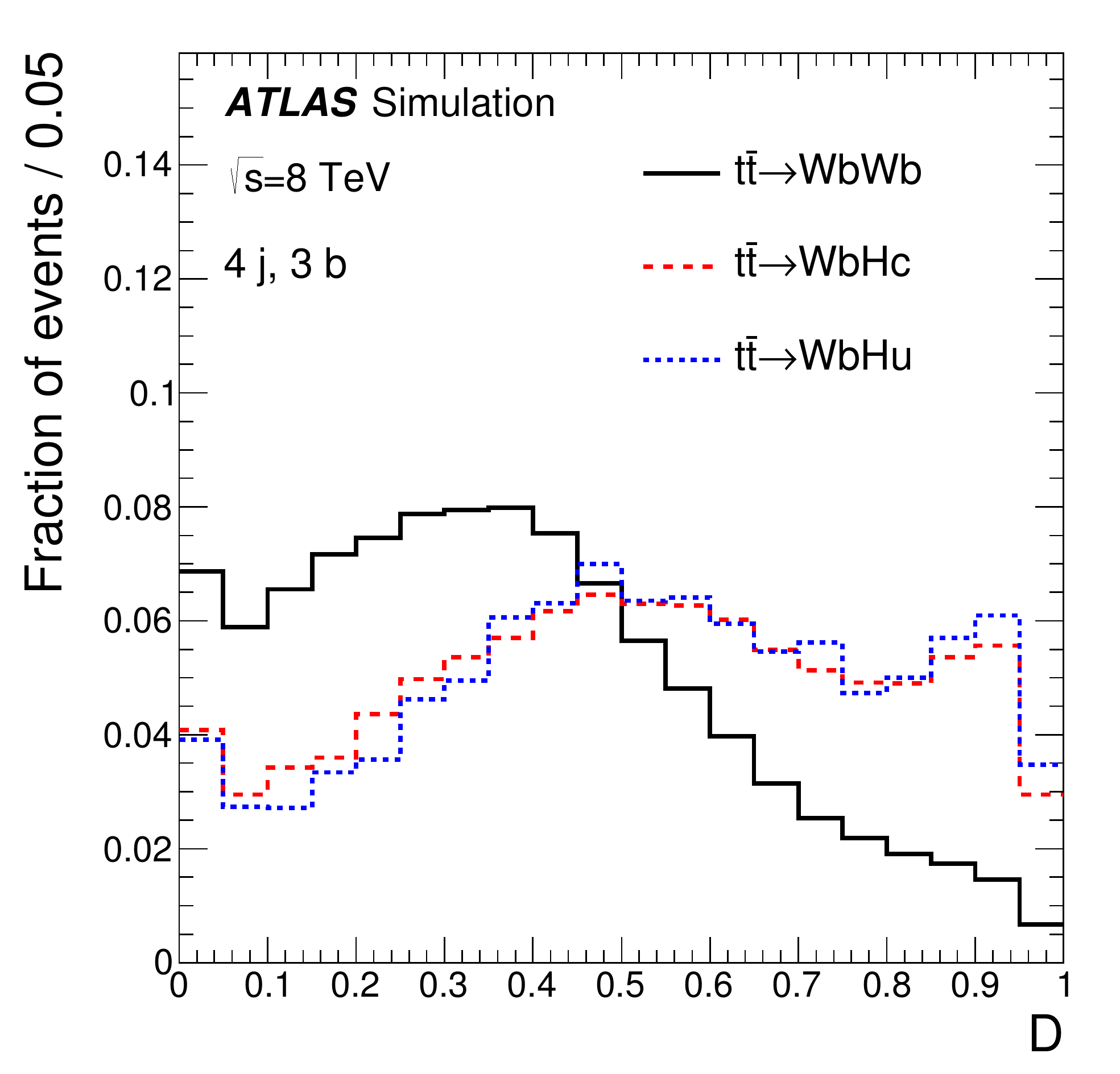}}
\subfloat[]{\includegraphics[width=0.33\textwidth]{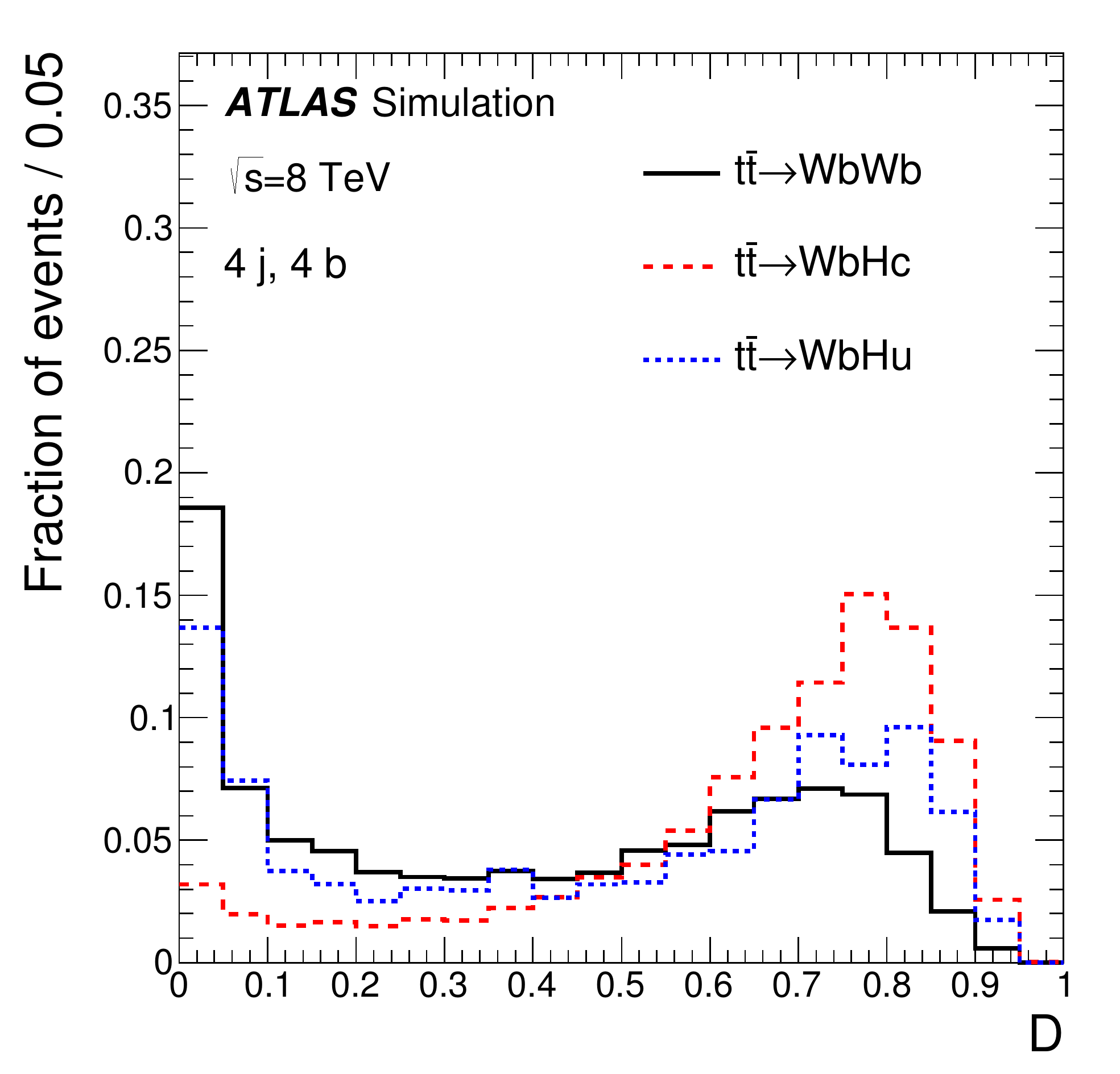}} \\
\subfloat[]{\includegraphics[width=0.33\textwidth]{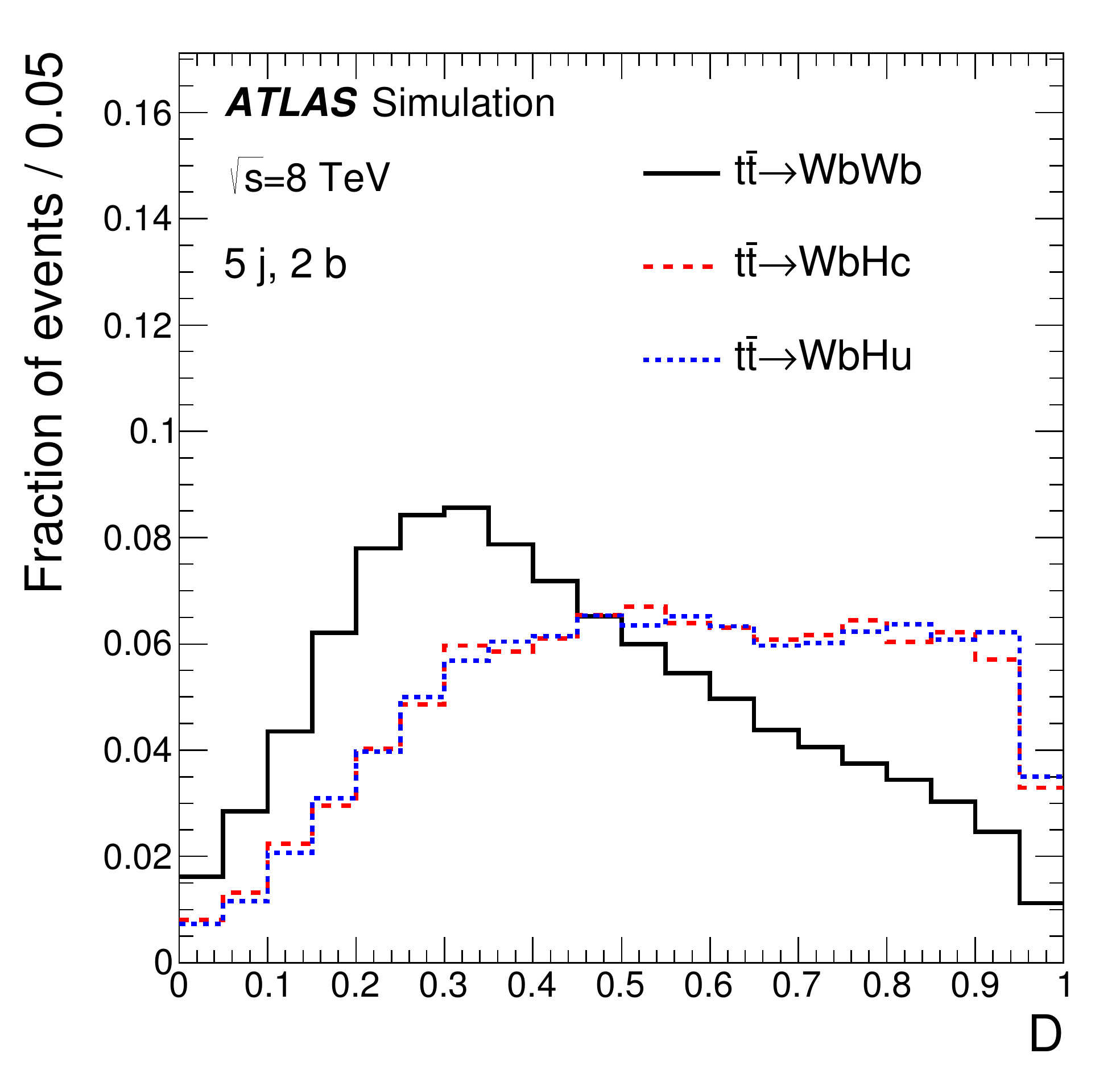}} 
\subfloat[]{\includegraphics[width=0.33\textwidth]{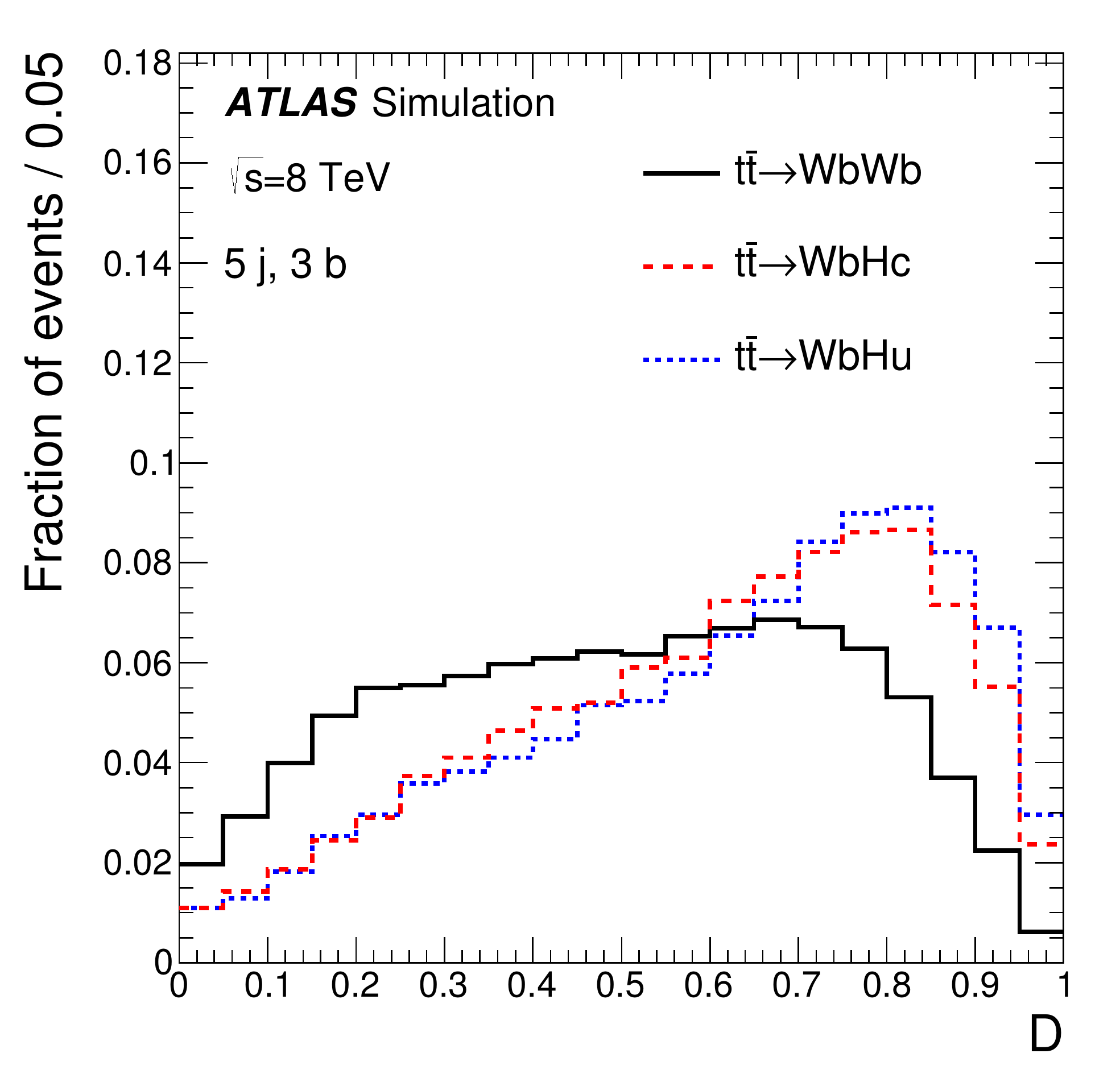}}
\subfloat[]{\includegraphics[width=0.33\textwidth]{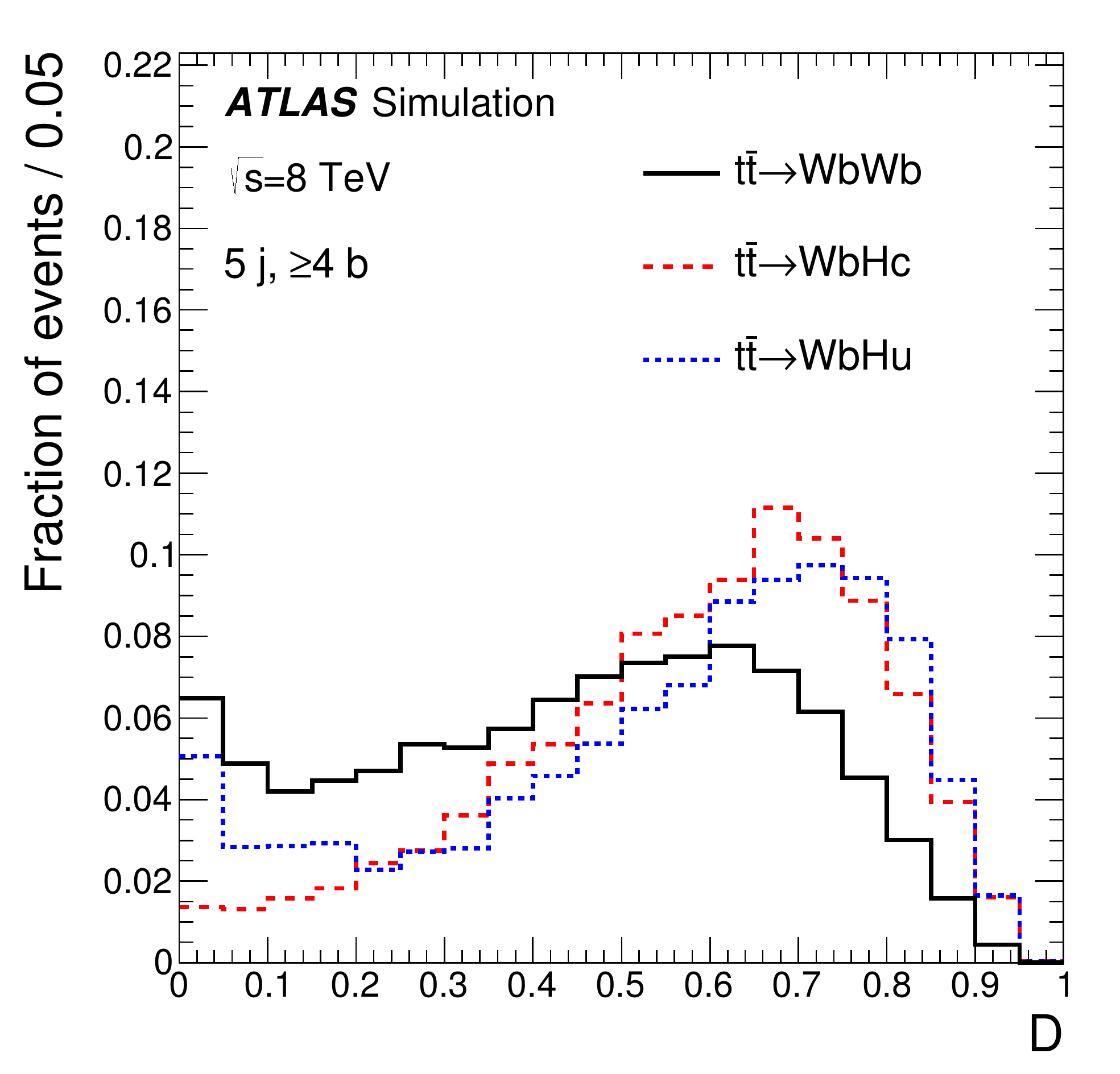}} \\
\subfloat[]{\includegraphics[width=0.33\textwidth]{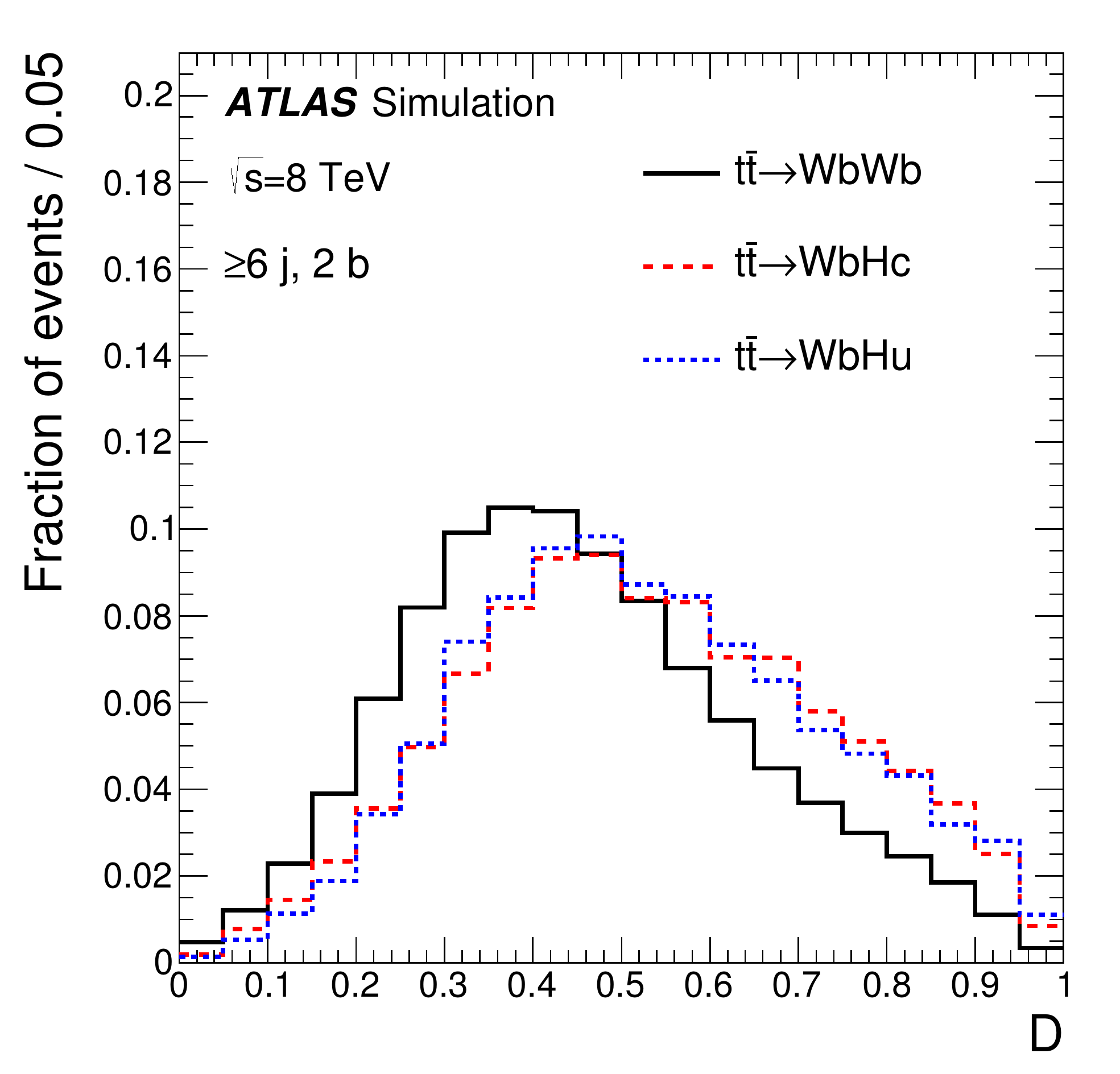}}
\subfloat[]{\includegraphics[width=0.33\textwidth]{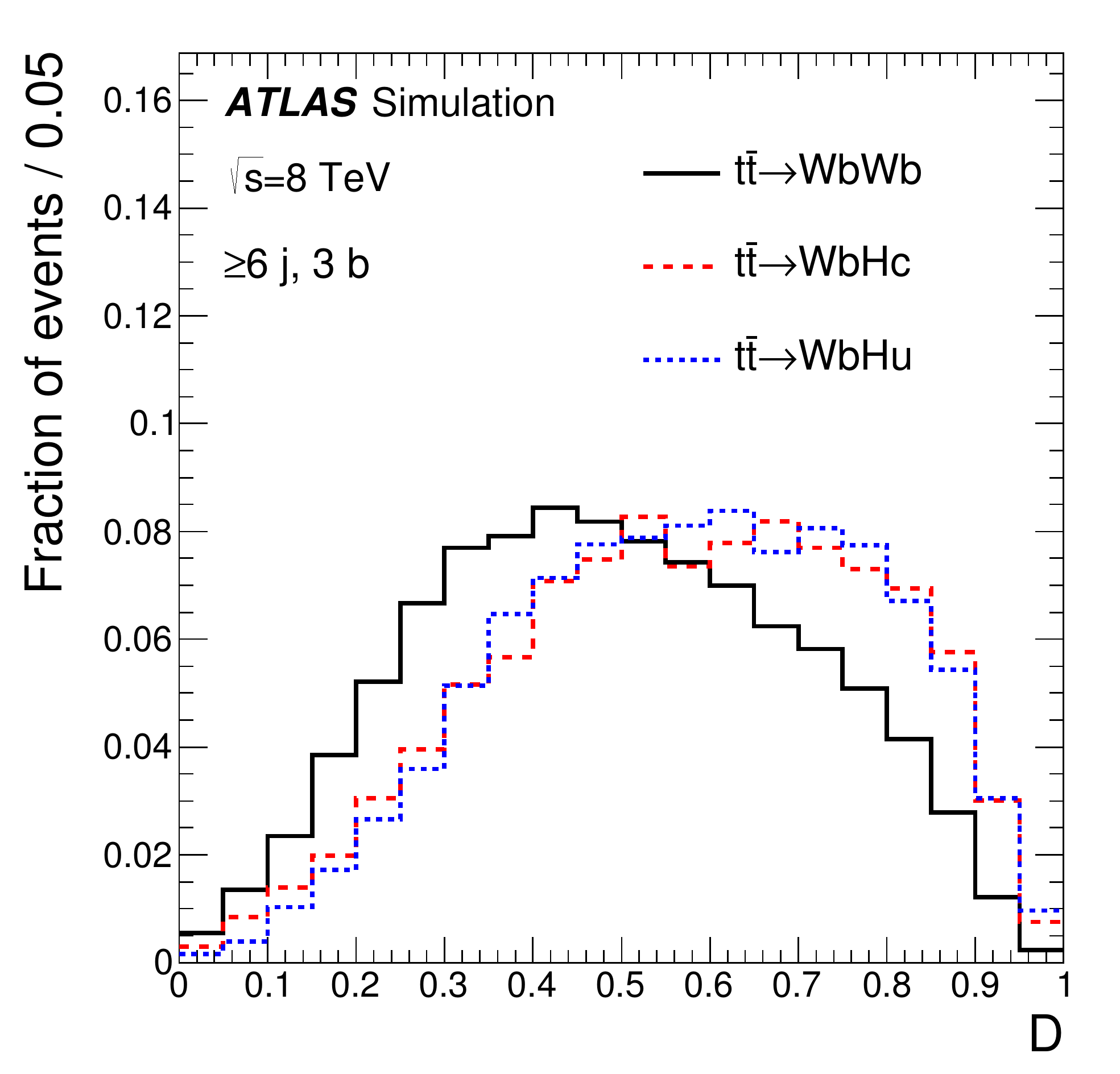}}
\subfloat[]{\includegraphics[width=0.33\textwidth]{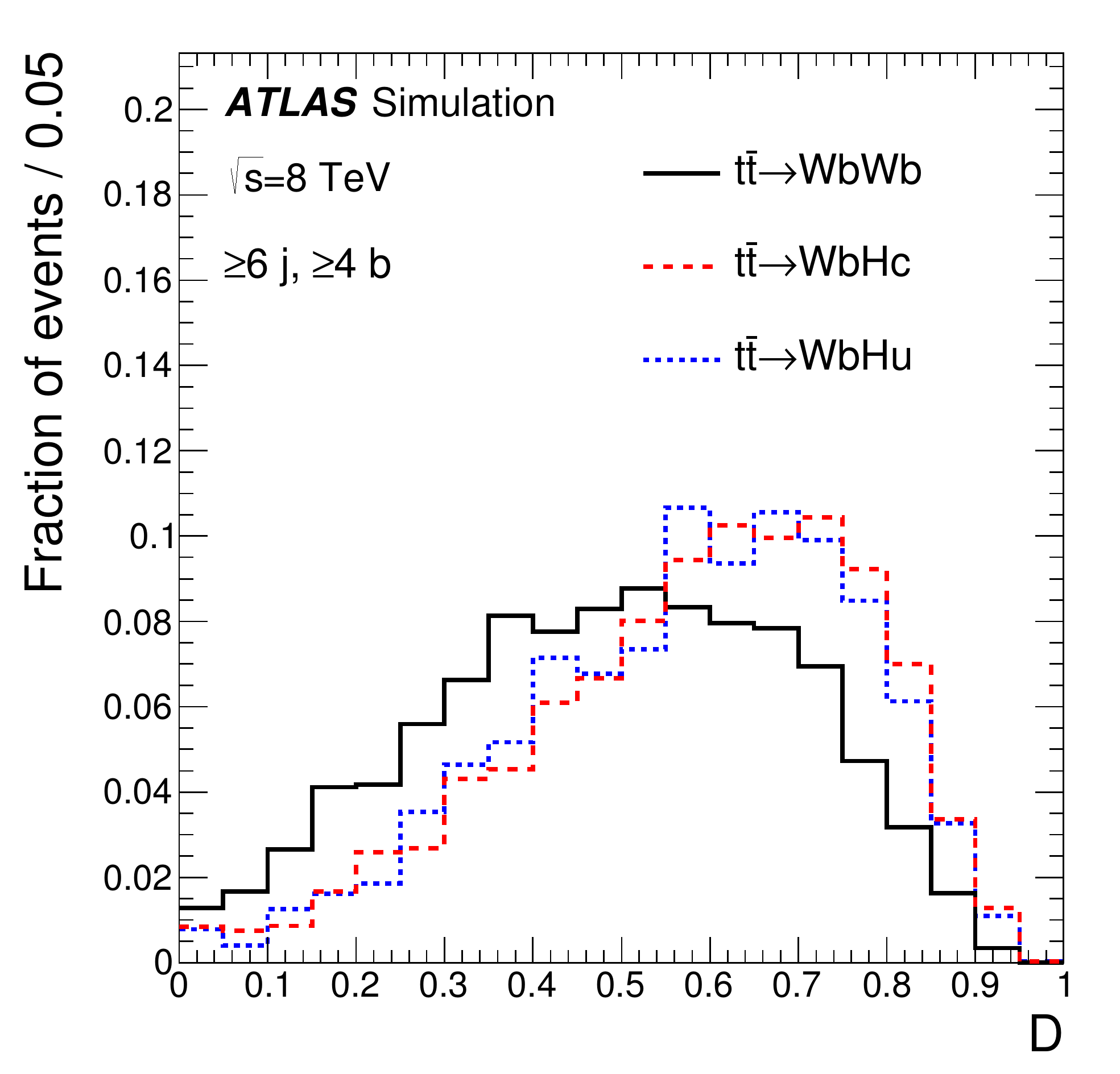}}  \\
\caption{Comparison of the shape of the $D$ discriminant distribution between the $\Hc$ (red dashed) and $\Hu$ (blue dotted) signals, 
and the $t\bar{t}\to WbWb$ background (black solid) in each of the channels considered in the analysis:
(a) (4 j, 2 b), (b) \mbox{(4 j, 3b)}, (c) (4 j, 4 b), (d) (5 j, 2 b), (e) (5 j, 3 b), (f) (5 j, $\geq$4 b), (g) ($\geq$6 j, 2 b), 
(h) ($\geq$6 j, 3 b), and (i) ($\geq$6 j, $\geq$4 b).} 
\label{fig:DV_Shape_comparison}
\end{center}
\end{figure*}

\section{Systematic uncertainties}
\label{sec:systematics}
				   
Several sources of systematic uncertainty are considered that can affect the normalisation of signal 
and background and/or the shape of their corresponding final discriminant distributions.  
Each source of systematic uncertainty is considered to be uncorrelated with the other sources.  
Correlations of a given systematic uncertainty are maintained across processes and channels.  
Table~\ref{tab:SystSummary} presents a list of all systematic uncertainties considered in the analysis 
and indicates whether they are taken to be normalisation-only, or to affect both shape and normalisation.

\begin{table*}[htbp]
\centering
\begin{tabular}{lcc}
\toprule\toprule
Systematic uncertainty & Type  & Components \\
\midrule
{\bf Luminosity}                  &  N & 1\\\midrule
{\bf Reconstructed Objects}                 &   & \\
Electron                  & SN & 5 \\
Muon                      &  SN & 6 \\
Jet reconstruction      & SN & 1\\ 
Jet vertex fraction         & SN    & 1\\
Jet energy scale            & SN & 22\\
Jet energy resolution       & SN & 1\\
Missing transverse momentum  & SN & 2\\ 
$b$-tagging efficiency      & SN & 6\\
$c$-tagging efficiency      & SN & 4\\
Light-jet tagging efficiency    & SN & 12\\ 
High-\pt\ tagging   & SN & 1 \\ \midrule
{\bf Background Model}                 &   & \\
$t\bar{t}$ cross section    &  N & 1\\
$t\bar{t}$ modelling: $\pt$ reweighting   & SN & 9\\
$t\bar{t}$ modelling: parton shower & SN & 3\\
$t\bar{t}$+HF: normalisation & N & 2 \\
$t\bar{t}$+$c\bar{c}$: $\pt$ reweighting  & SN & 2 \\
$t\bar{t}$+$c\bar{c}$: generator & SN & 4 \\
$t\bar{t}$+$b\bar{b}$: NLO shape & SN & 8 \\
$W$+jets normalisation      &  N & 3\\
$W$ $\pt$ reweighting     &  SN & 1\\
$Z$+jets normalisation      &  N & 3\\
$Z$ $\pt$ reweighting     &  SN & 1\\
Single top normalisation    &  N & 3\\
Single top model            &  SN & 1\\
Diboson normalisation  &  N & 3\\
$t\bar{t}V$ cross section   &  N & 1\\
$t\bar{t}V$ model           &  SN & 1\\ 
$t\bar{t}H$ cross section & N & 1 \\
$t\bar{t}H$ model       & SN & 2 \\ 
Multijet normalisation  &  N & 4\\ \midrule
{\bf Signal Model}                 &   & \\
$t\bar{t}$ cross section    &  N & 1\\
Higgs boson branching ratios & N & 3 \\
$t\bar{t}$ modelling: $\pt$ reweighting   & SN & 9\\
$t\bar{t}$ modelling: $\pt$ reweighting non-closure  & N & 1\\
$t\bar{t}$ modelling: parton shower & N & 1\\
\bottomrule\bottomrule
\end{tabular}
\caption{\label{tab:SystSummary} List of systematic uncertainties considered. 
An ``N'' means that the uncertainty is taken as affecting only the normalisation for all relevant 
processes and channels, whereas ``SN'' means that the uncertainty is 
taken on both shape and normalisation.
Some of the systematic uncertainties are split into several components for a more
accurate treatment.}
\end{table*}

The leading sources of systematic uncertainty vary depending on the analysis channel considered, but they 
typically originate from $\ttbar$+jets modelling (including $\ttbar$+HF) and $b$-tagging. 
For example, the total systematic uncertainty  in the background normalisation in the (4 j, 4 b) channel, 
which dominates the sensitivity in the case of the $\Hc$ search, is approximately 20\%, 
with the largest contributions originating from $\ttbar$+HF normalisation, $b$-tagging efficiency, 
$c$-tagging efficiency, light-jet tagging efficiency and $\ttbar$ cross section.
However, as shown in section~\ref{sec:result}, the fit to data in the nine analysis channels allows the overall background 
uncertainty to be reduced significantly, to approximately 4.4\%. 
The reduced uncertainty results from the significant constraints provided by the data on some
systematic uncertainties, as well as the anti-correlations among sources of systematic uncertainty resulting
from the fit to the data.
The total systematic uncertainty on the $\Hc$ signal normalisation in the (4 j, 4 b) channel is approximately 17\%, 
with similar contributions from uncertainties related to $b$-tagging and overall signal modelling. After the fit,
this uncertainty is reduced to 7.8\%.
Table~\ref{tab:SystSummary_WbHc} presents a summary of the sys\-te\-ma\-tic uncertainties for the $\Hc$ 
search and their impact on the normalisation of the signal and the main backgrounds in the (4 j, 4 b) channel.

The following sections describe each of the systematic uncertainties considered in the analyses. 

\begin{table*}[htbp]
\centering
\begin{tabular}{l | c c c c  | c c c c}
\toprule\toprule
 & \multicolumn{4}{c|}{Pre-fit} & \multicolumn{4}{c}{Post-fit} \\ 
 &  $WbHc$ & $t\bar{t}$+LJ & $t\bar{t}+c\bar{c}$ & $t\bar{t}+b\bar{b}$ &  $WbHc$ & $t\bar{t}$+LJ & $t\bar{t}+c\bar{c}$ & $t\bar{t}+b\bar{b}$ \\
\midrule
Luminosity  & $\pm 2.8 $  & $\pm 2.8 $  & $\pm 2.8 $  & $\pm 2.8 $  & $\pm 2.6$ & $\pm 2.6 $  & $\pm 2.6 $  & $\pm 2.6 $ \\ 
Lepton efficiencies  & $\pm 1.5 $  & $\pm 1.5 $  & $\pm 1.5 $  & $\pm 1.5 $  & $\pm 1.5$ &  $\pm 1.5 $  & $\pm 1.5 $  & $\pm 1.5 $ \\ 
Jet energy scale  & $\pm 3.3 $  & $\pm 2.9 $  & $\pm 2.3 $  & $\pm 5.8 $  & $\pm 1.4$ &  $\pm 1.2 $  & $\pm 1.8 $  & $\pm 4.1 $ \\ 
Jet efficiencies  & $\pm 1.2 $  & --  & $\pm 1.9 $  & $\pm 1.7 $  & $\pm 0.9$ &  --  & $\pm 1.4 $  & $\pm 1.2 $ \\
Jet energy resolution  & --  & $\pm 1.2 $  & $\pm 2.8 $  & $\pm 2.9 $  & -- &  -- & $\pm 1.0 $  & $\pm 1.1 $ \\ 
$b$-tagging eff.  & $\pm 7.9 $  & $\pm 5.5 $  & $\pm 5.2 $  & $\pm 10 $  & $\pm 5.7$ & $\pm 3.9 $  & $\pm 3.7 $  & $\pm 6.6 $ \\ 
$c$-tagging eff.  & $\pm 7.0 $  & $\pm 6.6 $  & $\pm 13 $  & $\pm 3.5 $  & $\pm 6.3$ &  $\pm 6.0 $  & $\pm 11 $  & $\pm 3.2 $ \\ 
Light-jet tagging eff.  & $\pm 0.8 $  & $\pm 18 $  & $\pm 3.2 $  & $\pm 1.5 $  & $\pm 0.6$ &  $\pm 13 $  & $\pm 2.3 $  & $\pm 1.1 $ \\
$t\bar{t}$: reweighting  & $\pm 5.9 $  & $\pm 2.7 $  & $\pm 4.2 $  & --  & $\pm 3.8$ &  $\pm 1.9 $  & $\pm 2.3 $  & -- \\ 
$t\bar{t}$: parton shower  & $\pm 5.4 $  & $\pm 4.8 $  & $\pm 10 $  & $\pm 4.9 $  & $\pm 1.7$ &  $\pm 1.5 $  & $\pm 6.5 $  & $\pm 3.1 $ \\ 
$t\bar{t}$+HF: normalisation  & --  & --  & $\pm 50 $  & $\pm 50 $  & -- &  --  & $\pm 32 $  & $\pm 16 $ \\ 
$t\bar{t}$+HF: modelling  & --  & --  & --  & $\pm 7.7 $  & -- &  --  & --  & $\pm 7.4 $ \\ 
Signal modelling  & $\pm 6.9 $  & --  & --  & --  & $\pm 6.9$ &  --  & --  & -- \\ 
Theor. cross sections  & $\pm 6.2 $  & $\pm 6.2 $  & $\pm 6.2 $  & $\pm 6.2 $  & $\pm 3.9$ &  $\pm 3.9 $  & $\pm 3.9 $  & $\pm 3.9 $ \\ 
\midrule                                                                                                                                                                                               
Total   & $\pm 17 $  & $\pm 22 $  & $\pm 54 $  & $\pm 53 $  & $\pm 7.8$ &  $\pm 14 $  & $\pm 28 $  & $\pm 15 $ \\ 
\bottomrule\bottomrule
\end{tabular}
\caption{$\Hcbb$ search: summary of the systematic uncertainties considered in the (4 j, 4 b) 
channel and their impact (in \%) on the normalisation of the signal and the main backgrounds, before and after the fit to data.
The $\Hc$ signal and the $t\bar{t}$+light-jets background are denoted by ``$WbHc$''  and ``$t\bar{t}$+LJ'' respectively.
Only sources of systematic uncertainty resulting in a normalisation change of at least 0.5\% are displayed.
The total post-fit uncertainty can differ from the sum in quadrature of individual sources due to the 
anti-correlations between them resulting from the fit to the data.}
\label{tab:SystSummary_WbHc}
\end{table*}

\subsection{Luminosity}
\label{sec:syst_lumi}

The uncertainty on the integrated luminosity is 2.8\%, affecting the overall normalisation of
all processes estimated from the simulation. It is estimated from a calibration of the luminosity 
scale derived from beam-separation scans performed in November 2012, following the same 
methodology as that detailed in ref.~\cite{Aad:2013ucp}.

\subsection{Reconstructed objects}
\label{sec:syst_objects}

Uncertainties associated with leptons arise from the reconstruction,
identification and trigger. These efficiencies are measured 
using tag-and-probe techniques on $Z\to \ell^+\ell^-$ ($\ell=e,\mu$) data
and simulated samples. The small differences found are corrected for in the simulation.
Negligible sources of uncertainty originate from the corrections
applied to adjust the lepton momentum scale and resolution in
the simulation to match those in data. 
The combined effect of all these uncertainties results in an overall normalisation 
uncertainty on the signal and background of approximately 1.5\%.

Uncertainties associated with jets arise from the efficiency of jet reconstruction 
and identification based on the JVF variable, as well as the jet energy scale
and resolution. The largest contribution results from the jet energy scale,
whose uncertainty dependence on jet $\pt$ and $\eta$ is split into 22 uncorrelated sources 
that are treated independently in the analysis. 
It affects the normalisation of signal and backgrounds by approximately 3--4\%
in the most sensitive search channels, (4 j, 3 b) and (4 j, 4 b),  and up to 12\% in
the channels with $\geq$6 jets.

Uncertainties associated with energy scales and resolutions of leptons and jets 
are propagated to $\met$. Additional uncertainties originating from the modelling 
of the underlying event, in particular its impact on the $\pt$ scale and resolution 
of unclustered energy, are negligibly small.

The leading uncertainties associated with reconstructed objects in this analysis originate from the mo\-del\-ling
of the $b$-, $c$-, and light-jet-tagging efficiencies in the simulation, which is corrected
to match the efficiencies measured in data control samples~\cite{BTaggingEfficiency,CLTaggingEfficiency}
through dedicated scale factors.
Uncertainties on these factors include a total of six independent sources
affecting $b$-jets and four independent sources affecting $c$-jets. 
Each of these uncertainties has a different jet-$\pt$ dependence.
Twelve sources of uncertainty affecting light jets are considered, which depend on jet $\pt$ and $\eta$. 
The above sources of systematic uncertainty are taken as uncorrelated between $b$-jets, 
$c$-jets, and light-jets. They have their largest impact in the (4 j, 4 b) channel,
resulting in 10\%, 13\%, and 18\% normalisation uncertainties on the
$\ttbb$,  $\ttcc$, and $\ttbar$+light-jets background associated with the uncertainties on 
the $b$-, $c$-, and light-jet-tagging scale factors, respectively.
An additional uncertainty is included due to the extrapolation of these scale factors to jets 
with $\pt$ beyond the kinematic reach of the data calibration samples used ($\pt>300\gev$ for $b$- and $c$-jets, and $\pt>750\gev$ for light-jets),
taken to be correlated among the three jet flavours. This uncertainty has a very small impact in this analysis (e.g. $<0.2\%$ on the signal and 
background normalisations in the (4 j, 4 b) channel).

\subsection{Background modelling}
\label{sec:syst_bkgmodeling}

A number of sources of systematic uncertainty affecting the modelling of $t\bar{t}$+jets are considered. 
A brief summary is provided below, with further details available in ref.~\cite{Aad:2015gra},
as the uncertainty treatment is identical.
An uncertainty of $+6.1\%$/$-6.4\%$ is assumed for the inclusive $t\bar{t}$ production
cross section~\cite{Czakon:2011xx}, including contributions from varying the factorisation and renormalisation 
scales, and uncertainties arising from the PDF, $\alpha_{\rm S}$, and the top quark mass.
Uncertainties associated with the reweighting procedure applied to $\ttbar$+light-jets 
and $\ttcc$ processes include the nine leading sources of uncertainty in the differential cross-section 
measurement at $\sqrt{s}=7\tev$~\cite{Aad:2014zka}.
Additional uncertainties assigned to the modelling of the $\ttcc$ background
include a 50\% normalisation uncertainty, the full differences between applying and not applying the
reweightings of the top quark and $\ttbar$ $\pt$ spectra, as well as smaller uncertainties 
associated with the choice of LO generator.
Uncertainties affecting the modelling of $\ttbb$ production include a normalisation uncertainty
of 50\% (taken to be uncorrelated with the same uncertainty assigned to the $\ttcc$ background) 
and shape uncertainties (including inter-category migration effects) associated with
the NLO prediction from {\sc Sherpa}+{\sc OpenLoops}, which is used for reweighting of the default {\sc Powheg-Box} $\ttbb$ prediction.
These include three different scale variations,  a different shower-recoil model scheme, and 
two alternative PDF sets (MSTW and NNPDF). Additional uncertainties are assessed for
the contributions to the $\ttbb$ background originating from multiple parton interactions 
or final-state radiation from top decay products, which are not part of the NLO prediction.
Finally, an uncertainty due to the choice of parton shower and hadronisation model 
is derived by comparing events produced by {\sc Powheg-Box} interfaced to {\sc Pythia} 
or {\sc Herwig}. This uncertainty is taken to be uncorrelated between the $\ttbar$+light-jets, 
$\ttcc$ and $\ttbb$ processes.

Uncertainties affecting the modelling of the $W$+jets background include a 7\% 
normalisation uncertainty for events with $\geq$4 jets coming from the data-driven normalisation procedure.
The corresponding normalisation uncertainty for $Z$+jets is 5\% for events with $\geq$2 jets.
In addition, a 24\% normalisation uncertainty is added in quadrature for each additional 
inclusive jet-multiplicity bin beyond the one where the background is normalised, based on a comparison among 
different algorithms for merging LO matrix elements and parton shower simulations~\cite{Alwall:2007fs}.
For example, $W$+jets events with exactly 4 jets, exactly 5 jets and $\geq 6$ jets are assigned normalisation 
uncertainties of 7\%, $ 7\% \oplus 24\% =25\%$ and $7\% \oplus 24\% \oplus 24\%=35\%$.
Finally, the full size of the $W$ and $Z$ boson $\pt$ correction, after symmetrisation, is taken as a systematic uncertainty. 
The above uncertainties are taken as uncorrelated between $W$+jets and $Z$+jets. 

Uncertainties affecting the modelling of the single-top-quark background include a 
$+5\%$/$-4\%$ uncertainty on the total cross section, which is estimated as a weighted average 
of the theoretical uncertainties on $t$-, $Wt$- and $s$-channel production~\cite{Kidonakis:2011wy,Kidonakis:2010ux,Kidonakis:2010tc}.
Similarly to the case of $W/Z$+jets, an additional 24\% normalisation uncertainty is added in quadrature for each additional 
inclusive jet-multiplicity bin above $\geq$3 jets.
An additional systematic uncertainty on $Wt$-channel production concerning the separation 
between $t\bar{t}$ and $Wt$ at NLO~\cite{Frixione:2008yi} is assessed by comparing
the nominal sample, which uses the so-called ``diagram subtraction'' scheme, with an alternative sample
using the ``diagram removal'' scheme.

Uncertainties on the diboson background normalisation include 5\% from the
NLO theoretical cross sections~\cite{Campbell:1999ah} and additional 24\% normalisation uncertainties 
added in quadrature for each additional inclusive jet-multiplicity bin above $\geq$2 jets.
Uncertainties on the $\ttbar V$ and $\ttbar H$ normalisations are 15\% and $+9\%$/$-12\%$ respectively,
from the uncertainties on their respective NLO theoretical cross sections~\cite{Campbell:2012dh,Garzelli:2012bn,Dittmaier:2011ti}. 
Additional small uncertainties arising from scale variations, which change the amount of initial-state radiation and thus the event
kinematics, are also considered.

Uncertainties on the data-driven multijet background estimate receive
contributions from the limited sample size in data, particularly at high jet and $b$-tag multiplicities, as 
well as from the uncertainty on the rate of fake leptons, estimated in 
different control regions (e.g. selected with a requirement on either the maximum $\met$ or $\mtw$). 
A combined normalisation uncertainty of 50\% due 
to all these effects is assigned, which is taken as correlated across jet
and $b$-tag multiplicity bins, but uncorrelated between electron and muon channels. 
No explicit shape uncertainty is assigned since the large statistical uncertainties associated with
the multijet background prediction, which are uncorrelated 
between bins in the final discriminant distribution, effectively cover all possible shape uncertainties. 

\subsection{Signal modelling}
\label{sec:syst_sigmodeling}

Several normalisation and shape uncertainties are taken into account for the $\Hq$ signal.
The uncertainty on the $\ttbar$ cross section (see above) also applies to the $\Hq$ signal
and is taken to be the same as, and fully correlated with, that assigned to the $\ttbar \to WbWb$ background. 
Uncertainties on the $H \to b\bar{b}$ branching ratio are taken into account
following the recommendation in ref.~\cite{Dittmaier:2011ti}: $\pm 1.1\%$ ($\Delta\alpha_{\rm S}$), $\pm 1.4\%$ ($\Delta m_b$) and $\pm 0.8\%$ (theory).
Additional modelling uncertainties originate from non-closure of the reweighting procedure applied to correct the distributions of 
top quark $\pt$ and $\ttbar$ system $\pt$ from {\sc Protos} to match those from the uncorrected 
{\sc Powheg-Box}+{\sc Pythia} simulation, and  the uncertainties associated
with the further reweighting in the same variables to match the differential cross-section measurements at $\sqrt{s}=7\tev$, 
taken to be fully correlated with those assigned to the $\ttbar$+light-jets background.
Finally, an uncertainty from the choice of parton shower and hadronisation model is estimated by comparing
the $\ttbar \to WbWb$ yields between {\sc Powheg-Box}+{\sc Pythia} and {\sc Powheg-Box}+{\sc Herwig} in the
channels with two $b$-tags, which are enriched in $\ttbar$+light-jets, and thus taken to be representative of
what would be the signal acceptance uncertainty due to differences in extra jet radiation and 
$b$-quark fragmentation between the two parton shower/hadronisation models.

\section{Statistical analysis}
\label{sec:stat_analysis}

The distributions of the final discriminants in each of the analysis channels 
considered are combined to test for the presence of a signal. 
The statistical analysis is based on a binned likelihood function $L(\mu,\theta)$ constructed as
a product of Poisson probability terms over all bins considered in the search.
In the case of several searches being combined, the product of Poisson probability terms is extended over all bins considered in all searches.
The function $L(\mu,\theta)$ depends
on the signal-strength parameter $\mu$,  defined as a multiplicative factor to the yield for $\Hq$ signal events
normalised to a reference branching ratio $\BR_{\rm ref}(t\to Hq)=1\%$,
and $\theta$, a set of nuisance parameters that encode the effect of systematic uncertainties
on the signal and background expectations and are implemented in the likelihood function as Gaussian or
log-normal priors with their width parameters corresponding to the size of the respective uncertainties. 
The relationship between $\mu$ and the corresponding $\BR(t\to Hq)$ is: 
\begin{equation}
\mu = \frac{\BR(t\to Hq)[1-\BR(t\to Hq)]}{\BR_{\rm ref}(t\to Hq)[1-\BR_{\rm ref}(t\to Hq)]}.
\label{eq:mu_br}
\end{equation}
For a given $\mu$ value, the SM $t\bar{t}\to WbWb$ background contribution is scaled accordingly 
in order to preserve the inclusive $t\bar{t}$ cross section. The corresponding multiplicative factor
would be $[1-\BR(t\to Hq)]^2$, with $\BR(t\to Hq)$ being a function of $\mu$ as can
be derived from equation~(\ref{eq:mu_br}): 
\begin{equation}
\BR(t\to Hq) = \frac{1-\sqrt{1-4\BR_{\rm ref}(t\to Hq)(1-\BR_{\rm ref}(t\to Hq))\mu}}{2}.
\label{eq:br_mu}
\end{equation}
Therefore, the total number of signal and background events in a given bin depends on $\mu$ and $\theta$. 
The best-fit $\BR(t\to Hq)$ is obtained by performing a binned likelihood fit to the data under the signal-plus-background
hypothesis, i.e. maximising the likelihood function $L(\mu,\theta)$ over $\mu$ and $\theta$.
The nuisance parameters $\theta$ allow variations of the expectations for signal and background
according to the corresponding systematic uncertainties, and their fitted values correspond to the deviations from
the nominal expectations that globally provide the best fit to the data.
This procedure allows a reduction of the impact of systematic uncertainties on the search sensitivity by taking
advantage of the highly populated background-dominated channels included in the likelihood fit.

The test statistic $q_\mu$ is defined as the profile likelihood ratio: 
$q_\mu = -2\ln(L(\mu,\hat{\hat{\theta}}_\mu)/L(\hat{\mu},\hat{\theta}))$,
where $\hat{\mu}$ and $\hat{\theta}$ are the values of the parameters that
maximise the likelihood function (with the constraint $0\leq \hat{\mu} \leq \mu$), and $\hat{\hat{\theta}}_\mu$ are the values of the
nuisance parameters that maximise the
likelihood function for a given value of $\mu$. 
Statistical uncertainties in each bin of the discriminant distributions are also taken into account via dedicated nuisance parameters in the fit.     
The test statistic $q_\mu$ is implemented in the {\sc RooFit} package~\cite{Verkerke:2003ir,RooFitManual} and
is used to measure the compatibility of the observed data with the background-only hypothesis 
by setting $\mu=0$ in the profile likelihood ratio: $q_0 = -2\ln(L(0,\hat{\hat{\theta}}_0)/L(\hat{\mu},\hat{\theta}))$.
The $p$-value (referred to as $p_0$) representing the compatibility of the data with the background-only hypothesis is estimated by integrating
the distribution of $q_0$ from background-only pseudo-experiments, approximated using the asymptotic formulae given in ref.~\cite{Cowan:2010js}, 
above the observed value of $q_0$. The observed $p_0$-value is checked for each explored signal scenario.
In the absence of any significant excess above the background expectation, upper limits on $\mu$, and thus on 
$\BR(t\to Hq)$ via equation~(\ref{eq:br_mu}), are derived by using $q_\mu$ in the CL$_{\rm{s}}$ method~\cite{Junk:1999kv,Read:2002hq}.
Values of $\BR(t\to Hq)$  yielding CL$_{\rm{s}}$$<$0.05,  where CL$_{\rm{s}}$ is computed using the asymptotic 
approximation~\cite{Cowan:2010js}, are excluded at $\geq$95\% CL.

\section{Results}
\label{sec:result}

This section presents the results obtained from the individual searches for $\Hq$, as well as their combination. 

\subsection{$H \to b\bar{b}$}
\label{sec:results_bb}

Following the statistical analysis discussed in section~\ref{sec:stat_analysis}, a binned likelihood fit under the signal-plus-background hypothesis 
is performed on the distributions of the final discriminant in the nine analysis channels considered. 
Figures~\ref{fig:prepostfit_unblinded_WbHc_2btagex}--\ref{fig:prepostfit_unblinded_WbHc_4btagin} show a comparison 
of the data and prediction in the final discriminant in each of the analysis channels, both pre- and 
post-fit to data, in the case of the $\Hc$ search.  The post-fit yields can be found in appendix~\ref{sec:prepostfit_yields_appendix}.
The best-fit branching ratio obtained is $\BR(t\to Hc)=[0.17 \pm 0.12\,({\rm stat.}) \pm 0.17\,({\rm syst.})]\%$,
assuming that $\BR(t\to Hu)=0$. 
A similar fit is performed for the $\Hu$ search, yielding $\BR(t\to Hu)=[-0.07 \pm 0.17\,({\rm stat.}) \pm 0.28\,({\rm syst.})]\%$,
assuming that $\BR(t\to Hc)=0$.  The different measured values for the two branching ratios is the result of the different sensitivities
of the $\Hc$ and $\Hu$ searches, as discussed in section~\ref{sec:event_categorisation}.

The large number of events in the analysis channels considered, together with their different background compositions, allows
the fit to place constraints on the combined effect of several sources of systematic uncertainty.
As a result, an improved background prediction is obtained with significantly reduced uncertainty, not only in the 
signal-depleted channels, but also in the most sensitive analysis channels for this search, (4 j, 3 b) and (4 j, 4 b).
The channels with two $b$-tags are used to constrain the leading uncertainties affecting the $\ttbar$+light-jets background prediction,
while the channels with $\geq$5 jets and $\geq$3 $b$-tags are sensitive to the uncertainties
affecting the $\ttbar$+HF background prediction. 
In particular, one of the main corrections applied by the fit is an increase 
of the $\ttbb$ normalisation by approximately 20\% relative to the nominal 
prediction by adjusting the corresponding nuisance parameter. This results
in an improved agreement between data and prediction in the (5 j, $\geq$4 b) and 
($\geq$6 j, $\geq$4 b)  channels, where the $\ttbb$ process provides a significant, 
or, in the latter case, dominant background contribution.\footnote{The overall 
change in $\ttbb$ normalisation can be different across channels due to the different 
impact of other nuisance parameters affecting $\ttbb$ modelling, such as that related 
to parton shower and hadronisation, which is changed by the fit by half of the prior
uncertainty.} In addition, the corresponding uncertainty is reduced from the initial 
50\% down to about 16\%.
This correction is in agreement with that found in ref.~\cite{Aad:2015gra}.
However, in contrast with ref.~\cite{Aad:2015gra}, the $\ttbb$ normalisation is not one of the leading 
uncertainties affecting this search, since the $\ttbb$ background is subdominant
in the (4 j, 3 b) channel, and it has  a very different final discriminant shape from that of the signal in the (\mbox{4 j}, \mbox{4 b}) channel.

As an illustration, figure~\ref{fig:ranking_bb_Hc} provides a summary of the leading 15 systematic uncertainties affecting the $\Hc$ search, 
quantifying their impact on the signal strength $\mu$, both before and after the fit, and displaying the constraints
provided by the data on the associated nuisance parameters. 
The pre-fit impact on $\mu$ is estimated by fixing the corresponding nuisance parameter at  $\rm{\theta_{0}} \pm \Delta\theta$,
where $\rm{\theta_{0}}$ is the nominal value of the nuisance parameter and $\Delta\theta$ is its pre-fit uncertainty, and performing 
the fit again. 
The difference between the default and modified $\mu$, $\Delta\mu$, represents the effect of the systematic uncertainty in question on $\mu$.
The same procedure is followed to estimate the post-fit impact on $\mu$,  but the corresponding nuisance parameter is instead fixed 
at  $\hat{\rm{\theta}} \pm \sigma_{\rm{\theta}}$,  where $\hat{\rm{\theta}}$ is the fitted value of the nuisance parameter and 
$\sigma_{\rm{\theta}}$ is its post-fit uncertainty.
For reference, $\Delta\mu=0.05$ corresponds to $\Delta\BR(t \to Hc)\simeq 0.05\%$.

Prior to the fit, the systematic uncertainties with the largest impact on $\mu$ are
the leading uncertainty for light-jet tagging and the uncertainty on the $\ttbar$ background associated with the choice of
parton shower and hadronisation models. The significant impact from light-jet tagging results from 
the large fraction of $\ttbar$+light-jets background present in the (4 j, 4 b) channel, which peaks at high values of the 
final discriminant, like the signal, and thus cannot be strongly constrained by the fit. 
Because of this, this uncertainty remains the leading one after the fit. 
In contrast, the uncertainty related to  $\ttbar$ modelling is significantly constrained by the fit since it has a large
impact ($\sim$5--16\%) on the $\ttbar$+light-jets background normalisation in the highly populated channels with two $b$-tags. 
As a result, this uncertainty is ranked only fourth in importance after the fit, becoming comparable to uncertainties 
such as the choice of renormalisation scale for $\ttbb$, the leading uncertainty for $c$-jet tagging and the $\ttcc$ normalisation.
Of these, the nuisance parameter associated with the choice of the renormalisation scale for $\ttbb$ 
is slightly pulled (by half of the prior uncertainty) to improve agreement with the data in the (4 j, 4 b) and (5 j, 4 b) channels. 
In these channels, this uncertainty causes variations of up to $\sim$5\% in the bin contents in some regions of the final discriminant, 
i.e. distorting its shape compared to that of the nominal prediction, but the sensitivity is not sufficient to constrain it significantly.
The leading uncertainty from $c$-tagging causes small (few percent) distortions in the shape of the background, and also 
cannot be constrained by the fit. In contrast, the fit is sensitive to the $\ttcc$ normalisation and the second-leading uncertainty 
for $c$-tagging,\footnote{The main effect of this uncertainty is a change in normalisation for the background with almost no effect on its shape
in the final discriminant.} through the comparison of data 
and predictions across channels with different $b$-tag multiplicity, yielding results in agreement with the nominal predictions
but with half the initial uncertainties. 

Other nuisance parameters have a smaller impact on the signal extraction and typically have small pulls or constraints.
One exception is the nuisance parameter associated with the $\ttbar$ cross section, which affects the signal extraction indirectly 
through the existing small fraction of non-$\ttbar$ background,
and after the fit is found to be consistent with the nominal prediction but is constrained owing to the large 
number of $\ttbar$ events. On the other hand, a slight pull is obtained for the nuisance parameter associated with one of the uncertainties
for the top quark $\pt$ and $\ttbar$ system $\pt$ reweightings, which is used by the fit to improve agreement between data and prediction 
in the channels with two $b$-tags but which has very small effect on the background prediction in the signal region.

In the absence of a significant excess in data above the background expectation, 95\% CL limits are set on $\BR(t\to Hc)$ and $\BR(t\to Hu)$.
The observed (expected) 95\% CL upper limits on the branching ratios 
are $\BR(t\to Hc)<0.56\%\,(0.42\%)$ and $\BR(t\to Hu)<0.61\%\,(0.64\%)$.
These upper limits can be translated into corresponding observed (expected) limits on the couplings 
of $|\lamHc|<0.14\,(0.12)$ and $|\lamHu|<0.15\,(0.15)$.

\begin{figure*}[htbp]
\begin{center}
\subfloat[]{\includegraphics[width=0.35\textwidth]{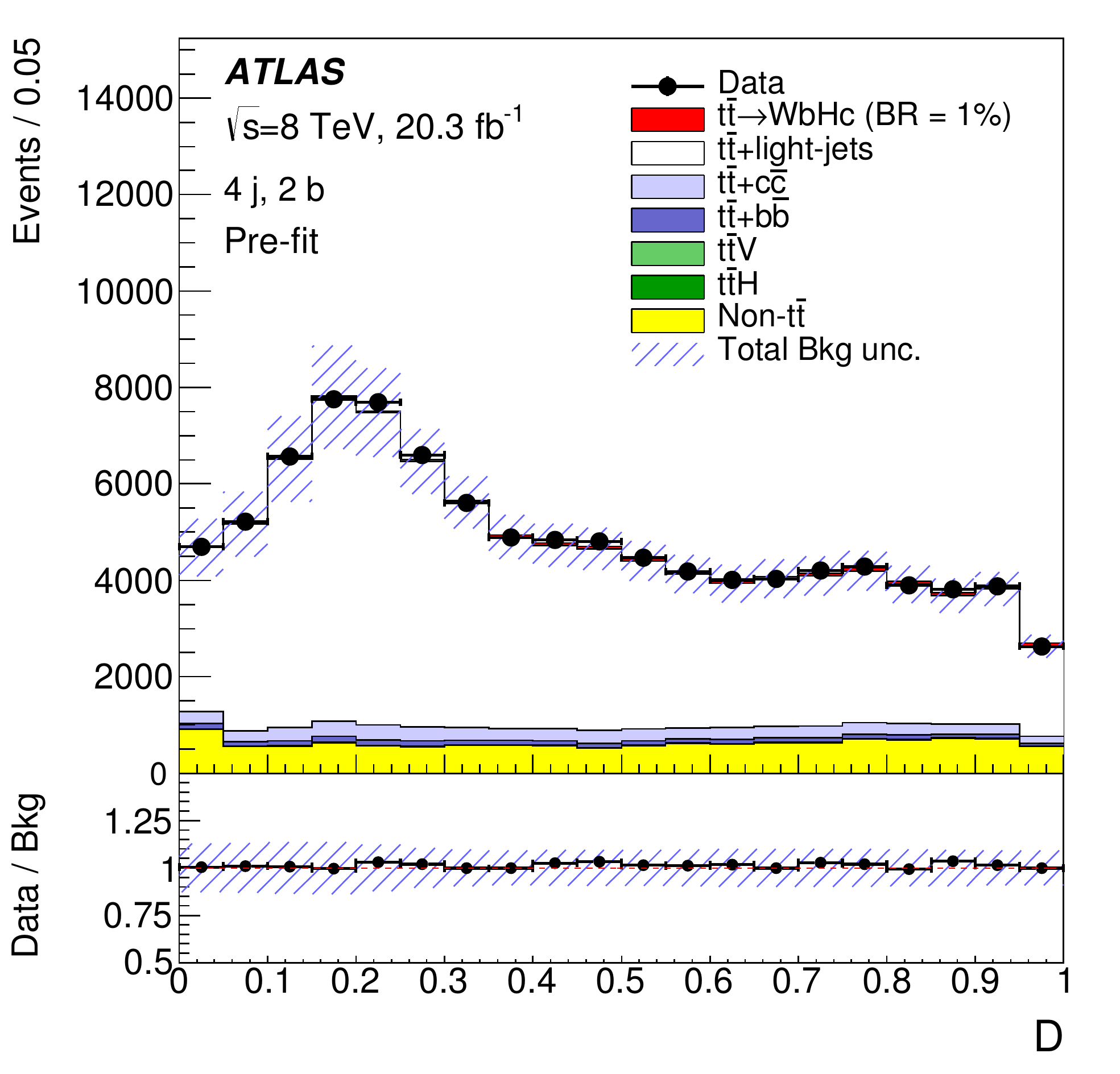}}
\subfloat[]{\includegraphics[width=0.35\textwidth]{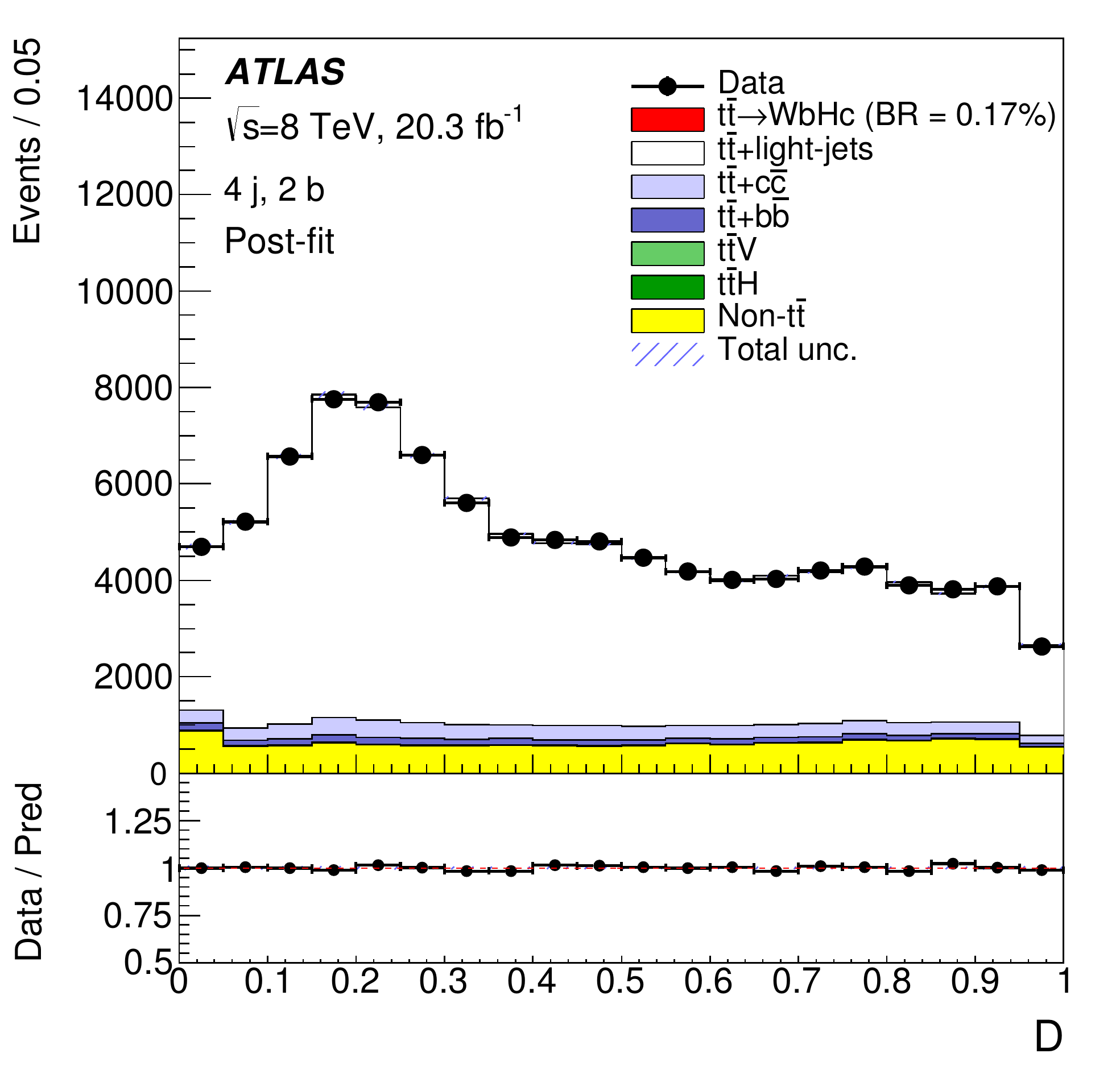}} \\
\subfloat[]{\includegraphics[width=0.35\textwidth]{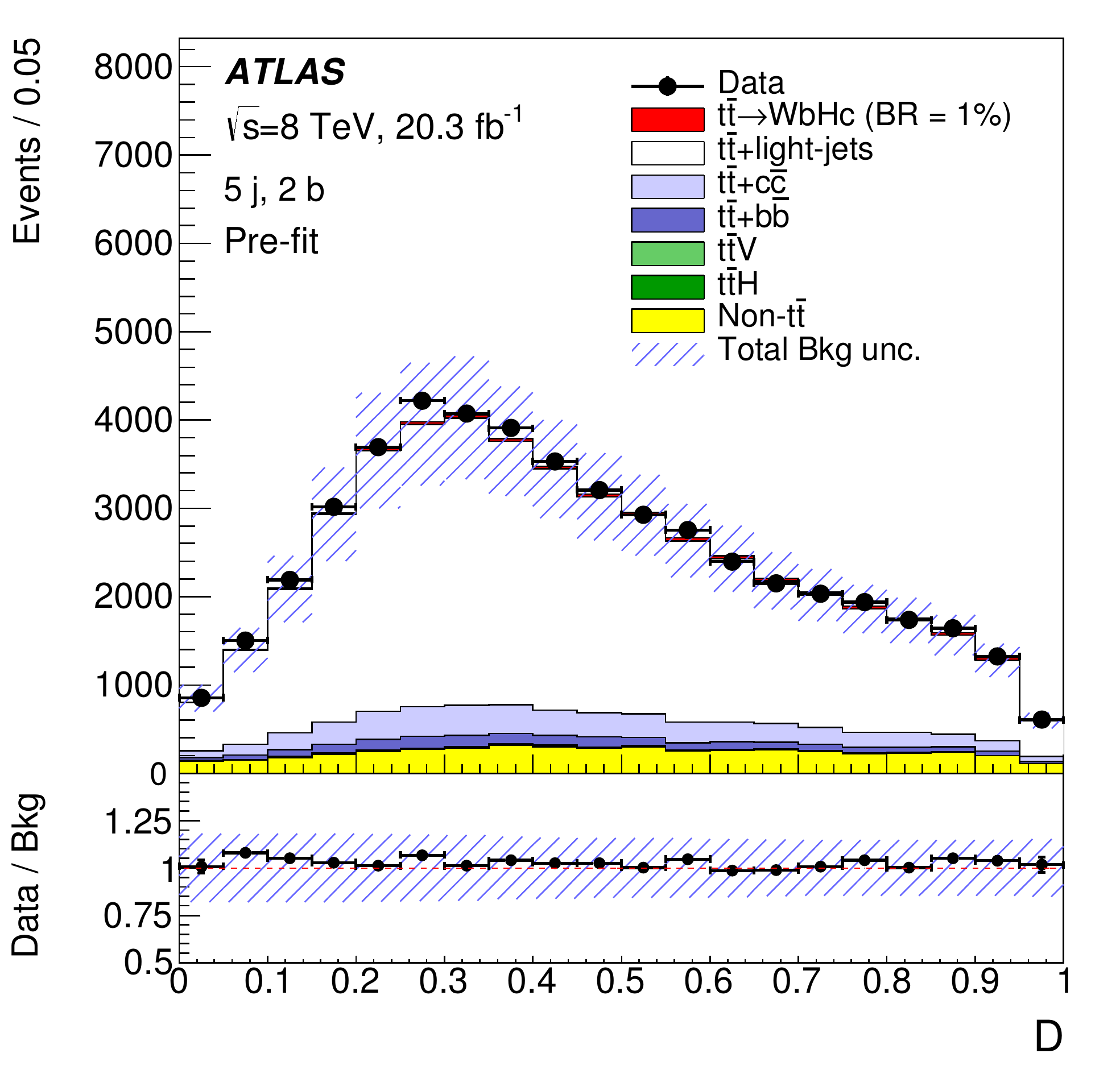}}
\subfloat[]{\includegraphics[width=0.35\textwidth]{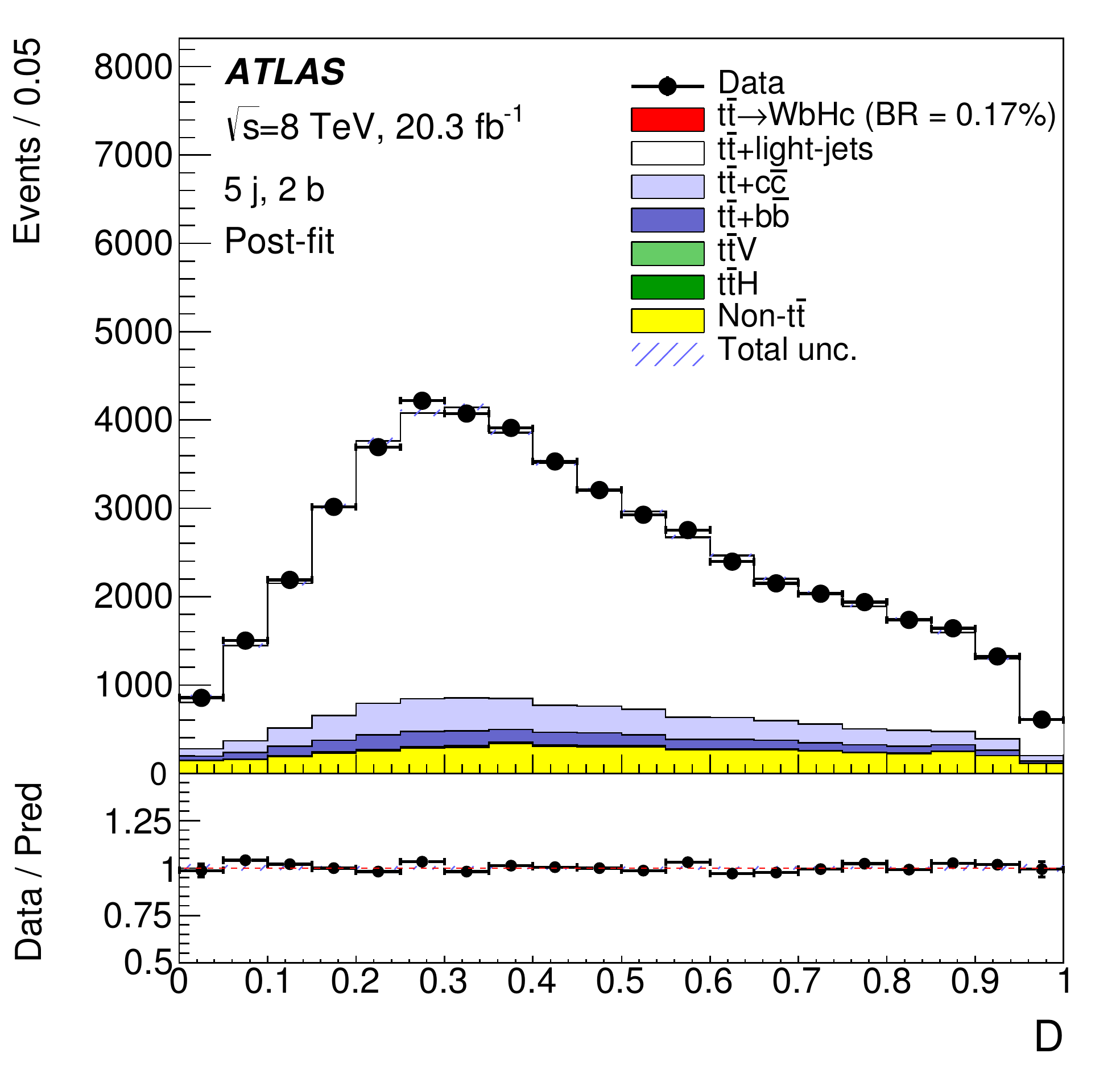}} \\
\subfloat[]{\includegraphics[width=0.35\textwidth]{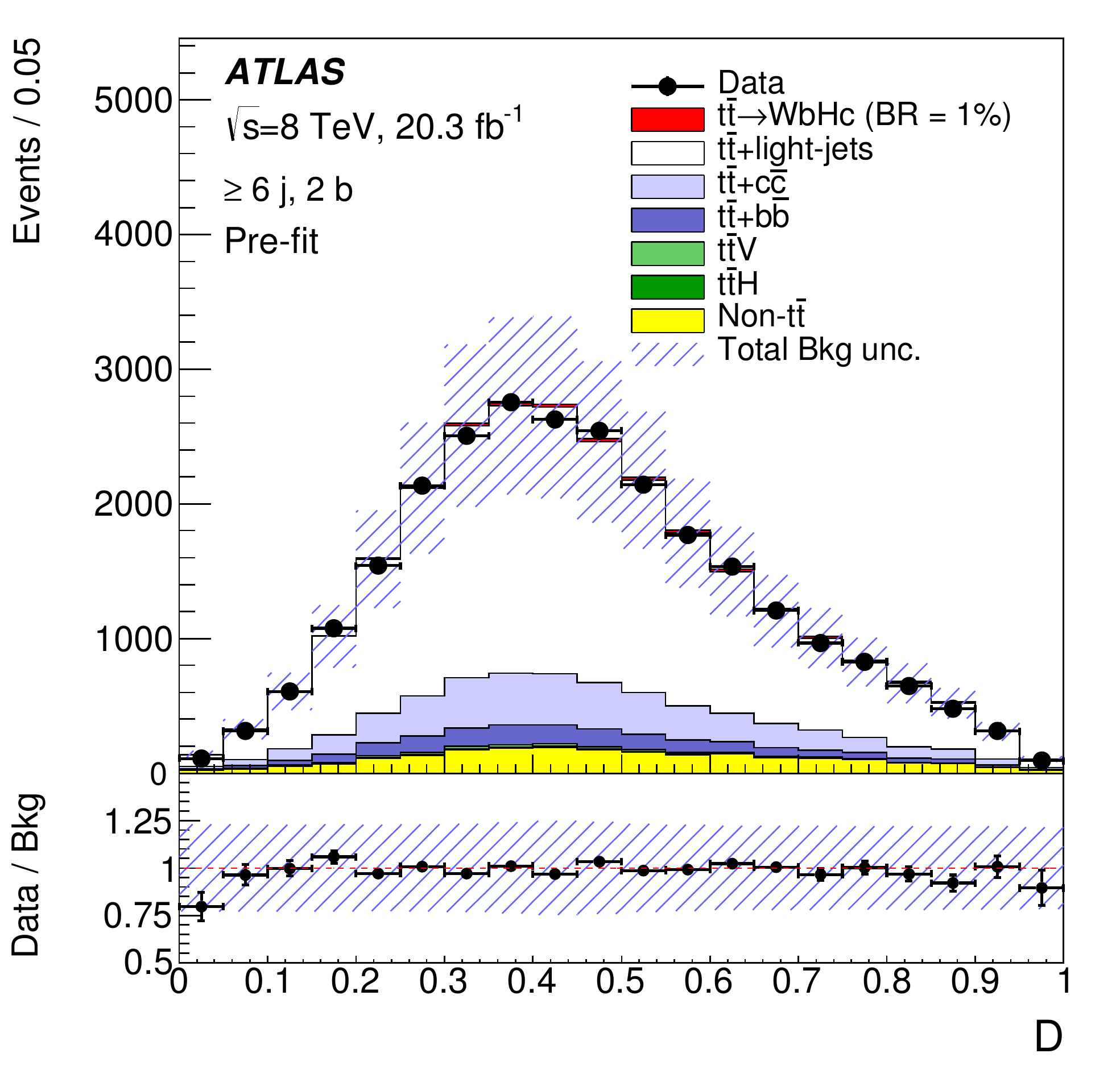}}
\subfloat[]{\includegraphics[width=0.35\textwidth]{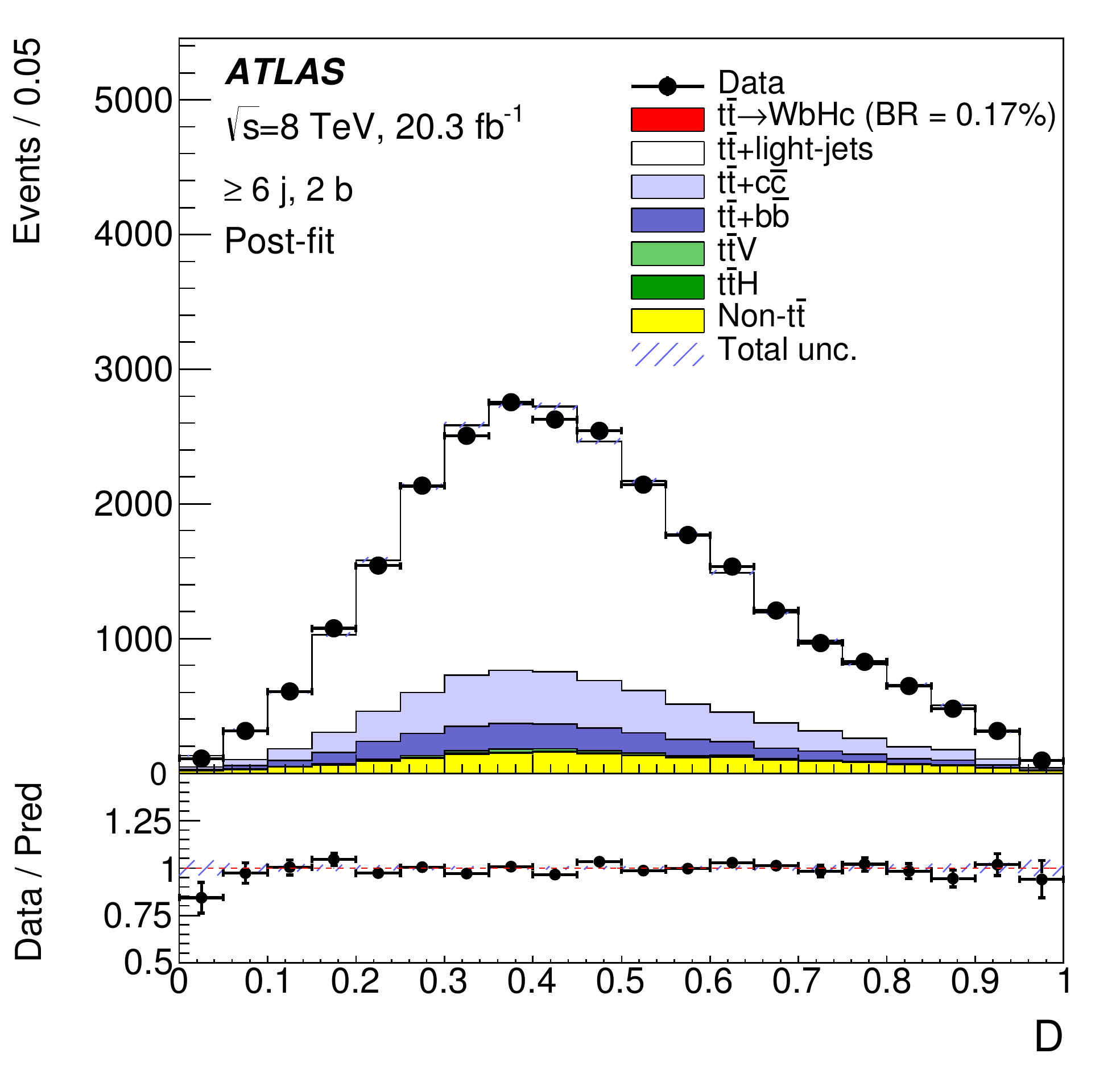}} \\
\caption{$\Hcbb$ search: comparison between the data and prediction for the distribution of the $D$ discriminant used in the (4 j, 2 b) channel 
(a) before the fit and (b) after the fit,  in the (5 j, 2 b) channel (c) before the fit and (d) after the fit, and in the ($\geq$6 j, 2 b) channel 
(e) before the fit and (f) after the fit. The fit is performed on data under the signal-plus-background hypothesis.  
In the pre-fit distributions the $\Hc$ signal (solid red) is normalised to $\BR(t\to Hc)=1\%$ and the $\ttbar \to WbWb$ background is normalised to the
SM prediction, while in the post-fit distributions both signal and $\ttbar \to WbWb$ background are normalised using the 
best-fit $\BR(t\to Hc)$.
The small contributions from $W/Z$+jets,  single top, diboson and multijet backgrounds are combined into a single background source 
referred to as ``Non-$\ttbar$''.
The bottom panels display the ratios of data to either the SM background prediction before the fit (``Bkg'')  or the total signal-plus-background
prediction after the fit (``Pred''). 
The hashed area represents the total uncertainty on the background.}
\label{fig:prepostfit_unblinded_WbHc_2btagex}
\end{center}
\end{figure*}

\begin{figure*}[htbp]
\begin{center}
\subfloat[]{\includegraphics[width=0.35\textwidth]{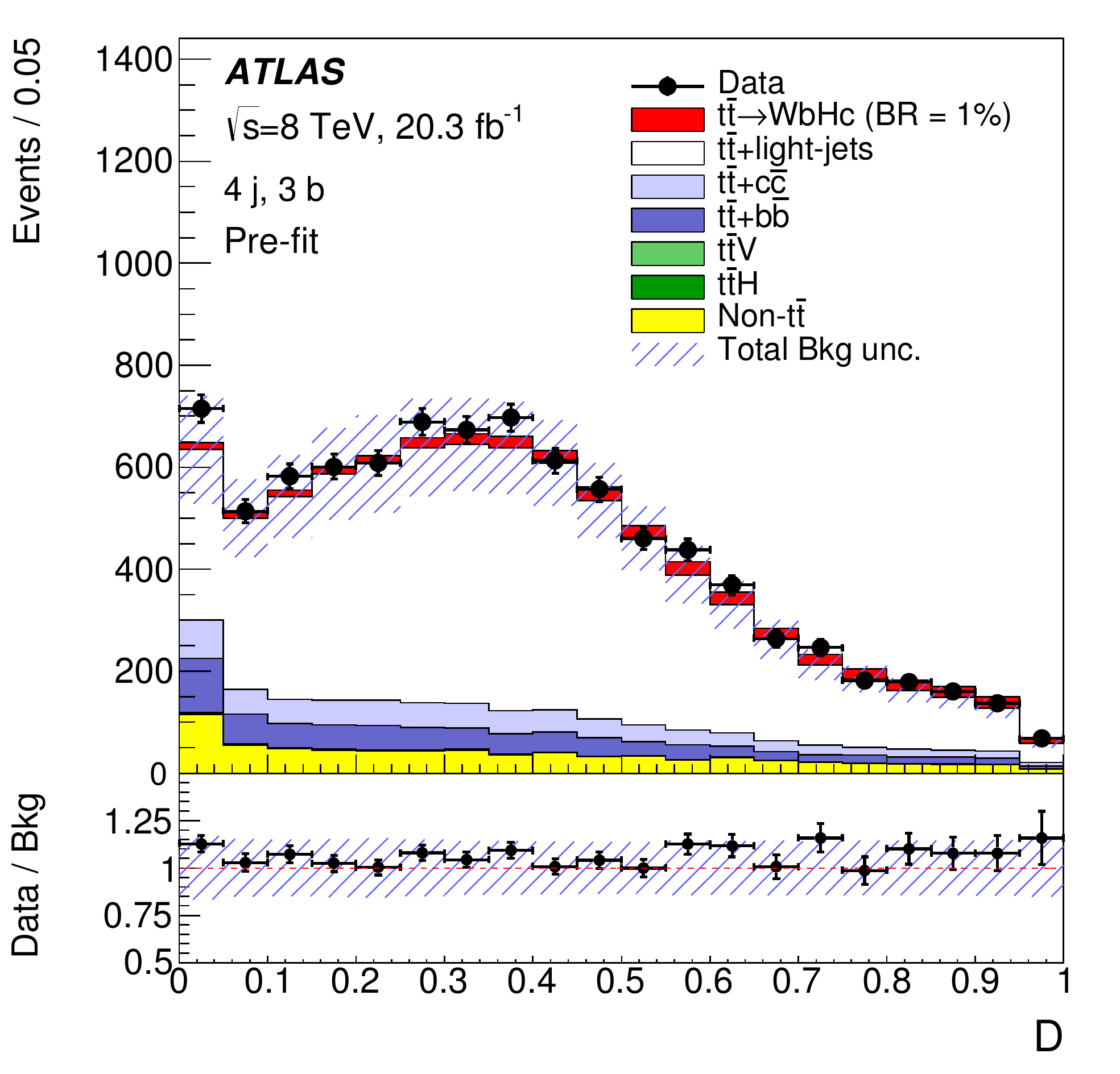}}
\subfloat[]{\includegraphics[width=0.35\textwidth]{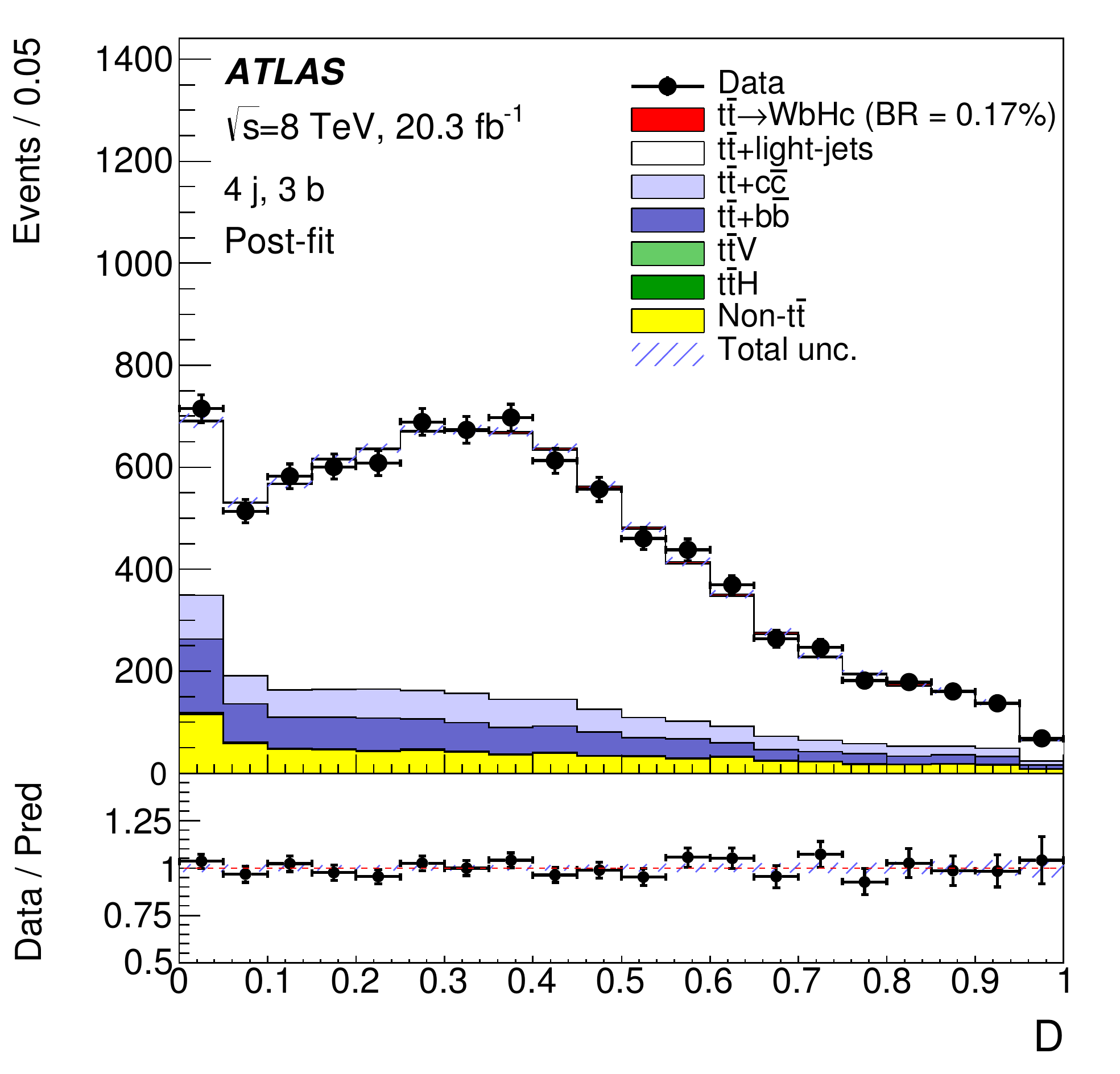}} \\
\subfloat[]{\includegraphics[width=0.35\textwidth]{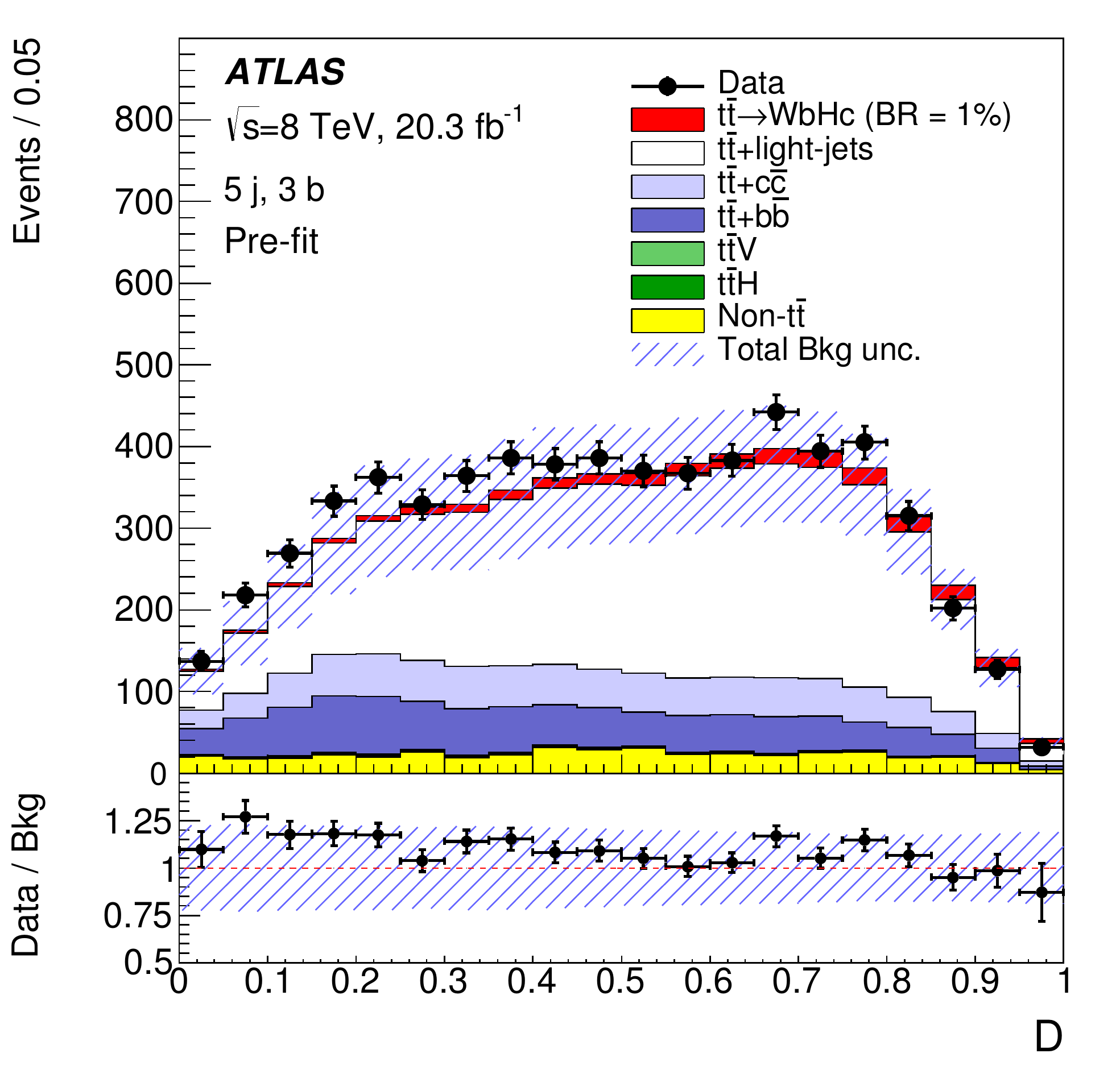}}
\subfloat[]{\includegraphics[width=0.35\textwidth]{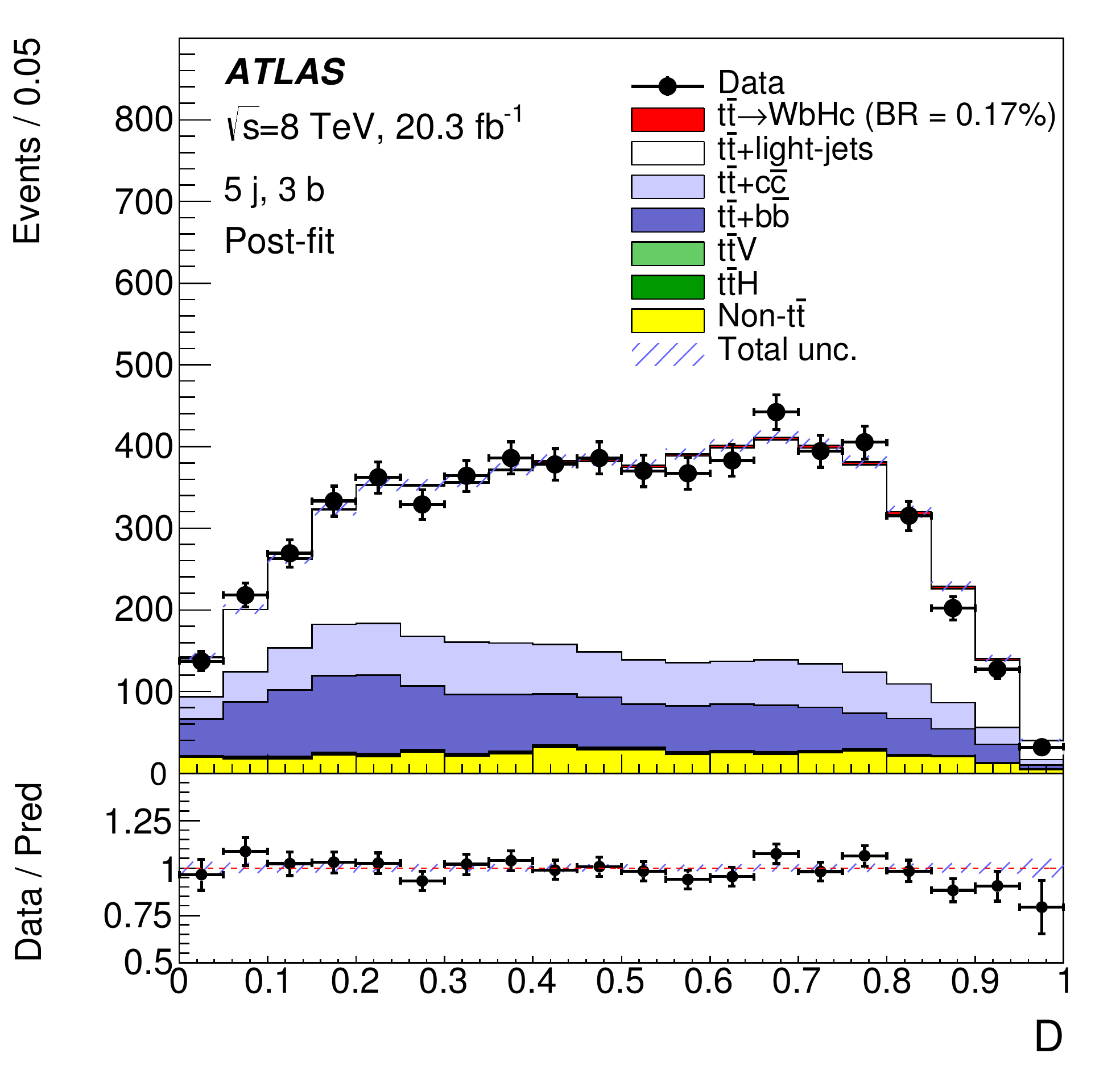}} \\
\subfloat[]{\includegraphics[width=0.35\textwidth]{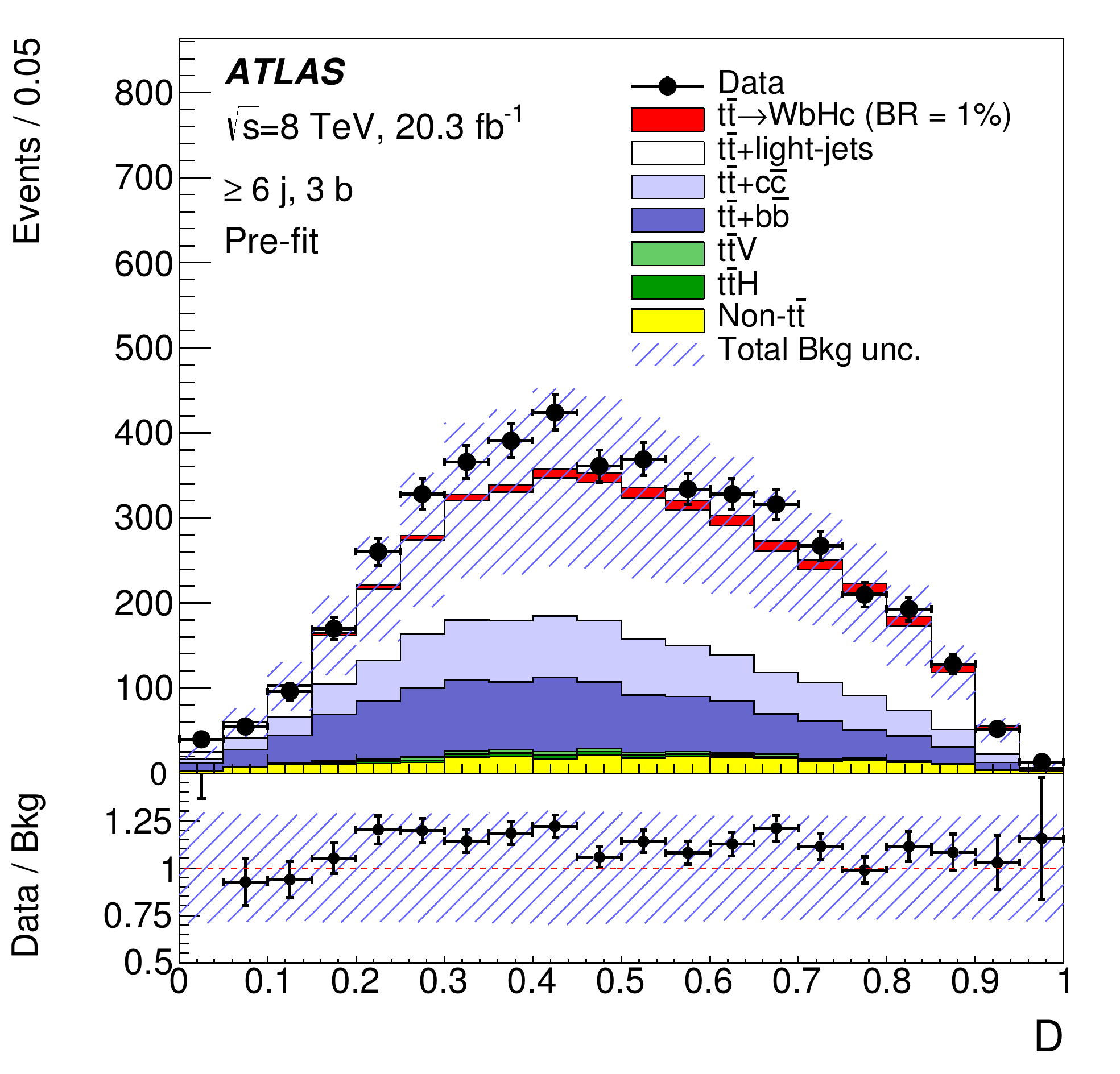}}
\subfloat[]{\includegraphics[width=0.35\textwidth]{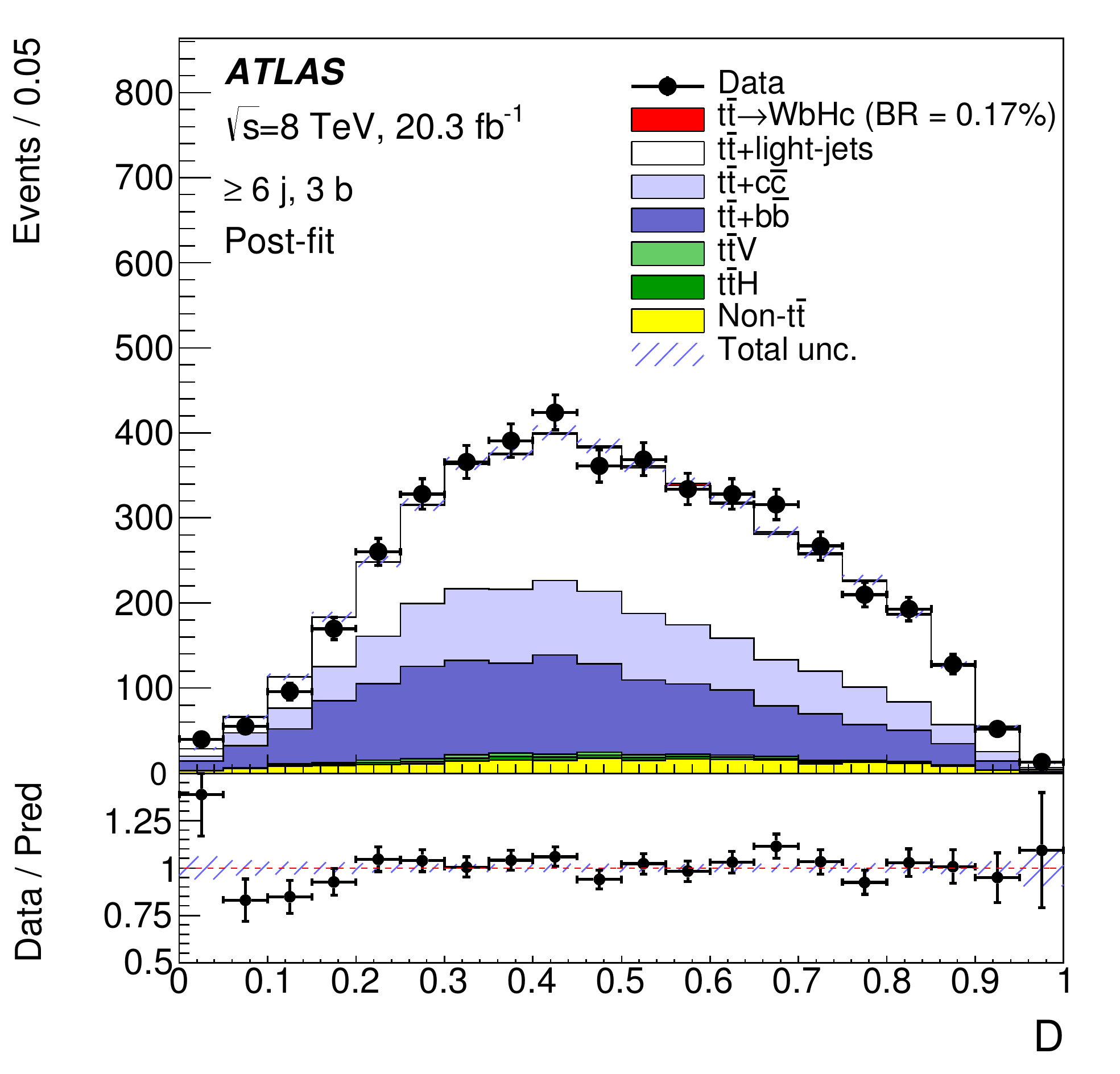}} \\
\caption{$\Hcbb$ search: comparison between the data and prediction for the distribution of the $D$ discriminant used in the (4 j, 3 b) channel 
(a) before the fit and (b) after the fit,  in the (5 j, 3 b) channel (c) before the fit and (d) after the fit, and in the ($\geq$6 j, 3 b) channel 
(e) before the fit and (f) after the fit. The fit is performed on data under the signal-plus-background hypothesis.  
In the pre-fit distributions the $\Hc$ signal (solid red) is normalised to $\BR(t\to Hc)=1\%$ and the $\ttbar \to WbWb$ background is normalised to the
SM prediction, while in the post-fit distributions both signal and $\ttbar \to WbWb$ background are normalised using the 
best-fit $\BR(t\to Hc)$.
The small contributions from $W/Z$+jets,  single top, diboson and multijet backgrounds are combined into a single background source 
referred to as ``Non-$\ttbar$''.
The bottom panels display the ratios of data to either the SM background prediction before the fit (``Bkg'')  or the total signal-plus-background
prediction after the fit (``Pred''). 
The hashed area represents the total uncertainty on the background.}
\label{fig:prepostfit_unblinded_WbHc_3btagex}
\end{center}
\end{figure*}

\begin{figure*}[htbp]
\begin{center}
\subfloat[]{\includegraphics[width=0.35\textwidth]{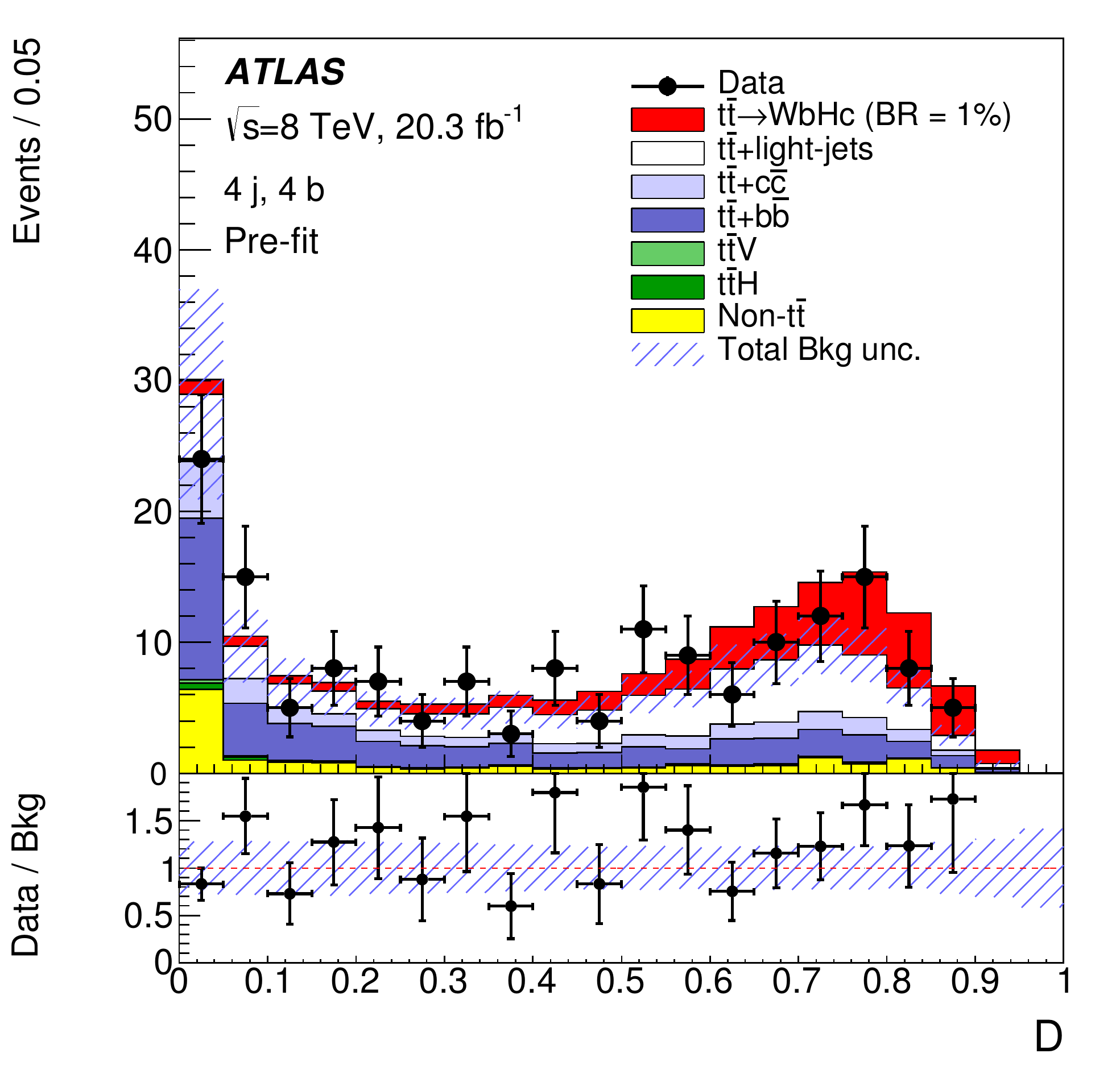}}
\subfloat[]{\includegraphics[width=0.35\textwidth]{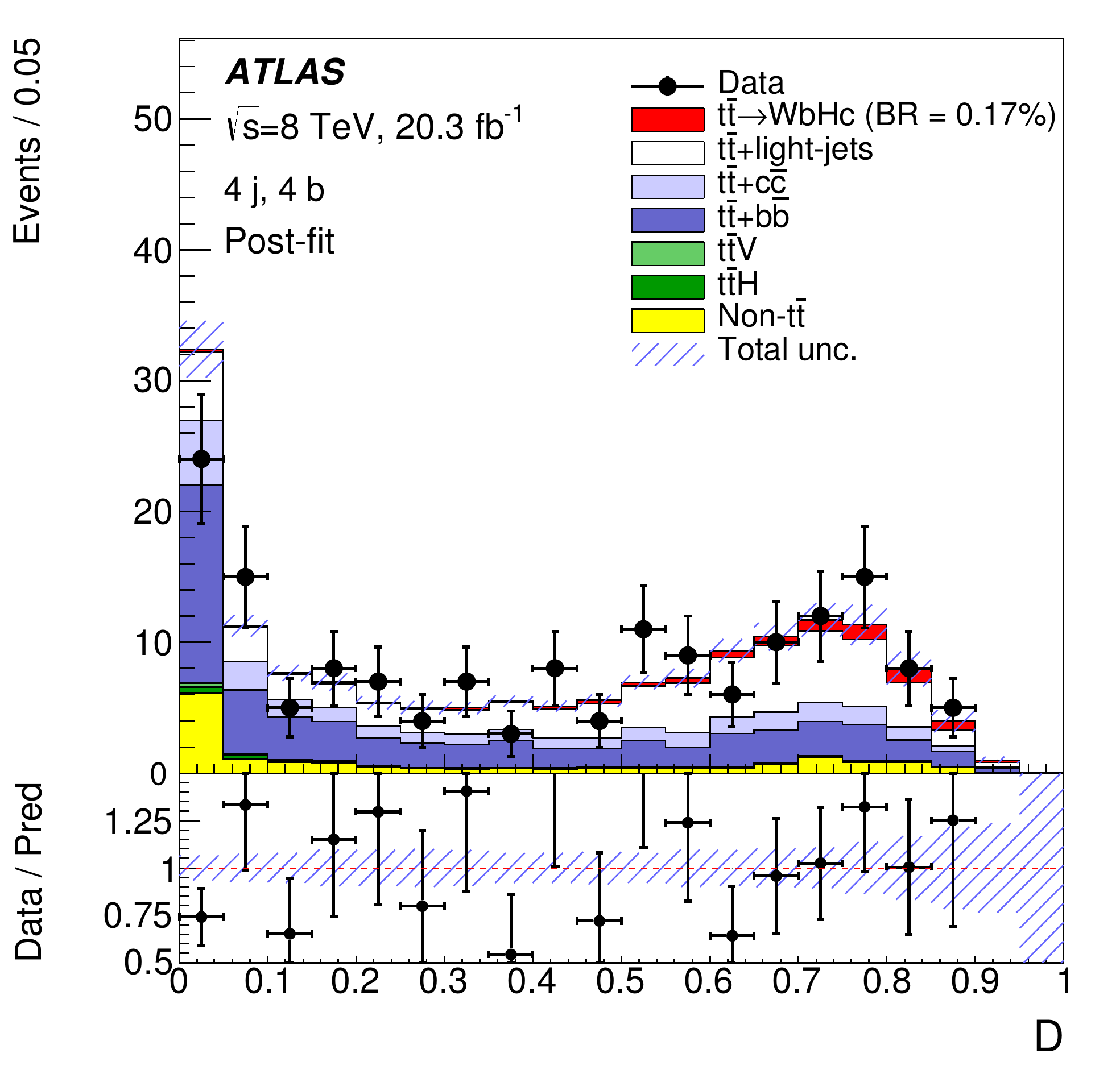}} \\
\subfloat[]{\includegraphics[width=0.35\textwidth]{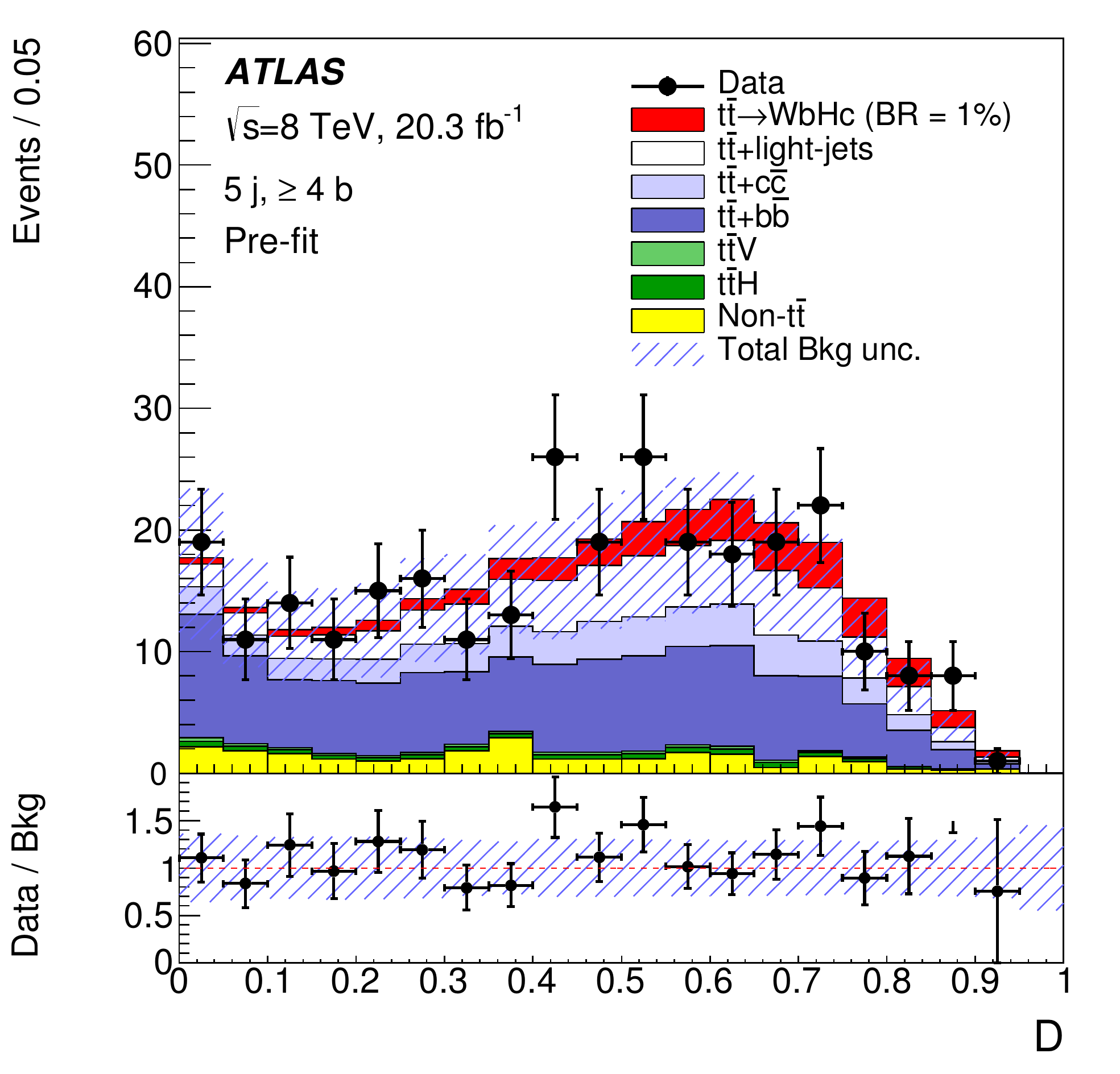}}
\subfloat[]{\includegraphics[width=0.35\textwidth]{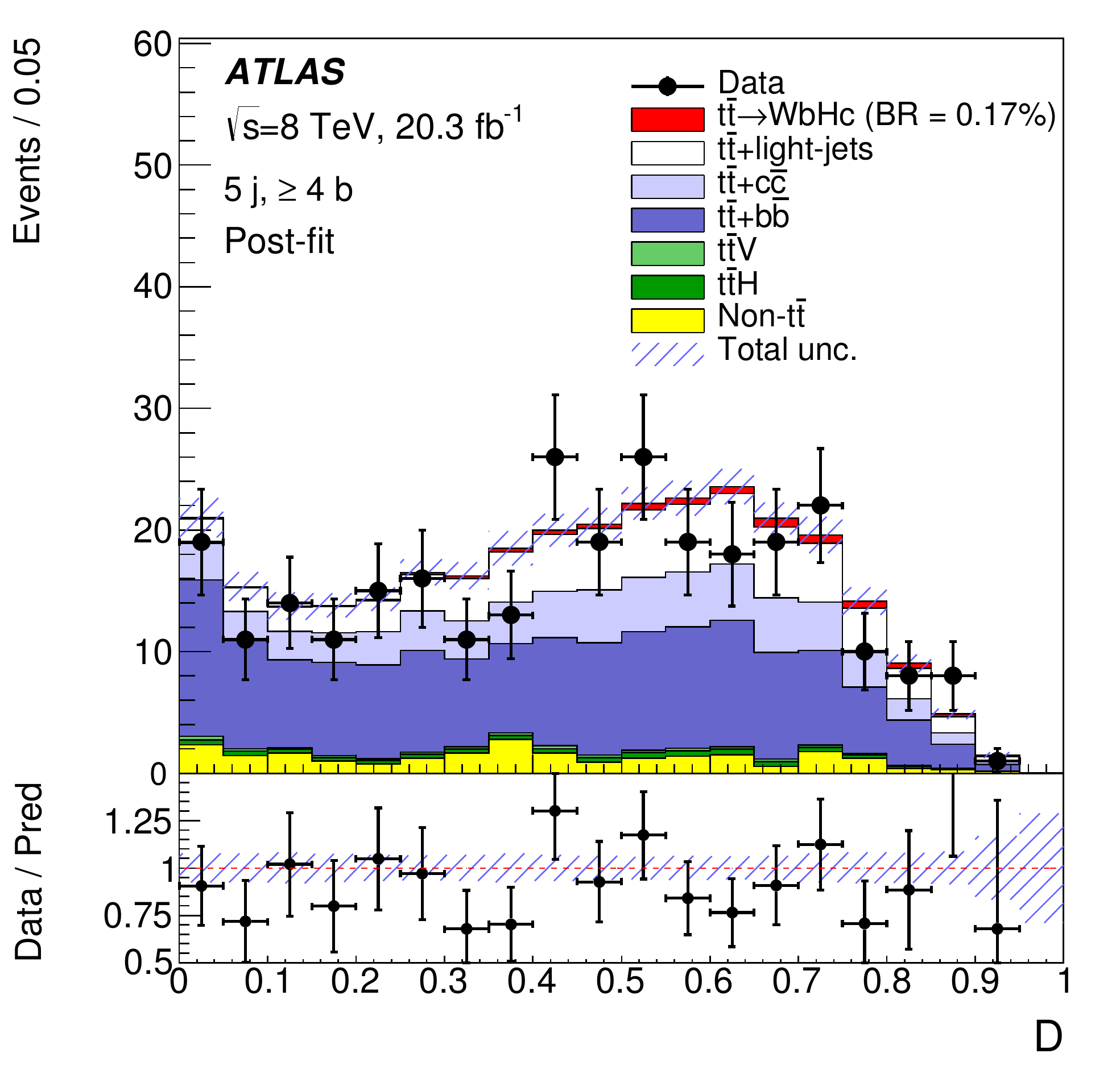}} \\
\subfloat[]{\includegraphics[width=0.35\textwidth]{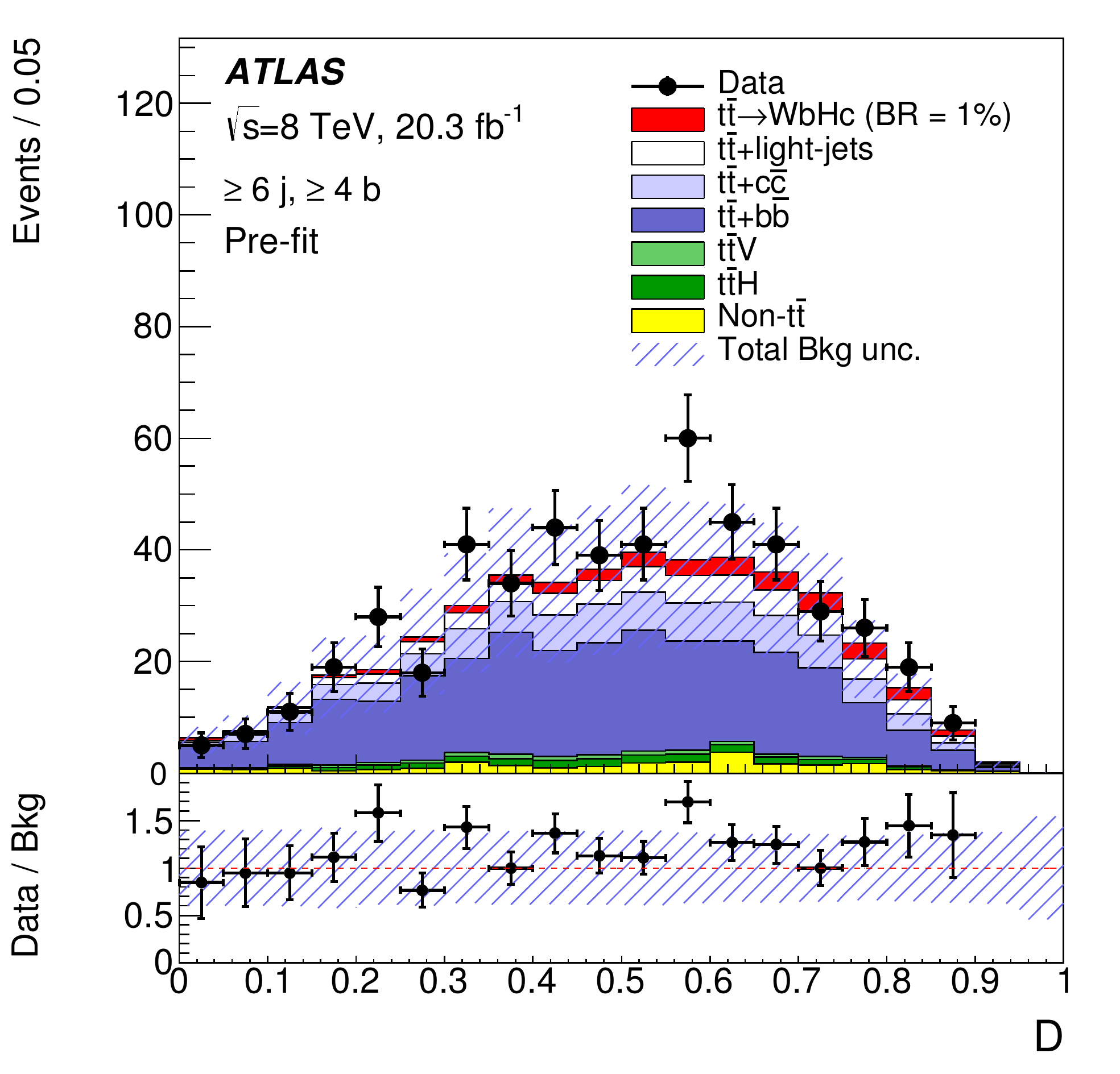}}
\subfloat[]{\includegraphics[width=0.35\textwidth]{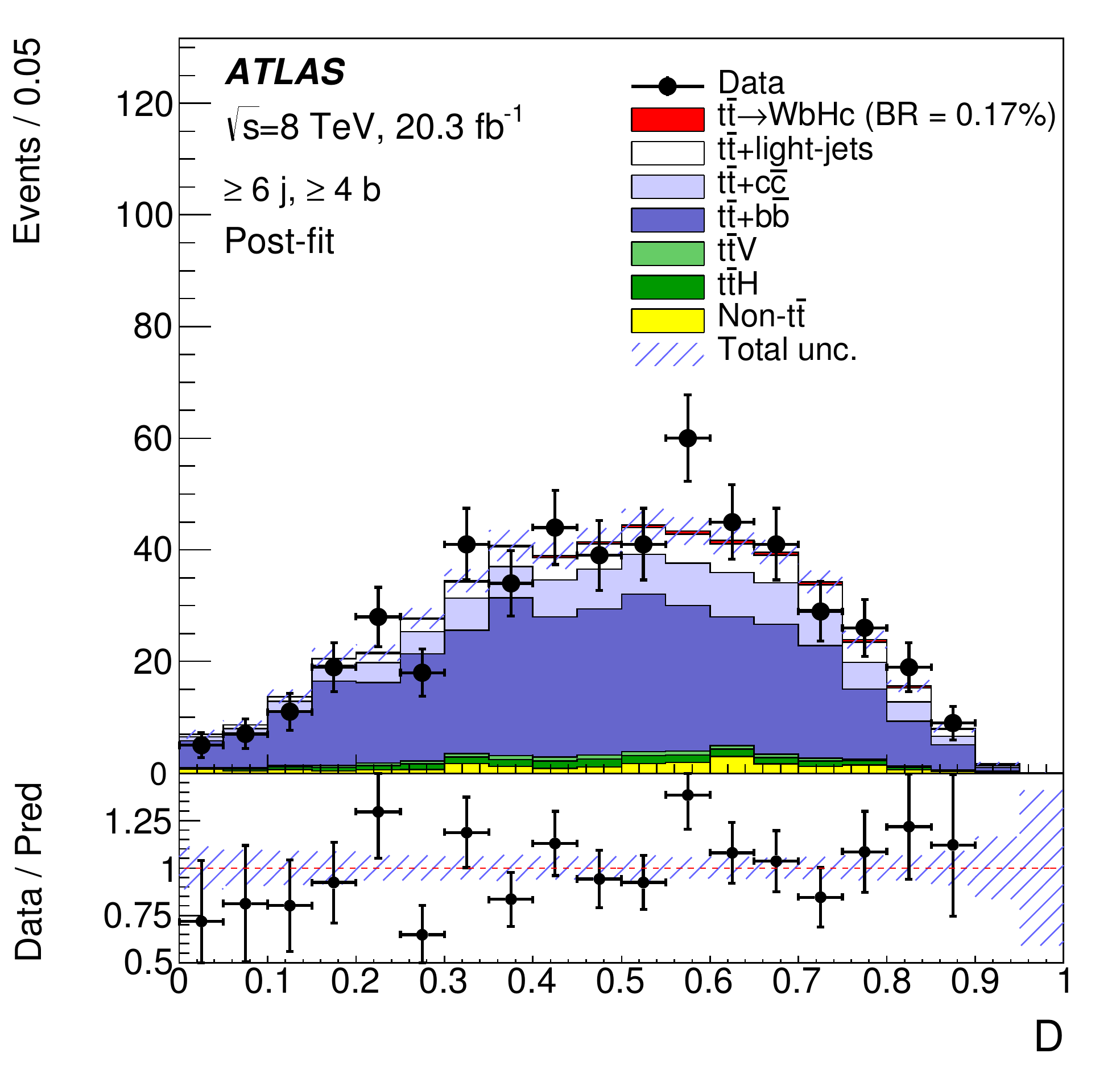}} \\
\caption{$\Hcbb$ search: comparison between the data and prediction for the distribution of the $D$ discriminant used in the (4 j, 4 b) channel 
(a) before the fit and (b) after the fit,  in the (5 j, $\geq$4 b) channel (c) before the fit and (d) after the fit, and in the ($\geq$6 j, $\geq$4 b) channel (e) before the fit and (f) after the fit. The fit is performed on data under the signal-plus-background hypothesis.  
In the pre-fit distributions the $\Hc$ signal (solid red) is normalised to $\BR(t\to Hc)=1\%$ and the $\ttbar \to WbWb$ background is normalised to the
SM prediction, while in the post-fit distributions both signal and $\ttbar \to WbWb$ background are normalised using the 
best-fit $\BR(t\to Hc)$.
The small contributions from $W/Z$+jets,  single top, diboson and multijet backgrounds are combined into a single background source 
referred to as ``Non-$\ttbar$''.
The bottom panels display the ratios of data to either the SM background prediction before the fit (``Bkg'')  or the total signal-plus-background
prediction after the fit (``Pred''). 
The hashed area represents the total uncertainty on the background.}
\label{fig:prepostfit_unblinded_WbHc_4btagin}
\end{center}
\end{figure*}

\begin{figure*}[htbp!]
\begin{center}
\includegraphics[width=0.60\textwidth]{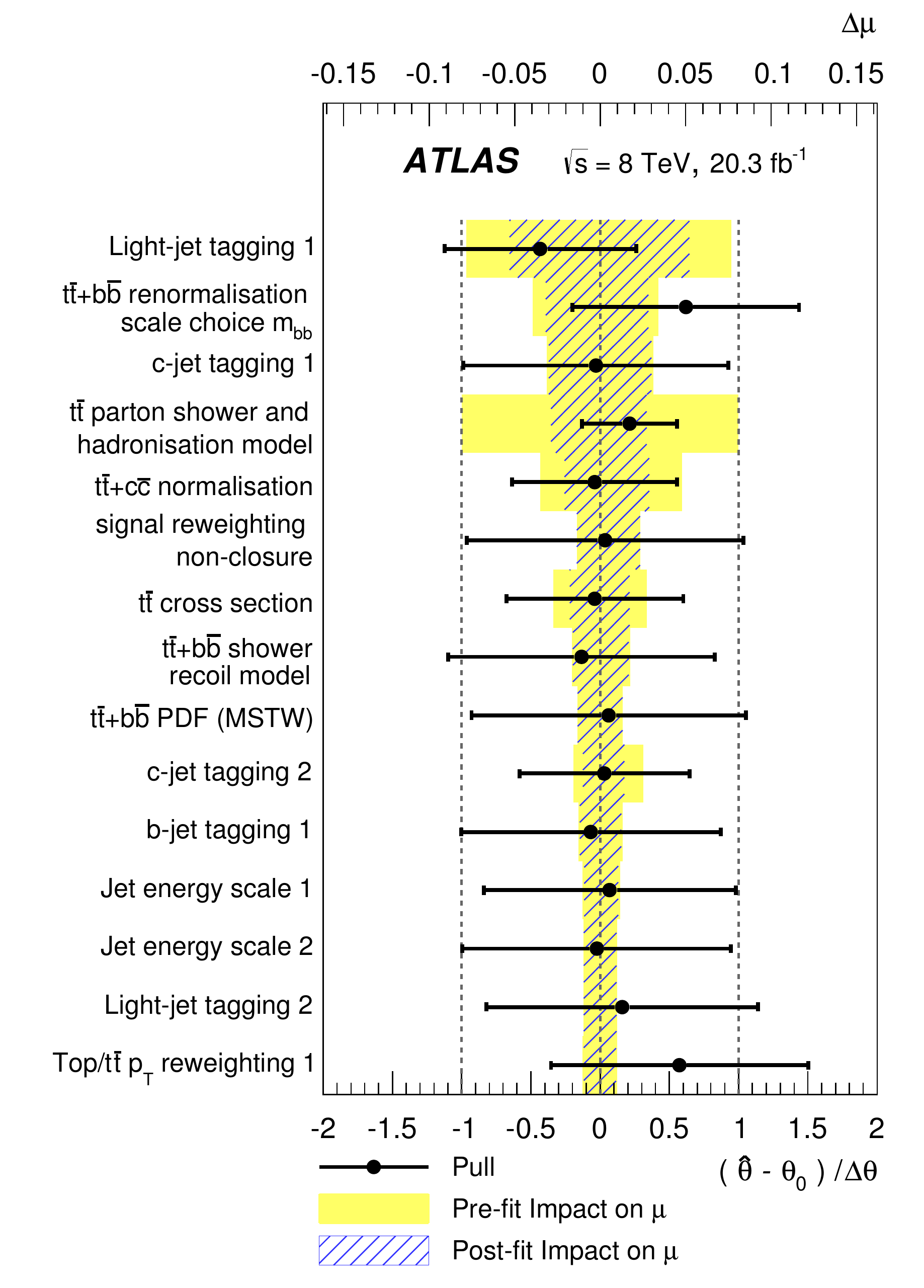}
\caption{\small {$\Hcbb$ search: the fitted values of the nuisance parameters for the most important sources
of systematic uncertainty and their impact on the measured signal strength. 
The points, which are drawn conforming to the scale of the bottom axis, show the deviation of each of the 
fitted nuisance parameters, $\hat{\rm{\theta}}$, from $\rm{\theta_{0}}$, which is the nominal value of that nuisance 
parameter, in units of the pre-fit standard deviation $\Delta\theta$.
The error bars show the post-fit uncertainties, 
$\sigma_{\theta}$, which are close to 1 if the data do not provide any further constraint on that uncertainty.
Conversely, a value of $\sigma_{\theta}$ much smaller than 1 indicates a significant reduction with respect to 
the original uncertainty. The nuisance parameters are sorted according to their post-fit effect on 
$\mu$ (hashed blue area), conforming to the scale of the top axis, with those with the largest impact at the top. }}
\label{fig:ranking_bb_Hc}
\end{center}
\end{figure*}

\subsection{$H \to \gamma\gamma$}
\label{sec:results_gg}

A search for $\Hqgg$ published by the ATLAS Collaboration uses a data set corresponding to an integrated luminosity 
of 4.5~fb$^{-1}$ at $\sqrt{s}= 7\tev$ and 20.3 fb$^{-1}$ at $\sqrt{s}=8\tev$~\cite{Aad:2014dya}.
The event selection requires at least two reconstructed photon candidates and additional requirements
to select $t \to Wb$ decays. Events are categorised into two channels, leptonic and hadronic, depending 
on the $W$ boson decay modes. The leptonic channel selects events with exactly one lepton ($e$ or $\mu$), 
at least two jets, and at least one $b$-tagged jet.  The hadronic channel selects events
with no reconstructed lepton, at least four jets, and at least one $b$-tagged jet. In both channels, additional
requirements are made to select events compatible with $\Hq$ production by exploiting the invariant masses
of the reconstructed top quark candidates. 
Finally, the diphoton mass ($m_{\gamma\gamma}$) distribution of the selected events is analysed using a sideband
technique in order to estimate the background in the signal region, defined to be $122\gev \leq m_{\gamma\gamma} \leq 129\gev$.

Based on the above strategy, this search has essentially no discrimination power between $\Hc$ and $\Hu$ signals,
because their selection acceptances  are very close, although not identical.
To facilitate combining it with the other searches discussed in this paper,  minor modifications 
to the inputs were made with respect to the published result, all having a negligible impact on the result.
They include updates to the $\ttbar$ cross-section uncertainty and the uncertainty model for Higgs branching ratios,
as well as the separate treatment of $\Hc$ and $\Hu$ signals taking into account their slightly different acceptances.
The best-fit branching ratios obtained are 
$\BR(t\to Hc)=[0.22 \pm 0.26\,({\rm stat.}) \pm 0.10\,({\rm syst.})]\%$ and $\BR(t\to Hu)=[0.23 \pm 0.27\,({\rm stat.}) \pm 0.10\,({\rm syst.})]\%$
under the assumptions that $\BR(t\to Hu)=0$ and $\BR(t\to Hc)=0$ respectively.
The observed (expected) 95\% CL upper limits on the branching ratios from ref.~\cite{Aad:2014dya} 
remain the best estimates, $\BR(t\to Hq)<0.79\%\,(0.51\%)$. The corresponding limits on the couplings are $|\lamHq|<0.17\,(0.14)$.
These limits can be understood as applying to the sum of the $t \to Hc$ and $t \to Hu$ decay modes, or only to one of them, 
if the other decay mode is assumed to have a branching ratio equal to zero. In the former case, $\BR(t\to Hq) \equiv \BR(t\to Hc)+\BR(t\to Hu)$ and
$|\lamHq| \equiv \sqrt{|\lamHc|^2+|\lamHu|^2}$.

\subsection{$H \to W^+W^-, \tau^+\tau^-$}
\label{sec:results_ML}

The $H \to WW^*$ and $H\to\tau\tau$  decay modes are predicted to have significant branching ratios, of 21.5\% and 6.3\% respectively.
The resulting signatures for signal events, $\Hq \to W^\pm W^\pm W^\mp bq$ and $\Hq \to W^\pm \tau^\pm \tau^\mp b q$,
can be effectively exploited to suppress backgrounds in the case of multilepton final states. 

Recently, the ATLAS Collaboration has published a search for $\ttbar H$ production with
$H \to WW^*$, $\tau\tau$ and $ZZ^*$~\cite{Aad:2015iha} that exploits several multilepton
signatures resulting from the leptonic decays of $W$ and $Z$ bosons and/or the presence
of $\tau$ leptons. That search considered five separate event categories depending on the
number of reconstructed electrons or muons and hadronic $\tau$ candidates, of which the
following three are considered for a reinterpretation in the context of the $\Hq$ search:
\begin{itemize}
\item $\twolnotau$: two same-charge light leptons ($e$ or $\mu$) with no hadronic $\tau$ candidates ($\tau_{\rm had}$)
and $\geq$4 jets with $\geq$1 $b$-tagged jets. This channel is sensitive to the process 
$\Hq \to \ell^\pm \ell^\pm qqqb2\nu$. To further improve the sensitivity, this category is further subdivided 
into six subcategories depending on the flavour of the leptons and the number of jets:
($ee$, $\mu\mu$, $e\mu$)$\times$(4 jets, $\geq$5 jets).
\item $\threel$: three light leptons with either $\geq$3 jets of which $\geq$2 are $b$-tagged, or $\geq$4 jets of which 
$\geq$1 are $b$-tagged. This channel is sensitive to the process  $\Hq \to \ell^\pm \ell^\pm \ell^\mp qb3\nu$. 
\item $\twolonetau$: two same-charge light leptons ($e$ or $\mu$), one $\tau_{\rm had}$ candidate,
and $\geq$4 jets with $\geq$1 $b$-tagged jets. This channel is sensitive to the process 
$\Hq \to \ell^\pm \ell^\pm \tau^\mp qb3\nu$. 
\end{itemize}
The two other categories considered in the $\ttbar H$ search in multilepton final states 
but not in this reinterpretation are: $4\ell$ (four light leptons with $\geq$2 jets and $\geq$1 $b$-tagged jets) and 
$1\ell 2\tau_{\rm had}$ (one light lepton and two opposite-charge $\tau_{\rm had}$ candidates, 
with $\geq$3 jets and $\geq$1 $b$-tagged jets), which have very small signal acceptance and/or
poor signal-to-background ratio due to the large number of jets misidentified as $\tau_{\rm had}$ 
candidates.
The minimum number of jets required in the $\twolnotau$ category is well matched to 
the number of partons expected from $\Hq \to W^\pm W^\pm W^\mp bq$ signal events, while the 
$3\ell$ and $\twolonetau$ categories effectively require one or two jets beyond leading
order. Because of this, and the gain in branching ratio resulting from only requiring two, as opposed to three,  
$W$ bosons decaying leptonically, the $\twolnotau$ category dominates the sensitivity.

The largest background in the most sensitive category, $2\ell$4j, 
is $\ttbar$ or single-top production with one of the leptons originating from a decay of 
a heavy-flavour hadron (``non-prompt lepton''), followed by $\ttbar W$ production.
Smaller contributions arise from $\ttbar(Z/\gamma^*)$, $\ttbar H$, and diboson (primarily $WZ$)
production, and dilepton events from $\ttbar$ and $Z/\gamma^*$ with the wrong charge sign 
measured for one electron. In the remaining categories the non-prompt 
lepton background decreases in importance, relative to the prompt-lepton contributions.
The prompt contributions are estimated using the simulation, as typically the relevant processes
(e.g., $\ttbar V$, $V = W, Z$) have not been measured at an accuracy exceeding
that of theoretical predictions.
The non-prompt lepton background predictions are computed or validated using data control regions 
with similar lepton kinematic selections but with fewer jets (two or three). Further details are provided in 
ref.~\cite{Aad:2015iha}. In the search for $t \bar t H$, these regions have almost no 
contamination from signal, but the same is not necessarily true for the $\Hq$ process, 
which is characterised by a lower jet multiplicity than the $t \bar t H$ process.
This could potentially lead to an overestimate of the non-prompt background in
this reinterpretation, particularly in the $\twolnotau$ categories, where the effect is more pronounced.
The $\threel$ category estimates the non-prompt lepton background in
the signal region by extrapolating, in terms of lepton quality, from a data sideband region that is 
sufficiently depleted in signal, using normalisation factors obtained from the simulation. 

The non-prompt lepton background estimate for the $\twolnotau$ categories is
validated using simulation and control regions with two or three jets in which the lower $\pt$
(subleading) lepton satisfies 10 GeV $< \pt <$ 30 GeV.  Because the number of non-prompt
leptons increases significantly as the lepton \pt requirement is lowered, this
region is dominated by the non-prompt contribution for any non-excluded value of
$\BR(t \to Hq)$.  To extract the non-prompt background yields in these regions, binned likelihood fits 
to the subleading lepton $\pt$ distribution are performed, accounting for non-prompt lepton backgrounds, 
sources of prompt leptons (normalised to the SM expectation), and a potential $\Hq$ contribution.  
The fitted $\Hq$ contribution is in all cases found to be compatible with zero.  The obtained best-fit non-prompt background
yields in these low-$\pt$ control regions are used to normalise the simulation and obtain estimates for the
background in the $\twolnotau$ signal regions.  These are found to be compatible with the
nominal data-driven predictions within the stated uncertainties.  The nominal predictions
of the non-prompt backgrounds and their associated uncertainties are therefore
used without modification in this reinterpretation.

\begin{figure*}[t]
\begin{center}
\includegraphics[width=0.60\textwidth]{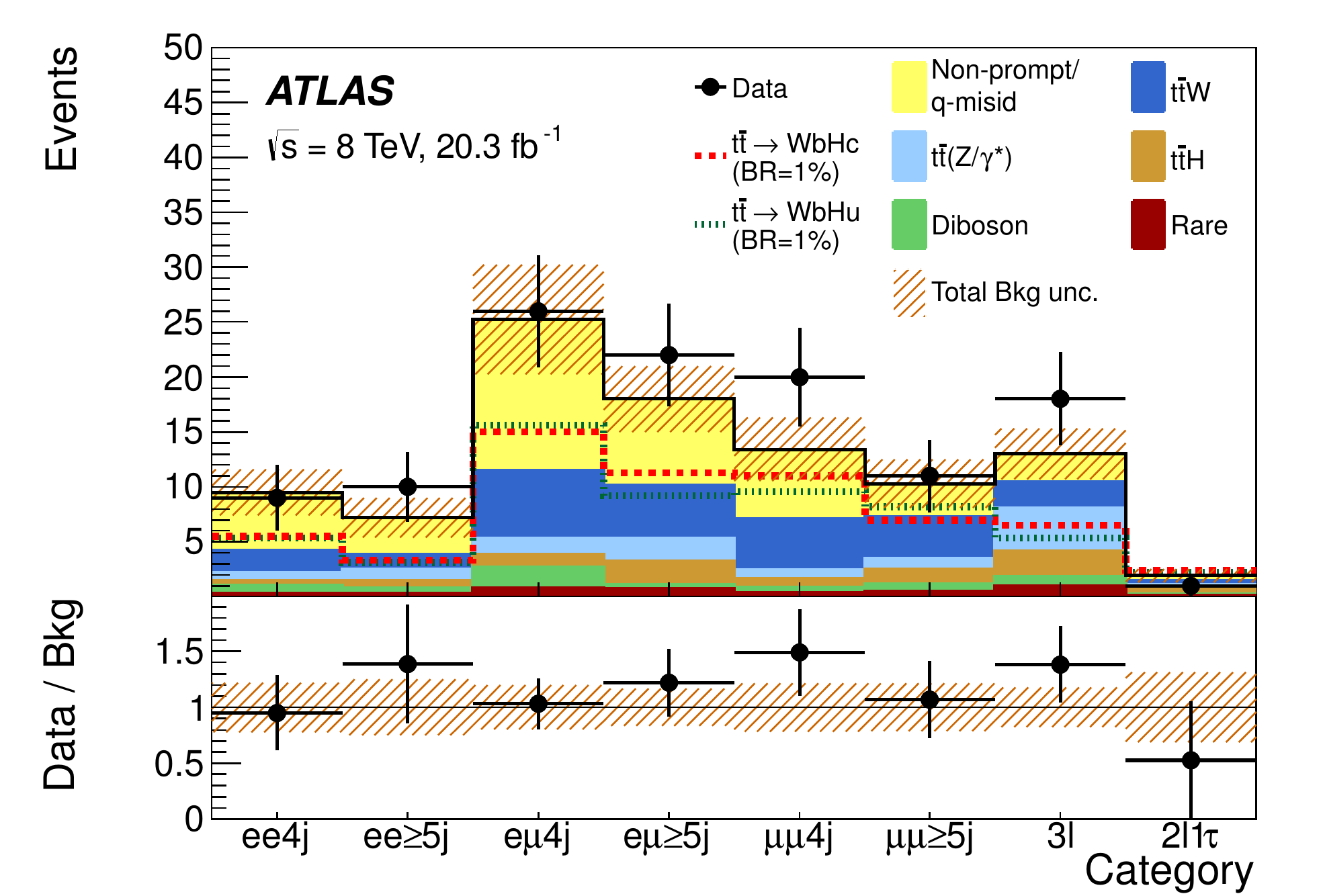}
\caption{$\HqML$ search: comparison between the data and background prediction for the yields in each of the analysis channels considered before the fit to data.
The expected $\Hc$ and $\Hu$ signals (dashed histograms) are shown separately normalised to $\BR(t\to Hq)=1\%$.
The sum of instrumental backgrounds originating from non-prompt leptons and lepton charge misidentification is denoted by 
``Non-prompt/q-misid''. The small contribution from rare processes such as 
$tZ$, $\ttbar WW$, triboson, $\ttbar\ttbar$ and $tH$ production are combined into a single background source denoted by ``Rare''. 
The bottom panel displays the ratio of data to the SM background (``Bkg'') prediction. 
The hashed area represents the total uncertainty on the background.} 
\label{fig:inclusive_category}
\end{center}
\end{figure*}

Figure~\ref{fig:inclusive_category} shows  the event yields in each of the categories, which are used as input to the 
statistical analysis discussed in section~\ref{sec:stat_analysis}. 
A table summarising the expected and observed yields can be found in appendix~\ref{sec:ML_yields_appendix}.
The best-fit branching ratio obtained is 
$\BR(t\to Hc)=[0.27 \pm 0.18\,({\rm stat.}) \pm 0.21\,({\rm syst.})]\%$ 
assuming that $\BR(t\to Hu)=0$. 
A similar fit is performed for the $\Hu$ search, yielding 
$\BR(t\to Hu)=[0.23 \pm 0.18\,({\rm stat.}) \pm 0.21\,({\rm syst.})]\%$,
assuming that $\BR(t\to Hc)=0$. 
The observed (expected) 95\% CL upper limits on the branching ratios 
are $\BR(t\to Hc)<0.79\%\,(0.54\%)$ and $\BR(t\to Hu)<0.78\%\,(0.57\%)$.
The corresponding observed (expected) limits on the couplings are $|\lamHc|<0.17\,(0.14)$ and $|\lamHu|<0.17\,(0.14)$ at 95\% CL.

\subsection{Combination of searches}
\label{sec:results_combo}

The three searches discussed in the sections~\ref{sec:results_bb}--\ref{sec:results_ML} are combined using the
statistical analysis discussed in section~\ref{sec:stat_analysis}. In this combination, the only systematic uncertainties 
taken to be fully correlated among the three searches are the $\ttbar$ cross section and the integrated luminosity for $\sqrt{s}=8\tev$ data.
The dominant uncertainties on the Higgs boson branching ratios primarily affect the $\Hqgg$ and
$\ttbar \to WbHq$, $H\to b\bar{b}$ searches.  Other uncertainties such as those associated with leptons, 
jet energy scale and $b$-tagging should be partially correlated between the $\ttbar \to WbHq$, $H\to \gamma\gamma$ search and the other two searches,
but the differences in treatment between analyses (different lepton selection and $\pt$ cuts, no uncertainty breakdown for jet energy scale and $b$-tagging  
in the $\ttbar \to WbHq$, $H\to \gamma\gamma$ search) make it difficult to account for correlations. 
However, given that the $\ttbar \to WbHq$, $H\to \gamma\gamma$ search
is completely dominated by the data statistics, the effect of this simplification is negligible. 
The uncertainties taken as correlated between the $\ttbar \to WbHq$, $H\to b\bar{b}$ and $\ttbar \to WbHq$, $H\to WW^*, \tau\tau$ searches 
include those associated with lepton isolation, the leading $b$-tagging and jet energy scale uncertainties, and the
reweighting of top quark $\pt$ and $\ttbar$ system $\pt$. The rest of the uncertainties are taken to be uncorrelated among the searches.
This correlation scheme closely follows the procedure adopted in the combination of $\ttbar H$ searches by ATLAS~\cite{Aad:2015gba}.

The first set of combined results is obtained for each branching ratio separately, setting the other branching ratio to zero.
The best-fit combined branching ratios are $\BR(t\to Hc)=[0.22 \pm 0.10\,({\rm stat.}) \pm 0.10\,({\rm syst.})]\%$ and 
$\BR(t\to Hu)=[0.16 \pm 0.11\,({\rm stat.}) \pm 0.12\,({\rm syst.})]\%$.  
The difference between the central values of $\BR(t\to Hc)$ and $\BR(t\to Hu)$ originates from the ability of the $H \to b\bar{b}$ search to 
probe both decay modes separately.
A comparison of the best-fit branching ratios for the individual searches and their combination can be found in figure~\ref{fig:summary_printnum_hc} for $\BR(t\to Hc)$ and figure~\ref{fig:summary_printnum_hu} for $\BR(t\to Hu)$.
Figure~\ref{fig:CLs_vs_BR_combo} shows the CL$_{\rm{s}}$ versus branching ratio for the combination. 
The observed (expected) 95\% CL combined upper limits on the branching ratios are 
$\BR(t\to Hc)<0.46\%\,(0.25\%)$ and $\BR(t\to Hu)<0.45\%\,(0.29\%)$. 
The corresponding observed (expected) upper limits on the couplings are $|\lamHc|<0.13\,(0.10)$ and $|\lamHu|<0.13\,(0.10)$.
A summary of the upper limits on the branching ratios obtained by the individual searches, as well as their combination, 
can be found in figure~\ref{fig:limits_combo_1D}.

\begin{figure*}[h!]
\begin{center}
\includegraphics[width=0.9\textwidth]{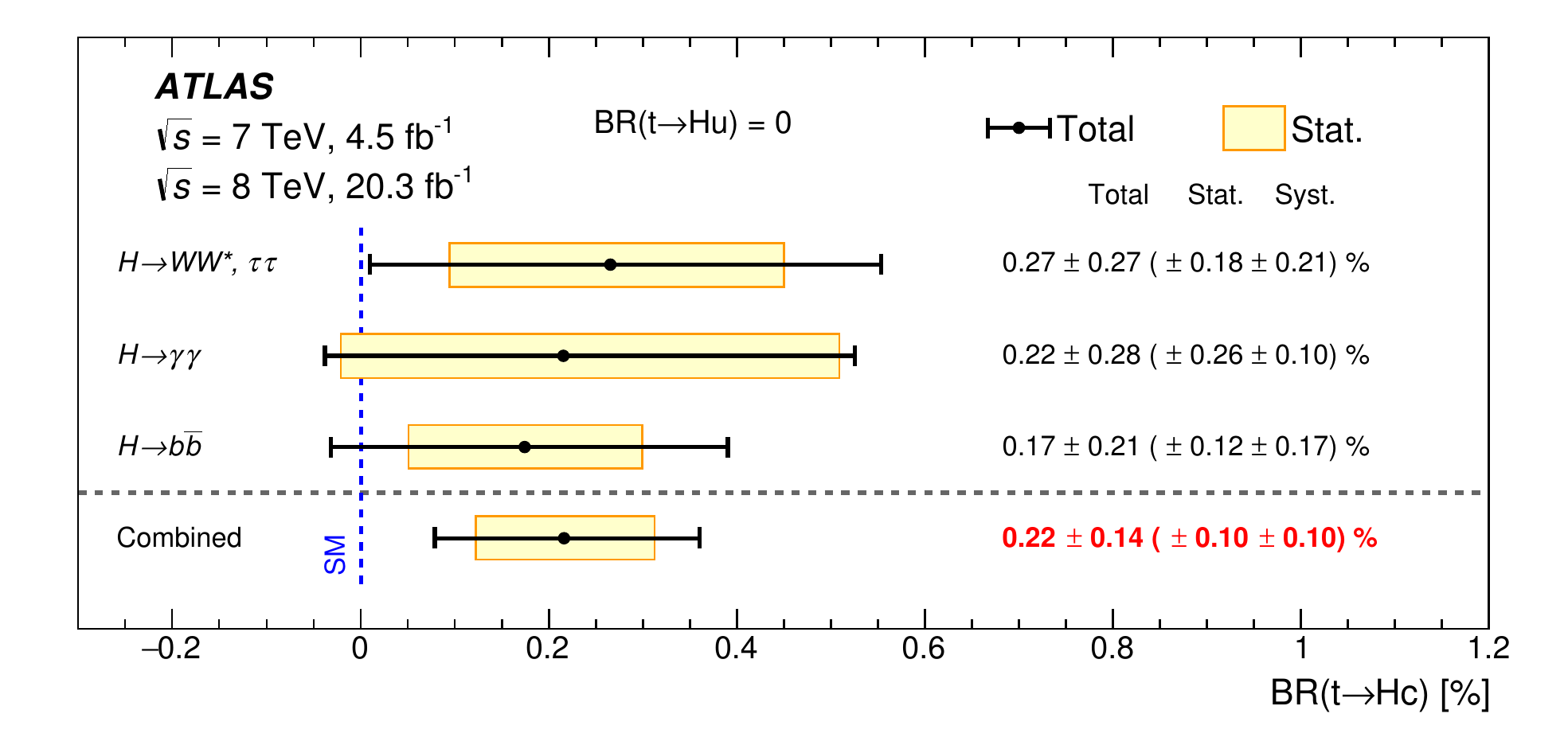}
\caption{\small {Summary of the best-fit $\BR(t\to Hc)$ for the individual searches as well as their combination,
assuming that $\BR(t\to Hu)=0$.}}
\label{fig:summary_printnum_hc} 
\end{center}
\end{figure*}

\begin{figure*}[h!]
\begin{center}
\includegraphics[width=0.9\textwidth]{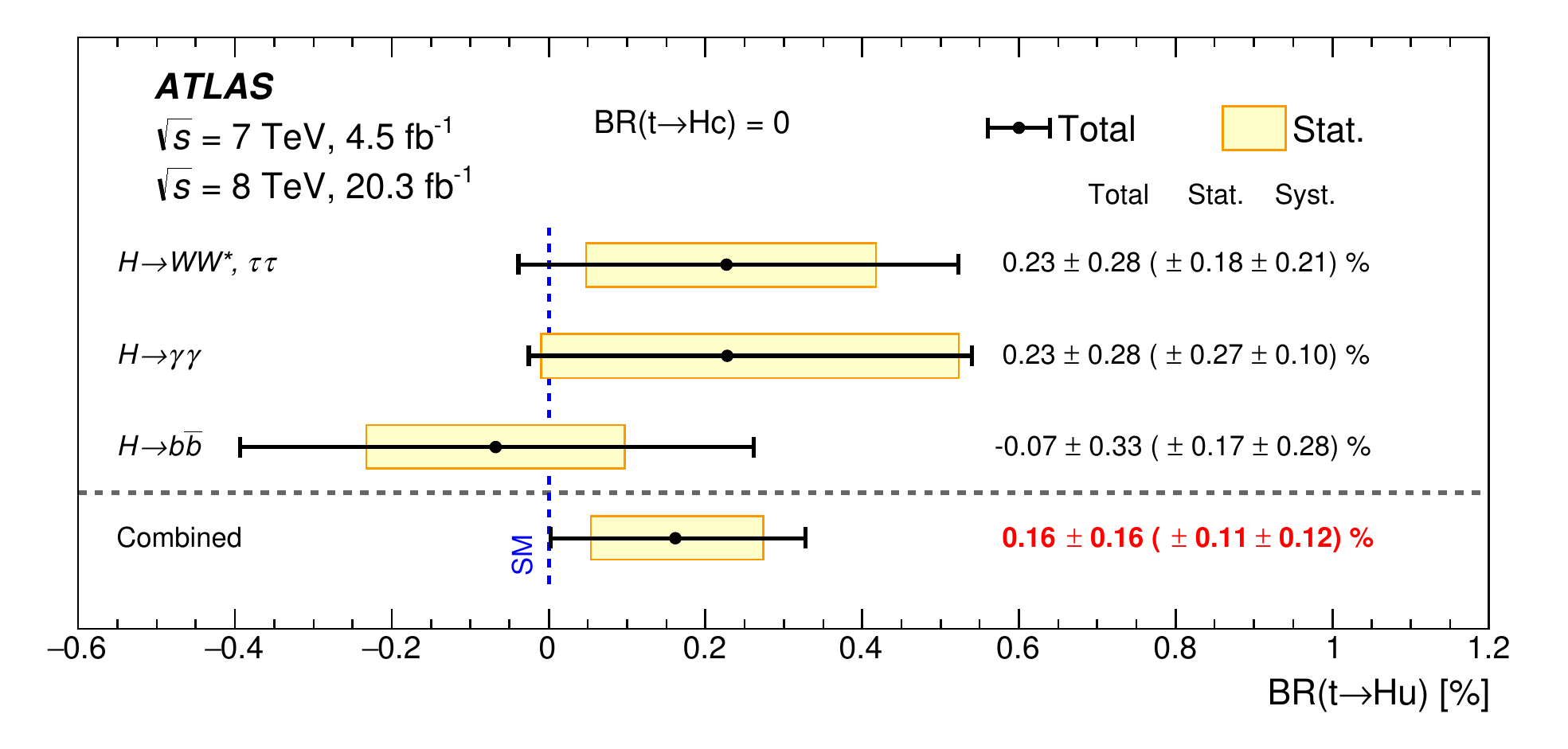}
\caption{\small {Summary of the best-fit $\BR(t\to Hu)$ for the individual searches as well as their combination,
assuming that $\BR(t\to Hc)=0$.}}
\label{fig:summary_printnum_hu} 
\end{center}
\end{figure*}

\begin{figure*}[htbp]
\begin{center}
\subfloat[]{\includegraphics[width=0.49\textwidth]{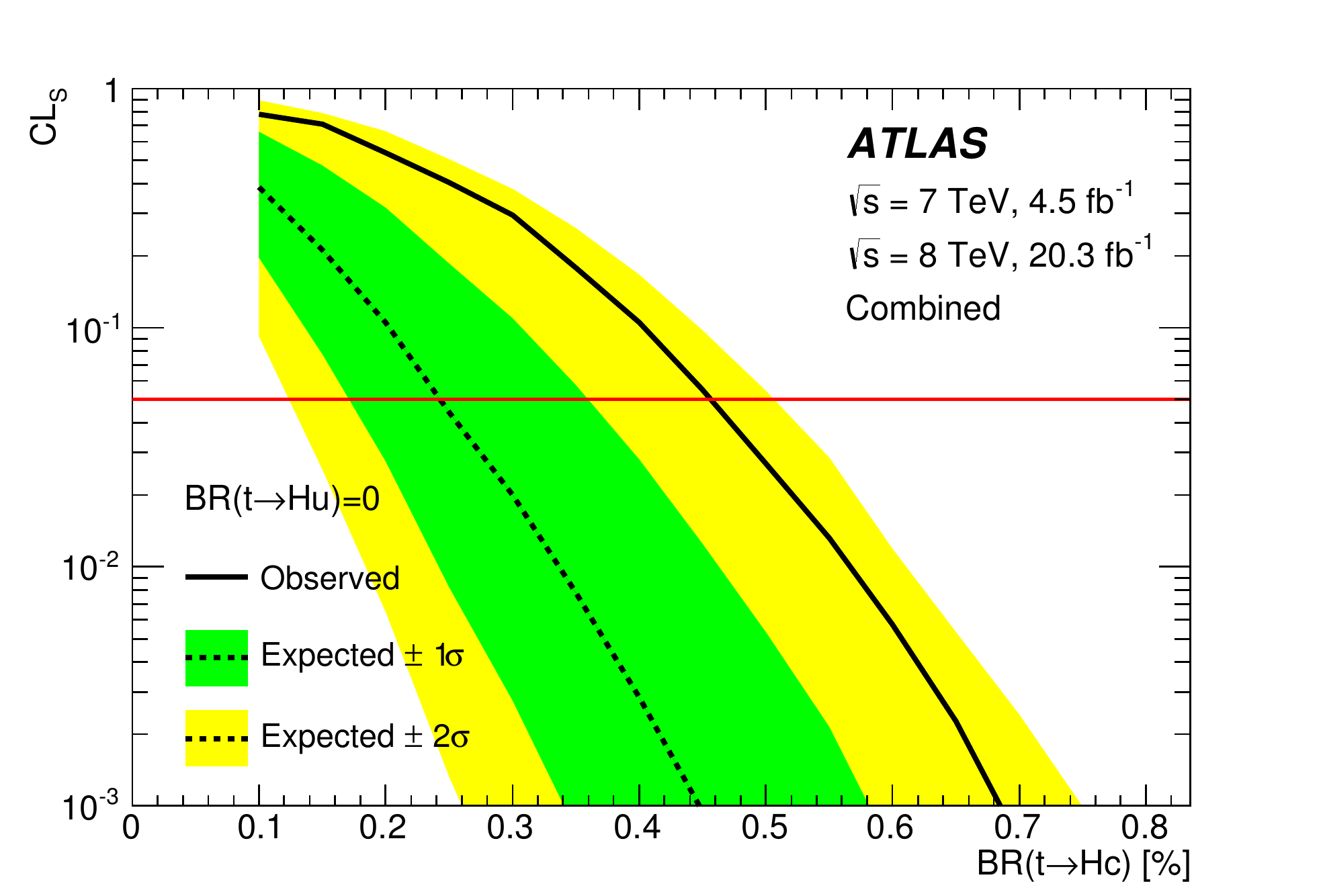}}
\subfloat[]{\includegraphics[width=0.49\textwidth]{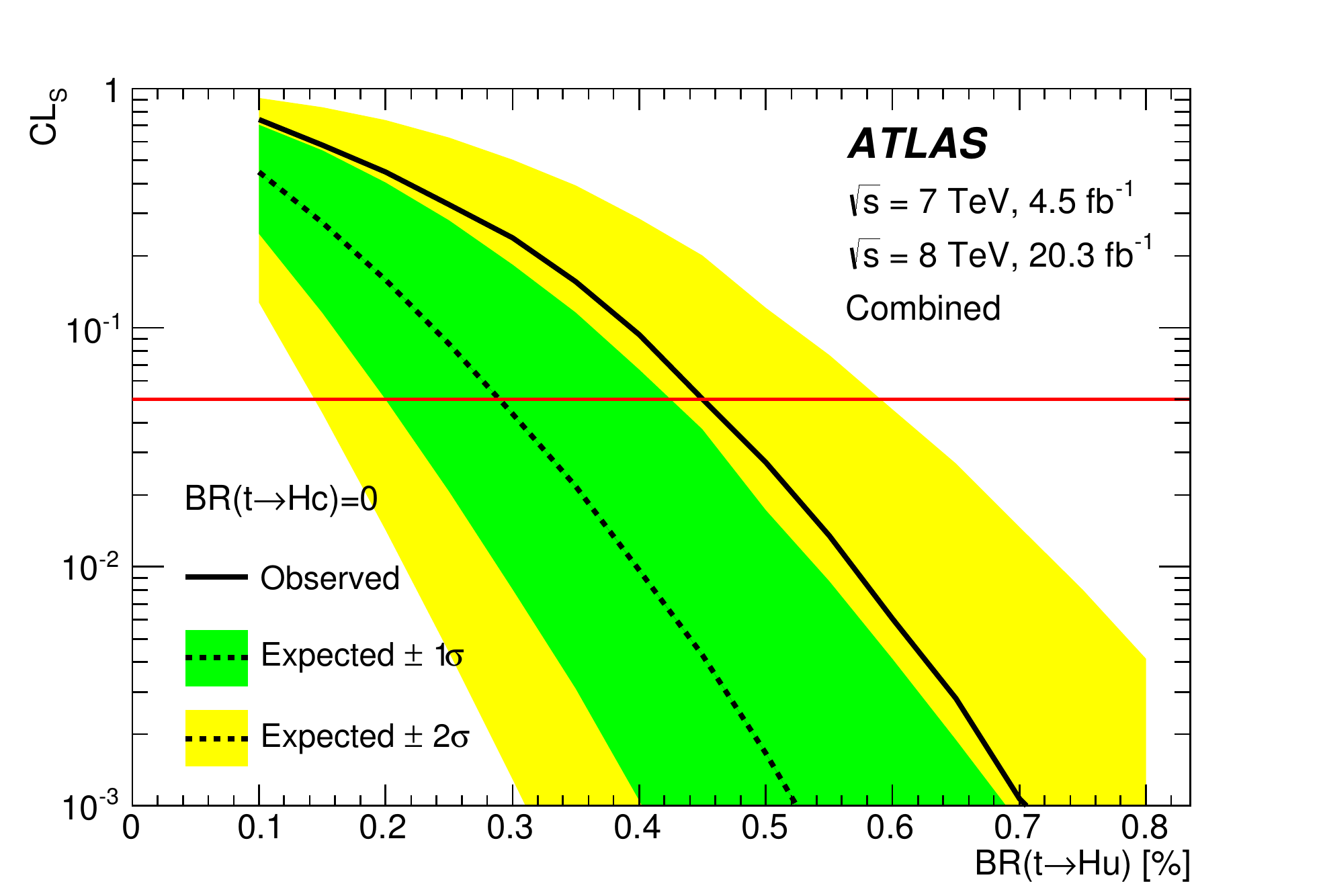}}
\caption{\small {(a) CL$_{\rm{s}}$ versus $\BR(t\to Hc)$ and (b) CL$_{\rm{s}}$ versus $\BR(t\to Hu)$ for the combination of the searches, 
assuming that the other branching ratio is zero.
The observed CL$_{\rm{s}}$ values (solid black lines) are compared to the expected (median) CL$_{\rm{s}}$ values under the background-only
hypothesis (dotted black lines). The surrounding shaded bands correspond to the 68\% and 95\% CL intervals around the expected CL$_{\rm{s}}$ values, 
denoted by $\pm 1\sigma$ and $\pm 2\sigma$, respectively.
The solid red line at CL$_{\rm{s}}$=0.05 denotes the value below which the hypothesis is excluded at 95\% CL.}}
\label{fig:CLs_vs_BR_combo} 
\end{center}
\end{figure*}

\begin{figure*}[htbp]
\begin{center}
\subfloat[]{\includegraphics[width=0.49\textwidth]{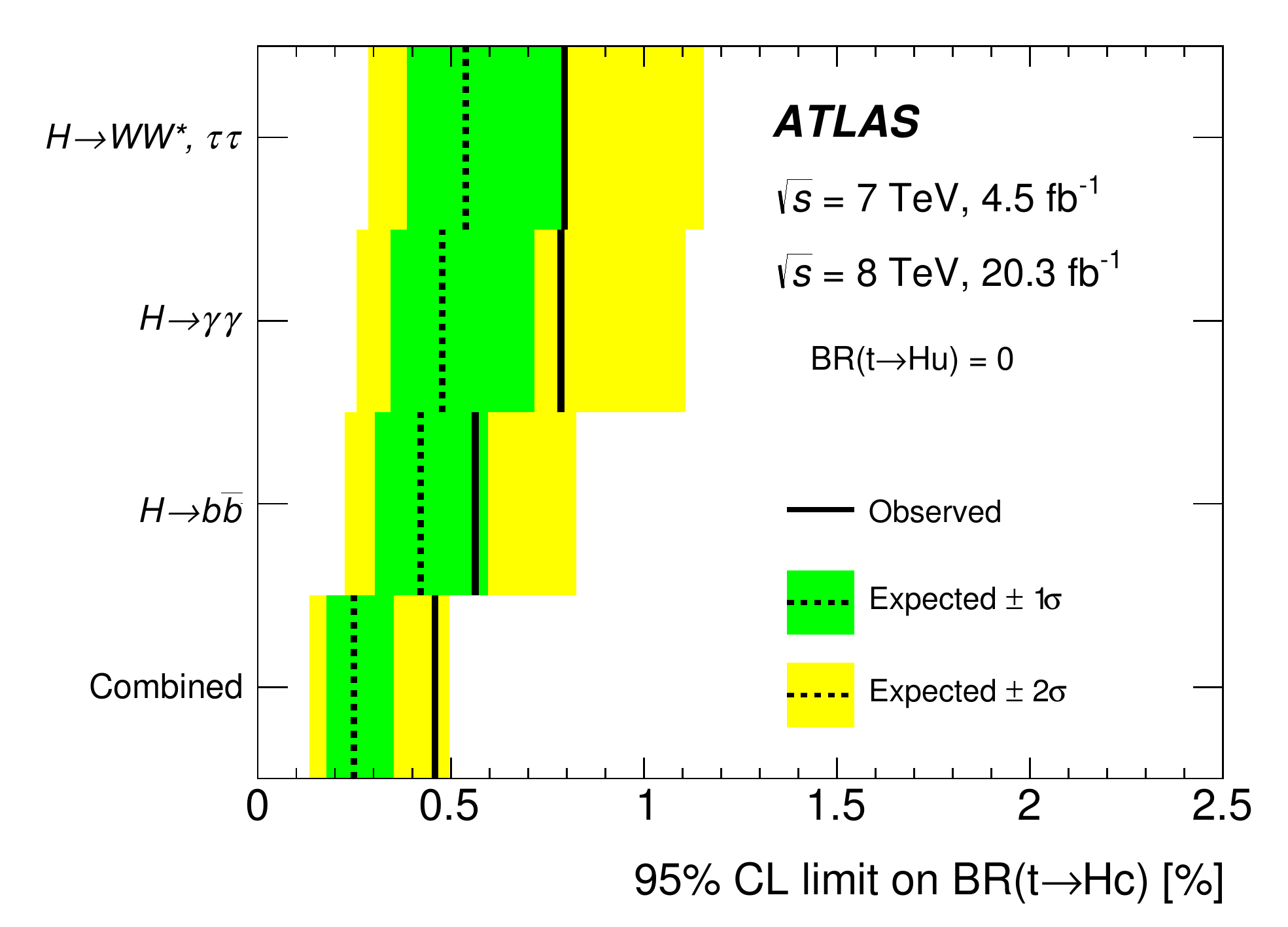}}
\subfloat[]{\includegraphics[width=0.49\textwidth]{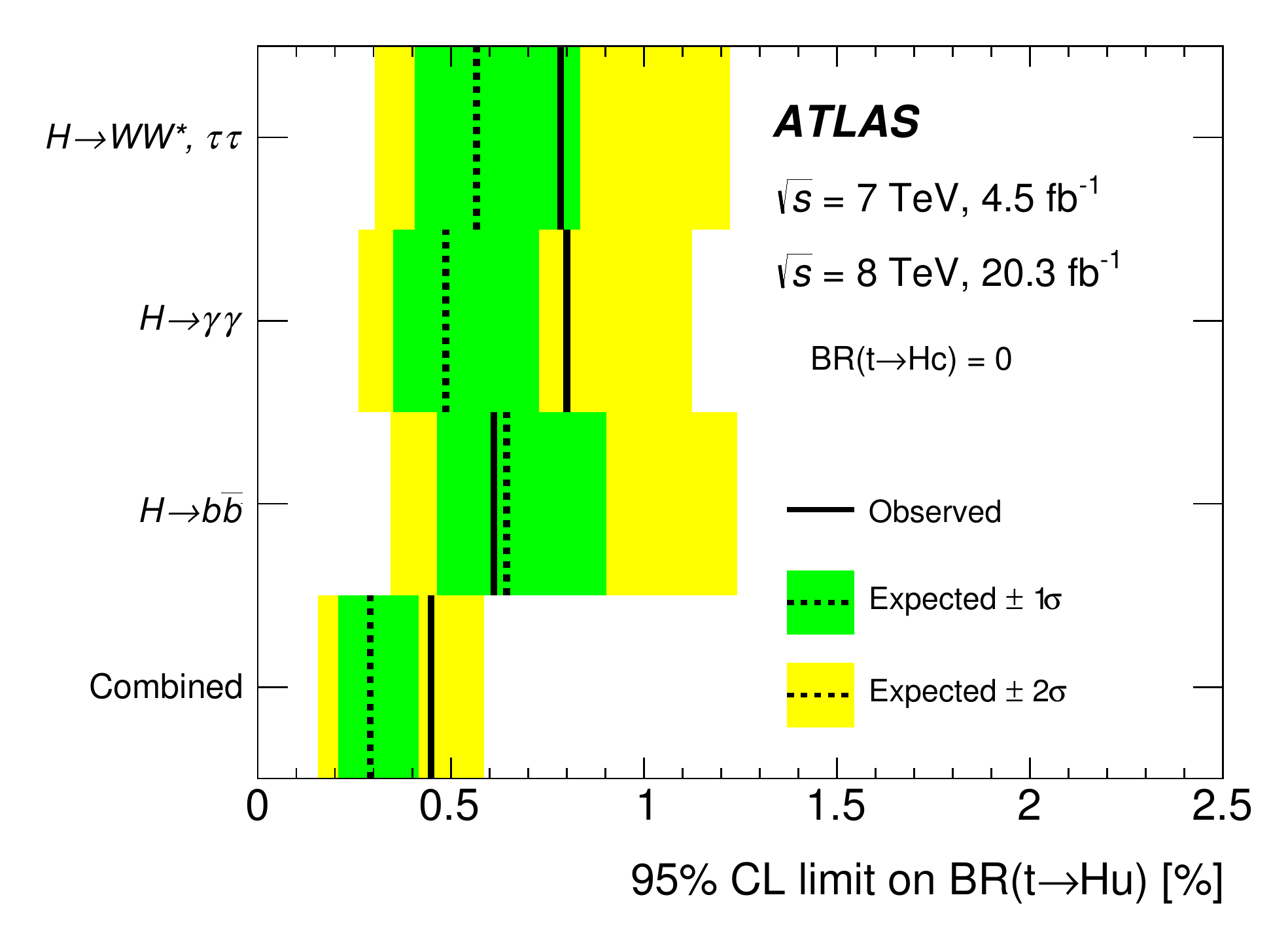}}
\caption{\small {95\% CL upper limits on (a) $\BR(t\to Hc)$ and (b) $\BR(t\to Hu)$ for the individual searches as well as their
combination, assuming that the other branching ratio is zero. The observed limits (solid lines) are compared to the 
expected (median) limits under the background-only
hypothesis (dotted lines). The surrounding shaded bands correspond to the 68\% and 95\% CL intervals around the expected limits, 
denoted by $\pm 1\sigma$ and $\pm 2\sigma$, respectively.
Because the asymptotic approximation is used in the calculation of CL$_{\rm{s}}$, 
the results of the $H \to\gamma\gamma$ search reported in this figure differ slightly from 
those published in ref.~\cite{Aad:2014dya}, which remain the most accurate results.}}
\label{fig:limits_combo_1D} 
\end{center}
\end{figure*}

A similar set of results can be obtained by  simultaneously probing both branching ratios. Although the $\ttbar \to WbHq$, $H\to \gamma\gamma$ and 
$t\bar{t}\to WbHq$, $H\to WW^*, \tau\tau$ searches are basically only sensitive to the sum of the two branching ratios, 
the  $\ttbar \to WbHq$, $H\to b\bar{b}$ search has different sensitivity to each of them, and a simultaneous fit can be performed.
The best-fit branching ratios obtained from the simultaneous fit are 
$\BR(t\to Hc)=[0.34 \pm 0.22\,({\rm stat.}) \pm 0.15\,({\rm syst.})]\%$ and 
$\BR(t\to Hu)=[-0.17 \pm 0.25\,({\rm stat.}) \pm 0.17\,({\rm syst.})]\%$, with a correlation coefficient of $-0.84$, 
as shown in figure~\ref{fig:nllscan_combo_2D}.
Figure~\ref{fig:limits_combo_2D}(a) shows the 95\% CL upper limits on the branching ratios in the $\BR(t\to Hu)$ versus $\BR(t\to Hc)$ plane. 
The corresponding upper limits on the couplings in the $|\lamHu|$ versus $|\lamHc|$ plane can be found in figure~\ref{fig:limits_combo_2D}(b).

\begin{figure*}[htbp]
\begin{center}
\includegraphics[width=0.49\textwidth]{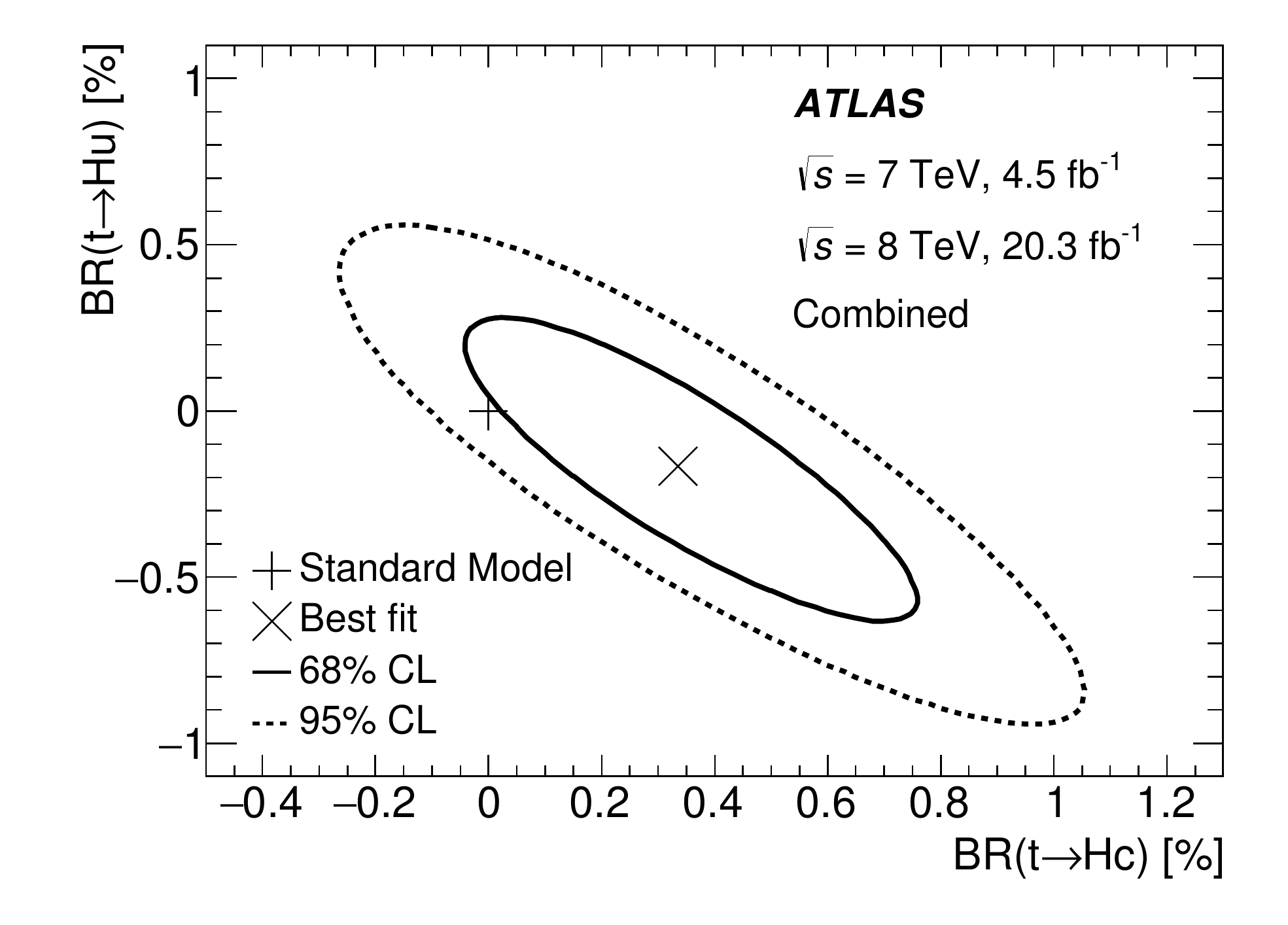}
\caption{\small {
Best-fit $\BR(t\to Hc)$ and $\BR(t\to Hu)$ and the corresponding 68\% CL (solid) and 95\% CL (dotted) regions for the combination of the searches.}}
\label{fig:nllscan_combo_2D} 
\end{center}
\end{figure*}

\begin{figure*}[htbp]
\begin{center}
\subfloat[]{\includegraphics[width=0.49\textwidth]{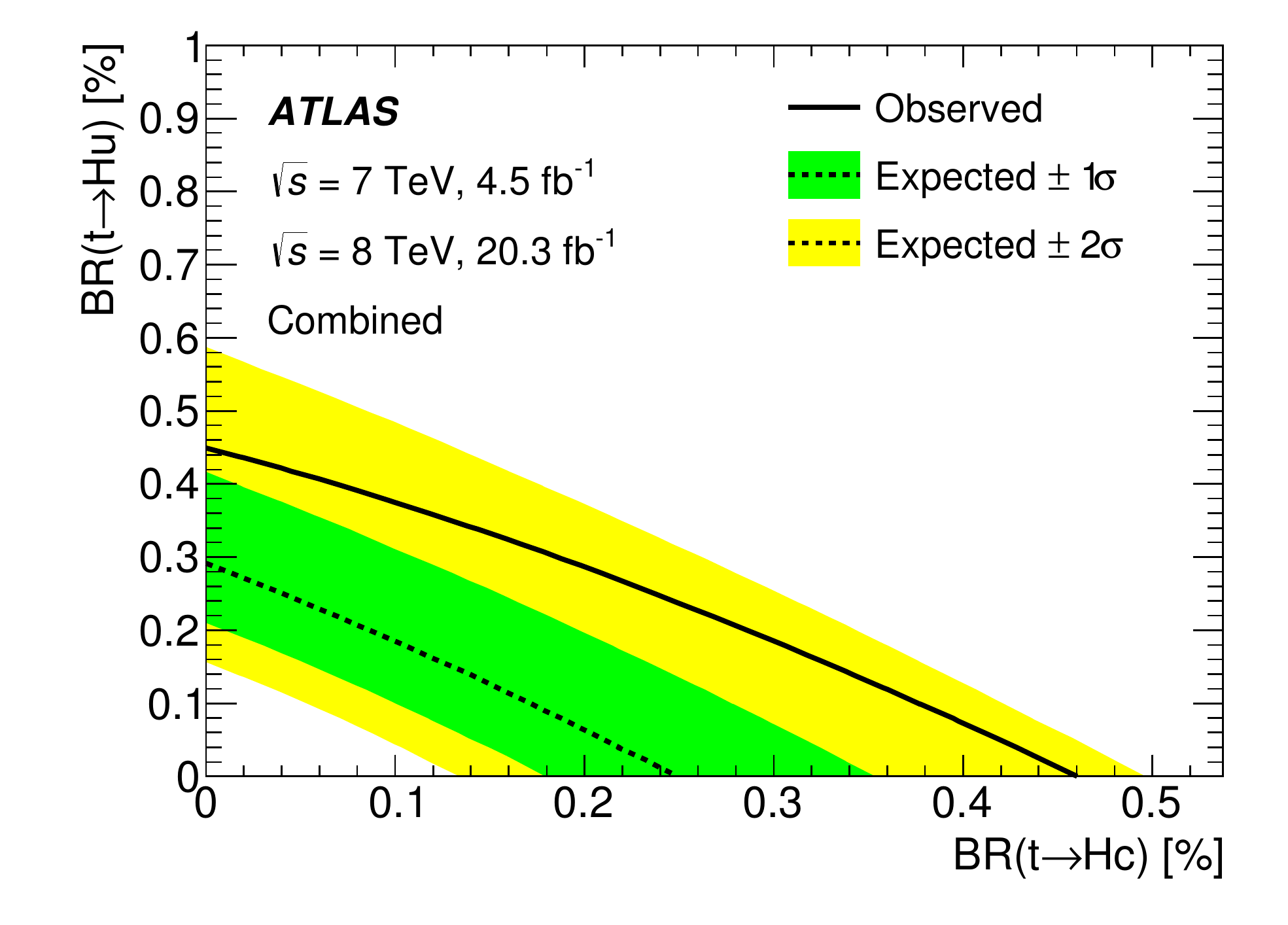}}
\subfloat[]{\includegraphics[width=0.49\textwidth]{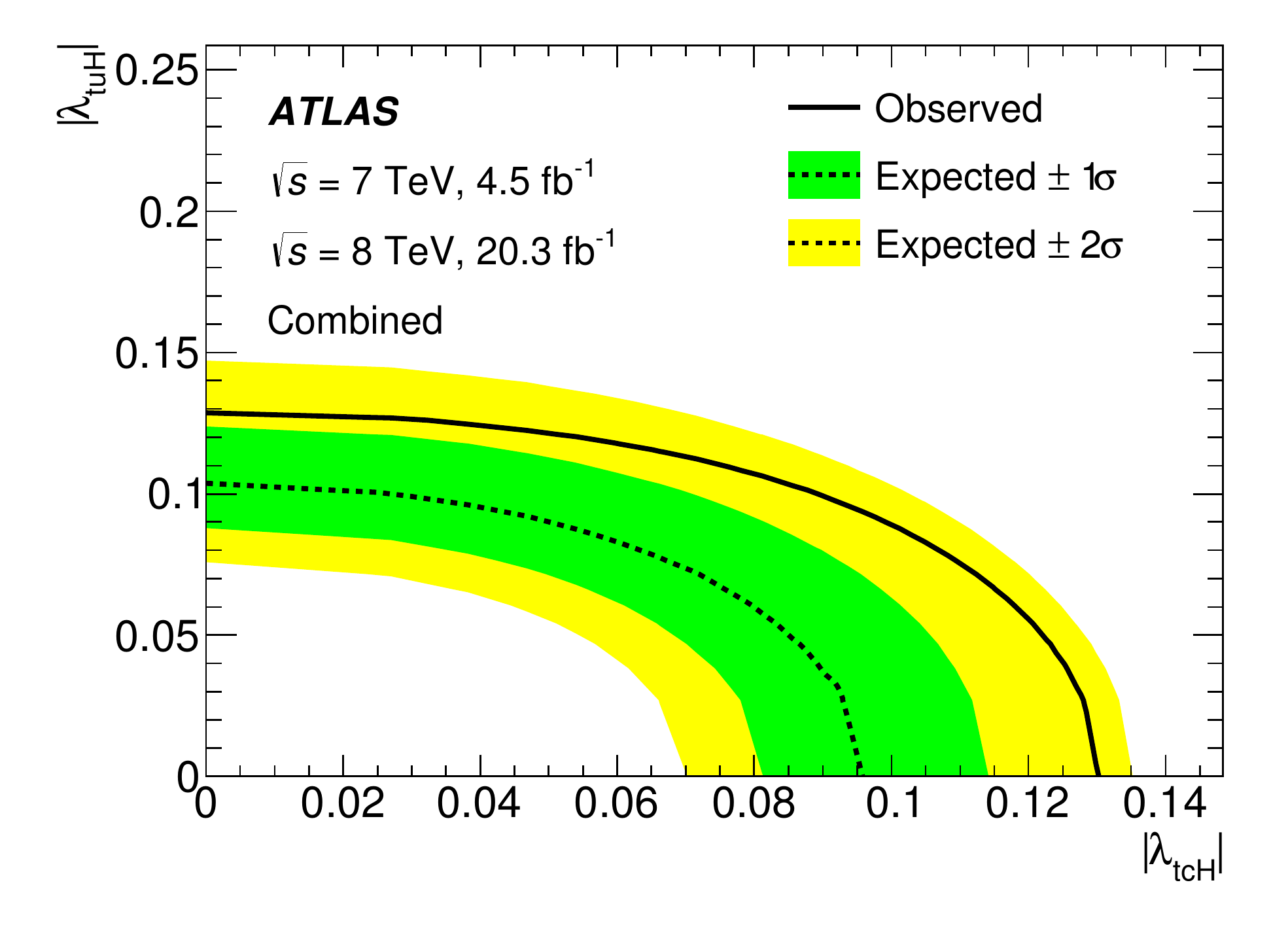}}
\caption{\small {95\% CL upper limits (a) on the plane of $\BR(t\to Hu)$ versus $\BR(t\to Hc)$ and (b) on the plane 
of $|\lamHu|$ versus $|\lamHc|$ for the combination of the searches. The observed limits (solid lines) are compared to the expected (median) limits under the background-only hypothesis (dotted lines). The surrounding shaded bands correspond to the 68\% and 95\% CL intervals around the expected limits, 
denoted by $\pm 1\sigma$ and $\pm 2\sigma$, respectively.}}
\label{fig:limits_combo_2D} 
\end{center}
\end{figure*}

\FloatBarrier

\section{Conclusion}
\label{sec:conclusion}
A search for flavour-changing neutral current decays of a top quark to an up-type quark ($q=u, c$) and the 
Standard Model Higgs boson,  where the Higgs boson decays to $b\bar{b}$, is presented. 
The analysis searches for top quark pair events in which one top quark decays to $Wb$, with the $W$ boson decaying leptonically, 
and the other top quark decays to $Hq$.
The search is based on $pp$ collisions at $\sqrt{s}=8\tev$ recorded in 2012 with the ATLAS detector at the CERN Large Hadron 
Collider and uses an integrated luminosity of 20.3 fb$^{-1}$.
Data are analysed in the lepton-plus-jets final state, characterised by an isolated electron or muon with moderately high transverse momentum 
and at least four jets. In this search the dominant background process is $\ttbar \to WbWb$.
To separate the signal from the background, the search exploits the high multiplicity of $b$-quark jets characteristic of signal events, 
and employs a likelihood discriminant that combines information from invariant mass distributions and the flavour of the jets.
No significant excess of events above the background expectation is found, and observed (expected) 95\% CL upper limits 
of 0.56\% (0.42\%) and 0.61\% (0.64\%) are derived for the $t\to Hc$ and $t\to Hu$ branching ratios respectively.

Results from other ATLAS searches are also summarised, including a previous search probing $H \to \gamma\gamma$ decays, and a 
reinterpretation of a search for $\ttbar H$ production in multilepton final states, which exploits the $H\to WW^*, \tau\tau$ decay modes. 
All searches discussed in this paper have comparable sensitivity, and thus their combination represents a significant improvement over the individual results.
The observed (expected) 95\% CL combined upper limits on the $t\to Hc$ and $t\to Hu$ branching ratios are 0.46\% (0.25\%) and 0.45\% (0.29\%) respectively.
The corresponding observed (expected) upper limits on the $|\lamHc|$ and $|\lamHu|$ couplings are 0.13 (0.10) and 0.13 (0.10) respectively. 
Upper limits in the $t\to Hc$ versus $t\to Hu$ branching ratio plane, as well as best-fit branching ratios, are also reported.
These are the most restrictive direct bounds on $tqH$ ($q=u,c$) interactions measured so far.

\section*{Acknowledgements}

We thank CERN for the very successful operation of the LHC, as well as the
support staff from our institutions without whom ATLAS could not be
operated efficiently.

We acknowledge the support of ANPCyT, Argentina; YerPhI, Armenia; ARC, Australia; BMWFW and FWF, Austria; ANAS, Azerbaijan; SSTC, Belarus; CNPq and FAPESP, Brazil; NSERC, NRC and CFI, Canada; CERN; CONICYT, Chile; CAS, MOST and NSFC, China; COLCIENCIAS, Colombia; MSMT CR, MPO CR and VSC CR, Czech Republic; DNRF, DNSRC and Lundbeck Foundation, Denmark; IN2P3-CNRS, CEA-DSM/IRFU, France; GNSF, Georgia; BMBF, HGF, and MPG, Germany; GSRT, Greece; RGC, Hong Kong SAR, China; ISF, I-CORE and Benoziyo Center, Israel; INFN, Italy; MEXT and JSPS, Japan; CNRST, Morocco; FOM and NWO, Netherlands; RCN, Norway; MNiSW and NCN, Poland; FCT, Portugal; MNE/IFA, Romania; MES of Russia and NRC KI, Russian Federation; JINR; MESTD, Serbia; MSSR, Slovakia; ARRS and MIZ\v{S}, Slovenia; DST/NRF, South Africa; MINECO, Spain; SRC and Wallenberg Foundation, Sweden; SERI, SNSF and Cantons of Bern and Geneva, Switzerland; MOST, Taiwan; TAEK, Turkey; STFC, United Kingdom; DOE and NSF, United States of America. In addition, individual groups and members have received support from BCKDF, the Canada Council, CANARIE, CRC, Compute Canada, FQRNT, and the Ontario Innovation Trust, Canada; EPLANET, ERC, FP7, Horizon 2020 and Marie Skłodowska-Curie Actions, European Union; Investissements d'Avenir Labex and Idex, ANR, Region Auvergne and Fondation Partager le Savoir, France; DFG and AvH Foundation, Germany; Herakleitos, Thales and Aristeia programmes co-financed by EU-ESF and the Greek NSRF; BSF, GIF and Minerva, Israel; BRF, Norway; the Royal Society and Leverhulme Trust, United Kingdom.

The crucial computing support from all WLCG partners is acknowledged
gratefully, in particular from CERN and the ATLAS Tier-1 facilities at
TRIUMF (Canada), NDGF (Denmark, Norway, Sweden), CC-IN2P3 (France),
KIT/GridKA (Germany), INFN-CNAF (Italy), NL-T1 (Netherlands), PIC (Spain),
ASGC (Taiwan), RAL (UK) and BNL (USA) and in the Tier-2 facilities
worldwide.

%

\clearpage
\appendix
\part*{Appendix}
\addcontentsline{toc}{part}{Appendix}
\section{Pre-fit and post-fit event yields in the $\Hqbb$ search}
\label{sec:prepostfit_yields_appendix}

Table~\ref{tab:Prefit_Yields_Unblind} presents the observed and predicted yields in each of the analysis channels 
for the $\Hqbb$ search before the fit to data. 
Table~\ref{tab:Postfit_Yields_Unblind_Hc} presents the observed and predicted yields in each of the analysis channels 
for the $\Hcbb$ search, after the fit to the data under the signal-plus-background hypothesis. 

\begin{table}
\small
\begin{center}
\begin{tabular}{l*{3}{c}}
\hline\hline
 & 4 j, 2 b & 4 j, 3 b & 4 j, 4 b\\
\hline
$\Hc$ & $890$ $\pm$ $100$ & $394$ $\pm$ $54$ & $41.6$ $\pm$ $7.2$\\
$\Hu$ & $851$ $\pm$ $98$ & $339$ $\pm$ $49$ & $3.81$ $\pm$ $0.71$\\
\hline
$t\bar{t}$+light-jets & $77400$ $\pm$ $8100$ & $6170$ $\pm$ $860$ & $53$ $\pm$ $12$\\
$t\bar{t}+c\bar{c}$ & $4900$ $\pm$ $2600$ & $680$ $\pm$ $370$ & $21$ $\pm$ $11$\\
$t\bar{t}+b\bar{b}$ & $1870$ $\pm$ $990$ & $680$ $\pm$ $370$ & $44$ $\pm$ $23$\\
$t\bar{t}V$ & $121$ $\pm$ $21$ & $15.5$ $\pm$ $2.9$ & $0.89$ $\pm$ $0.19$\\
$t\bar{t}H$ & $30.5$ $\pm$ $4.2$ & $12.7$ $\pm$ $1.9$ & $1.91$ $\pm$ $0.34$\\
$W$+jets & $4700$ $\pm$ $1600$ & $217$ $\pm$ $78$ & $5.4$ $\pm$ $2.0$\\
$Z$+jets & $1080$ $\pm$ $450$ & $50$ $\pm$ $22$ & $0.90$ $\pm$ $0.50$\\
Single top & $4900$ $\pm$ $1400$ & $340$ $\pm$ $100$ & $6.8$ $\pm$ $2.3$\\
Diboson & $212$ $\pm$ $75$ & $11.5$ $\pm$ $4.1$ & $0.24$ $\pm$ $0.11$\\
Multijet & $1540$ $\pm$ $550$ & $100$ $\pm$ $36$ & $3.4$ $\pm$ $1.2$\\
\hline
Total background & $96800$ $\pm$ $9600$          & $8300$ $\pm$ $1100$          & $138$ $\pm$ $32$         \\
\hline
Data & $98049$ & $8752$ & $161$\\
\hline\hline    
\end{tabular}
\vspace{0.2cm}

\begin{tabular}{l*{3}{c}}
\hline\hline
 & 5 j, 2 b & 5 j, 3 b & 5 j, $\geq$4 b\\
\hline
$\Hc$ & $483$ $\pm$ $96$ & $242$ $\pm$ $50$ & $35.1$ $\pm$ $7.7$\\
$\Hu$ & $473$ $\pm$ $95$ & $217$ $\pm$ $46$ & $8.4$ $\pm$ $2.0$\\
\hline
$t\bar{t}$+light-jets & $37600$ $\pm$ $6600$ & $3480$ $\pm$ $750$ & $61$ $\pm$ $18$\\
$t\bar{t}+c\bar{c}$ & $4300$ $\pm$ $2300$ & $810$ $\pm$ $460$ & $43$ $\pm$ $28$\\
$t\bar{t}+b\bar{b}$ & $1670$ $\pm$ $860$ & $890$ $\pm$ $470$ & $115$ $\pm$ $61$\\
$t\bar{t}V$ & $145$ $\pm$ $24$ & $26.5$ $\pm$ $4.5$ & $3.10$ $\pm$ $0.60$\\
$t\bar{t}H$ & $40.9$ $\pm$ $4.8$ & $22.3$ $\pm$ $2.9$ & $5.96$ $\pm$ $0.98$\\
$W$+jets & $1850$ $\pm$ $790$ & $131$ $\pm$ $57$ & $5.8$ $\pm$ $2.7$\\
$Z$+jets & $400$ $\pm$ $200$ & $29$ $\pm$ $14$ & $1.47$ $\pm$ $0.76$\\
Single top & $1880$ $\pm$ $740$ & $195$ $\pm$ $78$ & $8.3$ $\pm$ $3.1$\\
Diboson & $96$ $\pm$ $41$ & $8.0$ $\pm$ $3.5$ & $0.40$ $\pm$ $0.19$\\
Multijet & $450$ $\pm$ $160$ & $68$ $\pm$ $24$ & $8.3$ $\pm$ $3.0$\\
\hline
Total background & $48400$ $\pm$ $7800$          & $5700$ $\pm$ $1100$          & $252$ $\pm$ $75$         \\
\hline
Data & $49699$ & $6199$ & $286$\\
\hline\hline    
\end{tabular}
\vspace{0.2cm}

\begin{tabular}{l*{3}{c}}
\hline\hline
 & $\geq$ 6 j, 2 b & $\geq$6 j, 3 b & $\geq$6 j, $\geq$4 b\\
\hline
$\Hc$ & $267$ $\pm$ $68$ & $145$ $\pm$ $37$ & $31.1$ $\pm$ $8.3$\\
$\Hu$ & $259$ $\pm$ $67$ & $132$ $\pm$ $34$ & $10.3$ $\pm$ $2.8$\\
\hline
$t\bar{t}$+light-jets & $18800$ $\pm$ $4800$ & $2000$ $\pm$ $730$ & $52$ $\pm$ $40$\\
$t\bar{t}+c\bar{c}$ & $3700$ $\pm$ $2000$ & $850$ $\pm$ $500$ & $79$ $\pm$ $46$\\
$t\bar{t}+b\bar{b}$ & $1430$ $\pm$ $760$ & $970$ $\pm$ $520$ & $240$ $\pm$ $130$\\
$t\bar{t}V$ & $182$ $\pm$ $32$ & $44.6$ $\pm$ $8.1$ & $8.4$ $\pm$ $1.7$\\
$t\bar{t}H$ & $64.2$ $\pm$ $8.2$ & $39.8$ $\pm$ $5.4$ & $16.1$ $\pm$ $2.6$\\
$W$+jets & $880$ $\pm$ $440$ & $95$ $\pm$ $47$ & $8.5$ $\pm$ $4.5$\\
$Z$+jets & $180$ $\pm$ $100$ & $19$ $\pm$ $11$ & $1.5$ $\pm$ $0.9$\\
Single top & $840$ $\pm$ $410$ & $122$ $\pm$ $62$ & $11.9$ $\pm$ $6.2$\\
Diboson & $50$ $\pm$ $26$ & $6.0$ $\pm$ $3.0$ & $0.54$ $\pm$ $0.29$\\
Multijet & $176$ $\pm$ $62$ & $20.3$ $\pm$ $7.2$ & $0.93$ $\pm$ $0.50$\\
\hline
Total background & $26400$ $\pm$ $6100$          & $4200$ $\pm$ $1200$          & $420$ $\pm$ $160$         \\
\hline
Data & $26185$ & $4701$ & $516$\\
\hline\hline    
\end{tabular}
\vspace{0.2cm}

\end{center}
\vspace{-0.5cm}
\caption{
$\Hqbb$ search: predicted and observed yields in each of the analysis channels considered.
The prediction is shown before the fit to data. Also shown are the signal expectations for 
$\Hc$ and $\Hu$ assuming $\BR(t\to Hc)=1\%$ and $\BR(t\to Hu)=1\%$ respectively.
The $\ttbar \to WbWb$ background is normalised to the SM prediction.
The quoted uncertainties are the sum in quadrature of statistical and systematic uncertainties on the yields.
}
\label{tab:Prefit_Yields_Unblind}
\end{table} 

\begin{table}
\small
\begin{center}
\begin{tabular}{l*{3}{c}}
\hline\hline
 & 4 j, 2 b & 4 j, 3 b & 4 j, 4 b\\
\hline
$\Hc$ & $155$ $\pm$ $11$ & $68.5$ $\pm$ $5.0$ & $7.23$ $\pm$ $0.56$\\
\hline
$t\bar{t}$+light-jets & $77300$ $\pm$ $1700$ & $6240$ $\pm$ $190$ & $56.1$ $\pm$ $7.3$\\
$t\bar{t}+c\bar{c}$ & $5600$ $\pm$ $1500$ & $810$ $\pm$ $210$ & $23.4$ $\pm$ $6.0$\\
$t\bar{t}+b\bar{b}$ & $2420$ $\pm$ $360$ & $890$ $\pm$ $130$ & $54.1$ $\pm$ $7.5$\\
$t\bar{t}V$ & $122$ $\pm$ $19$ & $15.6$ $\pm$ $2.5$ & $0.90$ $\pm$ $0.15$\\
$t\bar{t}H$ & $30.9$ $\pm$ $3.6$ & $12.8$ $\pm$ $1.6$ & $1.92$ $\pm$ $0.26$\\
$W$+jets & $4900$ $\pm$ $1100$ & $231$ $\pm$ $55$ & $5.8$ $\pm$ $1.5$\\
$Z$+jets & $1040$ $\pm$ $390$ & $48$ $\pm$ $18$ & $0.78$ $\pm$ $0.32$\\
Single top & $5100$ $\pm$ $1000$ & $352$ $\pm$ $71$ & $7.1$ $\pm$ $1.5$\\
Diboson & $209$ $\pm$ $72$ & $11.6$ $\pm$ $4.0$ & $0.22$ $\pm$ $0.08$\\
Multijet & $1120$ $\pm$ $320$ & $75$ $\pm$ $22$ & $2.41$ $\pm$ $0.70$\\
\hline
Total & $98070$ $\pm$ $370$          & $8756$ $\pm$ $98$          & $159.9$ $\pm$ $7.4$         \\
\hline
Data & $98049$ & $8752$ & $161$\\
\hline\hline      
\end{tabular}
\vspace{0.2cm}

\begin{tabular}{l*{3}{c}}
\hline\hline
 & 5 j, 2 b & 5 j, 3 b & 5 j, $\geq$4 b\\
\hline
$\Hc$ & $85$ $\pm$ $10$ & $42.4$ $\pm$ $5.1$ & $6.17$ $\pm$ $0.77$\\
\hline
$t\bar{t}$+light-jets & $37600$ $\pm$ $1200$ & $3570$ $\pm$ $160$ & $66.4$ $\pm$ $8.1$\\
$t\bar{t}+c\bar{c}$ & $4700$ $\pm$ $1200$ & $970$ $\pm$ $240$ & $59$ $\pm$ $16$\\
$t\bar{t}+b\bar{b}$ & $2140$ $\pm$ $310$ & $1150$ $\pm$ $160$ & $144$ $\pm$ $17$\\
$t\bar{t}V$ & $145$ $\pm$ $23$ & $26.6$ $\pm$ $4.2$ & $3.12$ $\pm$ $0.50$\\
$t\bar{t}H$ & $41.0$ $\pm$ $4.5$ & $22.4$ $\pm$ $2.6$ & $5.98$ $\pm$ $0.74$\\
$W$+jets & $1940$ $\pm$ $560$ & $140$ $\pm$ $41$ & $6.0$ $\pm$ $1.8$\\
$Z$+jets & $380$ $\pm$ $170$ & $27.6$ $\pm$ $12.4$ & $1.38$ $\pm$ $0.63$\\
Single top & $2090$ $\pm$ $630$ & $219$ $\pm$ $66$ & $10.1$ $\pm$ $3.1$\\
Diboson & $93$ $\pm$ $39$ & $7.8$ $\pm$ $3.3$ & $0.37$ $\pm$ $0.16$\\
Multijet & $332$ $\pm$ $96$ & $46$ $\pm$ $13$ & $6.2$ $\pm$ $1.9$\\
\hline
Total & $49570$ $\pm$ $250$          & $6223$ $\pm$ $66$          & $308$ $\pm$ $11$         \\
\hline
Data & $49699$ & $6199$ & $286$\\
\hline\hline      
\end{tabular}
\vspace{0.2cm}

\begin{tabular}{l*{3}{c}}
\hline\hline
 & $\geq$6 j, 2 b & $\geq$6 j, 3 b & $\geq$6 j, $\geq$4 b\\
\hline
$\Hc$ & $46.1$ $\pm$ $4.2$ & $25.0$ $\pm$ $2.3$ & $5.41$ $\pm$ $0.52$\\
\hline
$t\bar{t}$+light-jets & $18590$ $\pm$ $800$ & $2080$ $\pm$ $140$ & $54.2$ $\pm$ $8.4$\\
$t\bar{t}+c\bar{c}$ & $3820$ $\pm$ $920$ & $980$ $\pm$ $240$ & $85$ $\pm$ $20$\\
$t\bar{t}+b\bar{b}$ & $1860$ $\pm$ $270$ & $1260$ $\pm$ $170$ & $320$ $\pm$ $35$\\
$t\bar{t}V$ & $178$ $\pm$ $27$ & $43.7$ $\pm$ $6.8$ & $8.3$ $\pm$ $1.3$\\
$t\bar{t}H$ & $64.0$ $\pm$ $7.2$ & $39.6$ $\pm$ $4.5$ & $15.9$ $\pm$ $1.8$\\
$W$+jets & $680$ $\pm$ $220$ & $75$ $\pm$ $24$ & $7.0$ $\pm$ $2.7$\\
$Z$+jets & $159$ $\pm$ $78$ & $16.8$ $\pm$ $8.3$ & $1.48$ $\pm$ $0.75$\\
Single top & $740$ $\pm$ $270$ & $108$ $\pm$ $40$ & $10.6$ $\pm$ $4.0$\\
Diboson & $48$ $\pm$ $23$ & $5.7$ $\pm$ $2.7$ & $0.51$ $\pm$ $0.25$\\
Multijet & $120$ $\pm$ $34$ & $13.9$ $\pm$ $4.0$ & $1.12$ $\pm$ $0.65$\\
\hline
Total & $26300$ $\pm$ $160$          & $4652$ $\pm$ $62$          & $508$ $\pm$ $22$         \\
\hline
Data & $26185$ & $4701$ & $516$\\
\hline\hline      
\end{tabular}
\vspace{0.2cm}

\end{center}
\vspace{-0.5cm}
\caption{
$\Hcbb$ search: predicted and observed yields in each of the analysis channels considered.
The background prediction is shown after the fit to data under the signal-plus-background hypothesis.
The quoted uncertainties are the sum in quadrature of statistical and systematic uncertainties on the yields, 
computed taking into account correlations among nuisance parameters and among processes.
}
\label{tab:Postfit_Yields_Unblind_Hc}
\end{table}

\section{Pre-fit event yields in the $\HqML$ search}
\label{sec:ML_yields_appendix}

Table~\ref{tab:Prefit_Yields_ML} presents the observed and predicted yields in each of the analysis channels for the $\HqML$ search before the fit to data.

\begin{table}
\small
\begin{center}
\begin{tabular}{lccc}
\hline\hline
 & $ee$4j & $e\mu$4j & $\mu\mu$4j\\
\hline
$t\bar t \to WbHc$ & 5.4 $^{+0.8}_{-0.6}$ & 15.0 $^{+1.7}_{-1.4}$ & 11.1
$^{+1.1}_{-1.2}$ \\
$t\bar t \to WbHu$ & 5.5 $\pm$ 0.7 & 15.9 $^{+1.7}_{-1.5}$ & 9.7
$^{+1.5}_{-1.0}$ \\
\hline
Non-prompt & 3.4 $\pm$ 1.7 & 12 $\pm$ 4 & 6.3 $\pm$ 2.6\\
$q$ mis-id & 1.8 $\pm$ 0.7 & 1.4 $\pm$ 0.6 & -- \\
$t\bar t W$ & 2.0 $\pm$ 0.4 & 6.2 $\pm$ 1.0 & 4.7 $\pm$ 0.9\\
$t\bar t (Z/\gamma^*)$ & 0.75 $\pm$ 0.20 & 1.5 $\pm$ 0.3 & 0.80 $\pm$ 0.22\\
Diboson & 0.7 $\pm$ 0.4 & 1.9 $\pm$ 1.0 & 0.53 $\pm$ 0.30\\
$t \bar t H$ & 0.44 $\pm$ 0.06 & 1.16 $\pm$ 0.14 & 0.74 $\pm$ 0.10\\
Rare & 0.25 $^{+0.04}_{-0.02}$ & 0.72 $\pm$ 0.05 & 0.34 $^{+0.04}_{-0.03}$ \\
\hline
Total background & 9.5 $\pm$ 2.1 & 25 $\pm$ 5 & 13.4 $\pm$ 2.9\\
\hline
Data & 9 & 26 & 20 \\
\hline\hline 
\end{tabular}
\vspace{0.2cm}

\begin{tabular}{lccc}
\hline\hline
 & $ee$$\geq$5j & $e\mu$$\geq$5j  & $\mu\mu$$\geq$5j \\
\hline
$t\bar t \to WbHc$ & 2.9 $\pm$ 0.6 & 10.2 $^{+2.0}_{-1.6}$ & 6.6
$^{+1.5}_{-1.3}$ \\
$t\bar t \to WbHu$ & 2.7 $\pm$ 0.8 & 8.2 $^{+1.6}_{-1.2}$ & 7.2
$^{+1.2}_{-1.4}$ \\
\hline
Non-prompt & 2.3 $\pm$ 1.2 & 6.7 $\pm$ 2.4 & 2.9 $\pm$ 1.4\\
$q$ mis-id & 1.1 $\pm$ 0.5 & 0.85 $\pm$ 0.35 & -- \\
$t\bar t W$ & 1.4 $\pm$ 0.4 & 4.8 $\pm$ 1.2 & 3.8 $\pm$ 0.9\\
$t\bar t (Z/\gamma^*)$ & 0.98 $\pm$ 0.26 & 2.1 $\pm$ 0.5 & 0.95 $\pm$ 0.25\\
Diboson &  0.47 $\pm$ 0.29 & 0.38 $\pm$ 0.30 & 0.7 $\pm$ 0.4\\
$t \bar t H$ & 0.73 $\pm$ 0.14 & 2.1 $\pm$ 0.4 & 1.41 $\pm$ 0.28\\
Rare & 0.27 $\pm$ 0.02 & 0.79 $^{+0.05}_{-0.04}$ & 0.38 $\pm$ 0.02\\
\hline
Total background & 7.2 $\pm$ 1.8 & 17 $\pm$ 3 & 10.0 $\pm$ 2.2\\
\hline
Data & 10 & 22 & 11 \\
\hline\hline 
\end{tabular}
\vspace{0.2cm}

\begin{tabular}{lcc}
\hline\hline
 & $3\ell$ & $2\ell 1\tau_{\rm had}$ \\
\hline
$t\bar t \to WbHc$ &  6.4 $\pm$ 0.9 & 2.4 $\pm$ 0.4\\
$t\bar t \to WbHu$ & 5.2 $^{+1.0}_{-0.8}$ & 2.1 $^{+0.5}_{-0.4}$\\
\hline
Non-prompt & 3.2 $\pm$ 0.7 & 0.4 $^{+0.6}_{-0.4}$\\
$q$ mis-id & -- & -- \\
$t\bar t W$ & 2.3 $\pm$ 0.7 & 0.38 $\pm$ 0.12 \\
$t\bar t (Z/\gamma^*)$ & 3.9 $\pm$ 0.8 & 0.37 $\pm$ 0.08\\
Diboson & 0.86 $\pm$ 0.55 & 0.12 $\pm$ 0.11\\
$t \bar t H$ & 2.34 $\pm$ 0.35 & 0.47 $\pm$ 0.08\\
Rare & 0.92 $^{+0.07}_{-0.06}$ & 0.10 $^{+0.02}_{-0.01}$\\
\hline
Total background & 13.7 $\pm$ 2.3 & 1.9 $\pm$ 0.6\\
\hline
Data & 18 & 1\\
\hline\hline 
\end{tabular}
\vspace{0.2cm}

\end{center}
\vspace{-0.5cm}
\caption{
$\HqML$ search: predicted and observed yields in each of the event categories considered.
The prediction is shown before the fit to data. Also shown are the signal expectations for 
$\Hc$ and $\Hu$ assuming $\BR(t\to Hc)=1\%$ and $\BR(t\to Hu)=1\%$ respectively.
The quoted uncertainties are the sum in quadrature of statistical and systematic uncertainties on the yields.
}
\label{tab:Prefit_Yields_ML}
\end{table}


\FloatBarrier

\printbibliography

%
\newpage 
\begin{flushleft}
{\Large The ATLAS Collaboration}

\bigskip

G.~Aad$^{\rm 85}$,
B.~Abbott$^{\rm 113}$,
J.~Abdallah$^{\rm 151}$,
O.~Abdinov$^{\rm 11}$,
R.~Aben$^{\rm 107}$,
M.~Abolins$^{\rm 90}$,
O.S.~AbouZeid$^{\rm 158}$,
H.~Abramowicz$^{\rm 153}$,
H.~Abreu$^{\rm 152}$,
R.~Abreu$^{\rm 116}$
% ...
\bigskip
\\
% ...
$^{11}$ Institute of Physics, Azerbaijan Academy of Sciences, Baku, Azerbaijan\\
% ...
$^{85}$ CPPM, Aix-Marseille Universit{\'e} and CNRS/IN2P3, Marseille, France\\
% ...
$^{90}$ Department of Physics and Astronomy, Michigan State University, East Lansing MI, United States of America\\
% ...
$^{107}$ Nikhef National Institute for Subatomic Physics and University of Amsterdam, Amsterdam, Netherlands\\
% ...
$^{113}$ Homer L. Dodge Department of Physics and Astronomy, University of Oklahoma, Norman OK, United States of America\\
% ...
$^{116}$ Center for High Energy Physics, University of Oregon, Eugene OR, United States of America\\
% ...
$^{151}$ Institute of Physics, Academia Sinica, Taipei, Taiwan\\
$^{152}$ Department of Physics, Technion: Israel Institute of Technology, Haifa, Israel\\
$^{153}$ Raymond and Beverly Sackler School of Physics and Astronomy, Tel Aviv University, Tel Aviv, Israel\\
% ...
$^{158}$ Department of Physics, University of Toronto, Toronto ON, Canada\\
% ...
\end{flushleft}



%

\end{document}